\pdfoutput=1
\documentclass[prl,reprint, epsfig,amssymb,longbibliography,amsmath]{revtex4-1}

\usepackage{blkarray}

\usepackage{amssymb,amsmath,bm,epsfig,graphicx,subfigure,hyperref}

\usepackage[utf8]{inputenc}
\usepackage[vietnamese, english]{babel}
\usepackage{chngcntr}
\usepackage{xcolor}
\usepackage{braket}
\usepackage{physics}
\usepackage{sidecap}
\usepackage{lipsum}

\begin{document}

\title {Obstruction and Interference in Low Energy Models for Twisted Bilayer Graphene}
\author{\foreignlanguage{vietnamese}{Võ Tiến Phong}}
\author{E. J. Mele}
\email{mele@physics.upenn.edu}
\affiliation{Department of Physics and Astronomy, University of Pennsylvania, Philadelphia PA 19104}
\date{\today}

\begin{abstract}
The electronic bands of twisted bilayer graphene (TBLG) with a large-period moiré superlattice fracture to form narrow Bloch minibands that are spectrally isolated  by forbidden energy gaps from remote dispersive bands. When these gaps are sufficiently large, one can study a band-projected Hamiltonian that correctly represents the dynamics within the minibands. This inevitably introduces nontrivial geometrical constraints that arise from the assumed form of the projection. Here we show that this choice has a profound consequence in a low-energy experimentally-observable signature which therefore can be used to tightly constrain the analytic form of the appropriate low-energy theory. We find that this can be accomplished by a careful analysis of the electron density produced by backscattering of Bloch waves from an impurity potential localized on the moiré superlattice scale.  We provide numerical estimates of the effect that can guide experimental work to clearly discriminate between competing models for the low-energy band structure.
\end{abstract}

\maketitle

 Twisted van der Waals heterostructures with large-period moiré supelattices are versatile platforms for exploring narrow-band physics, and the role of interactions in ground state selection and its excitations. Famously, in magic-angle graphene  with a rotation angle $\sim 1^{\rm \circ},$ the narrow bands near charge neutrality are tuned to a nearly flat condition, and various fractional band fillings are found to support interaction-driven insulating, superconducting, and magnetic states of matter \cite{CFF18, CFD18, YC19, LSY19, SS19}. To study the role of these interactions, it is a practical necessity to develop effective low-energy models that faithfully represent the spectral and topological properties. This is, however, a nontrivial task because the topology of the resulting effective model depends crucially on the choice of symmetries one retains in the low-energy projection. Indeed, in the current literature for twisted bilayer graphene (TBLG), there are broadly three classes of such models that are either Wannier-representable (described by hopping processes between basis states exponentially localized on Wyckoff centers) \cite{KV18, KF18}, fundamentally non-Wannier representable (prevented by a topological momentum space obstruction in the band structure) \cite{ZP18,PZ18}, or topologically fragile (an obstruction exists but can be removed by adding a few extra bands) \cite{ZP18,SW19, APY19}. In this work, we propose an experimentally-observable signature which can be used to distinguish between these incompatible models. Motivated by a recent experimental demonstration that the interference pattern of the backscattering of Bloch waves from an impurity can carry information about the Berry phase \cite{DG19}, we propose to distinguish between these different models by carefully analyzing the dark-field reconstruction of the induced change in the local density of states (LDOS) by the presence of a localized impurity on the moiré scale.

\begin{figure}[!h]
\includegraphics[scale=0.186]{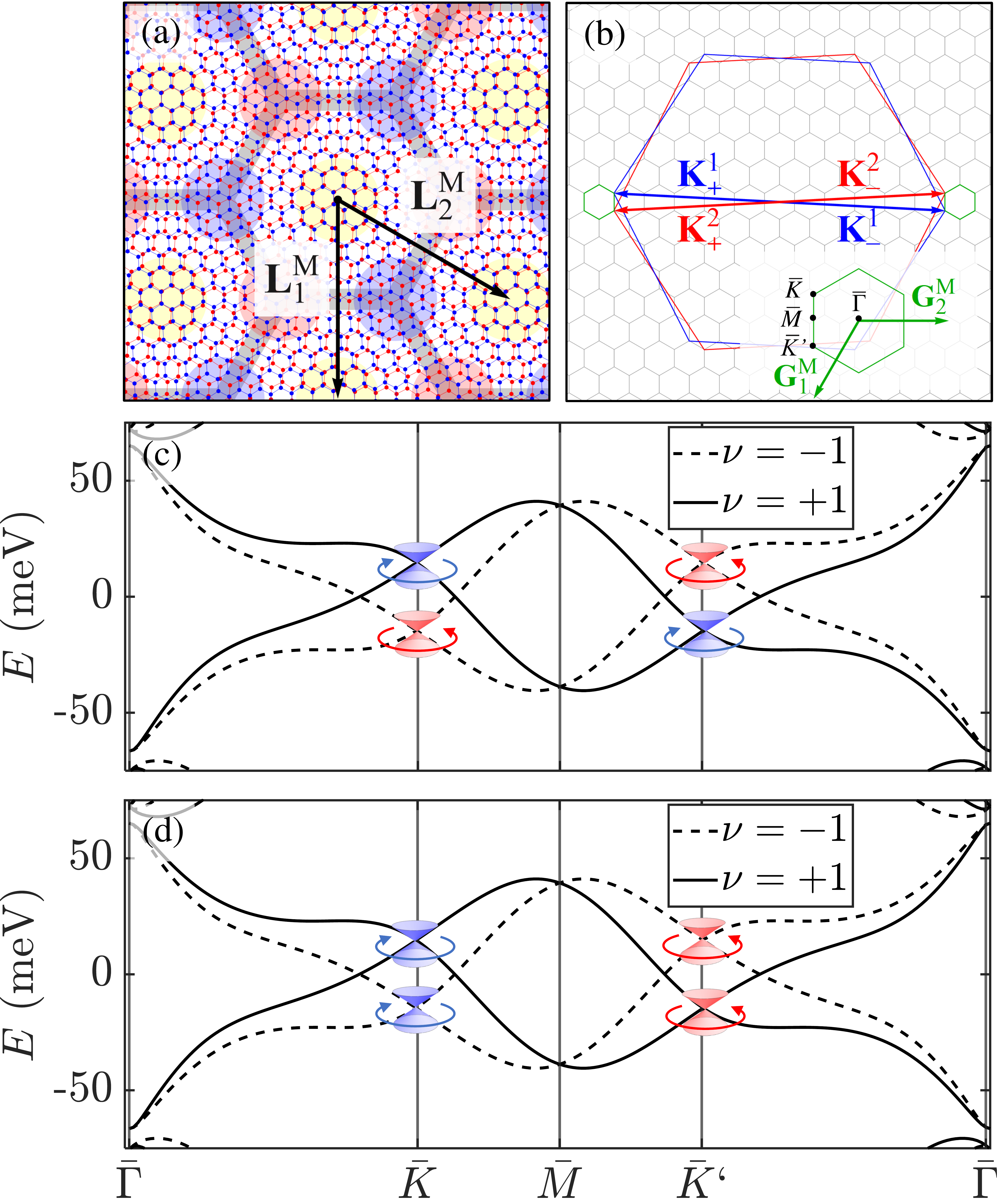}
\caption{\textbf{Real-space, momentum-space, and spectral representation of TBG.} (a) Lattice structure of TBLG formed by starting with an $AA$-stacked bilayer system and then twisting one layer relative to the other by an angle $\theta.$ The resulting structure is a long-wavelength moiré pattern that has $AA$ regions where the two layers are approximately aligned (yellow-shaded),  $AB$ regions  where the $A$ sublattice of the top layer is aligned with the $B$ sublattice of the bottom layer (red-shaded), and $BA$ regions where the $B$ sublattice of the top layer is aligned with the $A$ sublattice of the bottom layer (blue-shaded). The emergent moiré pattern forms a honeycomb lattice with lattice vectors $\mathbf{L}_1^\text{M}$ and $\mathbf{L}_2^\text{M}.$ (b) Mini-Brillouin zone (mBZ) of TBLG formed by the momentum-mismatch of the original Brillouin zones. The blue and red hexagons show the monolayer BZs rotating in opposite directions. The mBZ is shown in green with high-symmetry points labeled with an overline, $\bar{K},$ $\bar{K}',$ $\bar{M},$ and $\bar{\Gamma},$ and reciprocal lattice vectors indicated as $\mathbf{G}_1^\text{M}$ and $\mathbf{G}_2^\text{M}.$ (c)-(d) Band structure of TBLG  numerically calculated along high-symmetry lines from the continuum model with $\theta = 2^\circ,$ $w_\text{AA} = 79.7$ meV, $w_\text{AB} = 97.5$ meV, and $\hbar v_F/a = 2135.4$ meV. We also apply an interlayer bias $V = 200$ meV to shift the Dirac cones within a single valley in opposite directions in energy to improve visibility. The dash and solid lines are energy bands for the $ -$ and $+$ valleys respectively. We observe two Dirac cones in each valley at $\bar{K}$ and $\bar{K}'$. We indicate the chirality of the band crossings schematically by blue and red cones. Dirac cones in the same valley have the same chirality in (c), corresponding to the topology of the continuum model, and opposite chirality in (d), corresponding to the topology of a two-orbital tight-binding model. Dirac cones in different valleys are related by $T$ symmetry. }
\label{fig1}
\end{figure}

In TBLG, the emergent long-wavelength moiré structure contains $AA,$ $AB,$ and $BA$ regions as shown in Fig.~\ref{fig1}a. In the small-angle limit, the band structure of TBLG at low energies is dominated by strong hybridization of the monolayer Dirac cones induced by interlayer coherence \cite{dSPN07, SCV10, M10, BM11, KF18, GW19}. The interlayer hopping is conventionally modeled as a smooth local matrix-valued potential acting on the layer and sublattice degrees of freedom that interpolates between the $AA$ and $AB/BA$ registries \cite{dSPN07, M10, BM11, GW19}. In such a \textit{continuum} theory, microscopic symmetries of the moiré lattice structure are neglected in favor of emergent symmetries that are approximately preserved at long wavelengths \cite{AMV18, PZ18, ZP18,  SW19}. Additionally, because the two microscopic valleys are usually well-separated in momentum space as shown in Fig.~\ref{fig1}b, a continuum theory typically assumes they are not mixed, introducing a $U_\nu(1)$ valley symmetry when the Hamiltonian is decoupled into two independent valley sectors that are related by time-reversal $T$ symmetry. In a single valley, the projected Hamiltonian breaks $T$ symmetry, but retains composite $C_2T$ symmetry and $D_3$ point symmetries. When both valleys are considered together, their superposition recovers the full $D_6$ symmetry group of the lattice as well its $T$ symmetry. While none of these symmetries are generally exact, one expects them to be good approximations as long as the relevant physics is insensitive to microscopic details. Valley projection symmetry is of particular importance in our analysis, and will be assumed throughout \cite{KV18}.

Diagonalizing the (spinless) continuum Hamiltonian leads to two energy bands per valley near charge neutrality that are spectrally isolated from the rest of the band structure, which form the so-called flat bands. Much work has been devoted to characterizing their topology in hope of identifying a compact effective low-energy description that includes only two orbitals per valley \cite{KF18,PZ18, ZP18, APY19, SW19, PZSV19}. This turns out to be a delicate task. If one were to enforce the emergent symmetries of the continuum model, then  two Dirac cones form within a single valley with the same chirality. Heuristically, this can be understood as inherited  from the chirality of the original two {\it monolayer} Dirac cones in the same valley. More precisely, by having opposite mirror eigenvalues for the two bands at an $\bar{M}$ point, the form the Hamiltonian written in a two-component chiral representation near $\bar{K}$ and $\bar{K}'$ is forced by $M_y$ symmetry to have the same phase winding \cite{PZ18, ZP18}, as illustrated in Fig.~\ref{fig1}c. This observation prevents the construction of a local two-orbital tight-binding model defined on a honeycomb lattice which would require two Dirac cones in a single valley with opposite chirality,  as indeed is found for the band structure of monolayer graphene.

Alternatively, one might neglect the emergent symmetries altogether and argue that a generic sample of TBLG  usually has no exact $M_y$ symmetry because the twist center that determines the microscopic point symmetries is never under control experimentally. Then, one can posit exponentially-localized Wannier orbitals centered at the $AB$ and $BA$ regions. In this case, the resulting tight-binding model constructed from these Wannier orbitals will indeed carry opposite chirality, as illustrated in Fig.~\ref{fig1}d. This approach is appealing because it yields a simple two-orbital model that serves as the starting point for many studies investigating electron interactions in TBLG \cite{FZS18, VF18, LN18, TCS18, ZSF19, DMX19}. However, it comes at the cost of relieving the $C_2 T$ symmetry protection of the  linear band crossings.

 When a localized impurity is present in a solid, backscattering from the impurity potential produces a characteristic interference pattern in LDOS known as Friedel oscillations. The period of these radial oscillations encode information about the Fermi surface. As a result, Friedel oscillations have  been used successfully to reconstruct band dispersion \cite{CLE93}. Recently, it has been proposed and experimentally demonstrated that these oscillations also encode {\it Berry-phase information} \cite{DD17, DG19}, rendering them a crucial diagnostic tool for examining momentum-space topology. As this is the crucial feature distinguishing the low energy models outlined above, here we adapt this insight to study Friedel oscillations in TBLG as a probe of the chiral structure of its flat bands.

First, we consider a valley-polarized two-orbital model that describes the low-energy spectrum of TBLG. This model is formally equivalent to the  model of monolayer graphene, and as such, it can be represented using two exponentially-localized Wannier orbitals per valley centered at the $AB$ and $BA$ regions. In the Bloch basis, the Hamiltonian  expanded to linear order in momentum around the zone corners is
\begin{equation}
\label{eq: Eq1}
\begin{split}
\mathcal{H} \left( \bar{\mathbf{K}} + \mathbf{q} \right) &= -\hbar \tilde{v}_F \mathbf{q} \cdot \boldsymbol{\sigma}^*, \\
\mathcal{H} \left( \bar{\mathbf{K}}' + \mathbf{q} \right) &= -\hbar \tilde{v}_F \mathbf{q} \cdot \boldsymbol{\sigma},
\end{split}
\end{equation}
where $\boldsymbol{\sigma} = \left( \sigma_x, \sigma_y \right)$ are Pauli matrices, $\tilde{v}_F$ is the renormalized velocity that depends on interlayer hopping amplitudes $w_\text{AA}$ and $w_\text{AB},$ and $\mathbf{q}$ is measured from the respective zone corners. Written in the chiral representation, $\mathcal{H}\left( \mathbf{k} \right) = \mathbf{d} \left( \mathbf{k} \right) \cdot \boldsymbol{\sigma}$ for some vector-valued function $\mathbf{d}(\mathbf{k}),$ the chirality of some two-fold degenerate point $\mathbf{k}_c$ is defined as the integer winding number of $\mathbf{d}(\mathbf{k})$ as $\mathbf{k}$ goes around any simple closed  loop that contains $\mathbf{k}_c.$ In equation~\eqref{eq: Eq1}, $\mathbf{d}(\mathbf{q}) = \left(q_x, -q_y  \right)$  near $\bar{\mathbf{K}}$ and $\mathbf{d}(\mathbf{q}) = \left(q_x, q_y  \right)$  near $\bar{\mathbf{K}}',$ which shows that the Dirac cones carry opposite chirality in this model. For Bloch wavefunctions with wavevectors near a Dirac cone, the chirality defines the relative phase between the two sublattices. This phase difference will be crucial in the consideration of scattering processes induced by the presence of an impurity, especially in the detection of the relative chirality between two Dirac cones.

\begin{figure}[!h]
\includegraphics[scale=0.2]{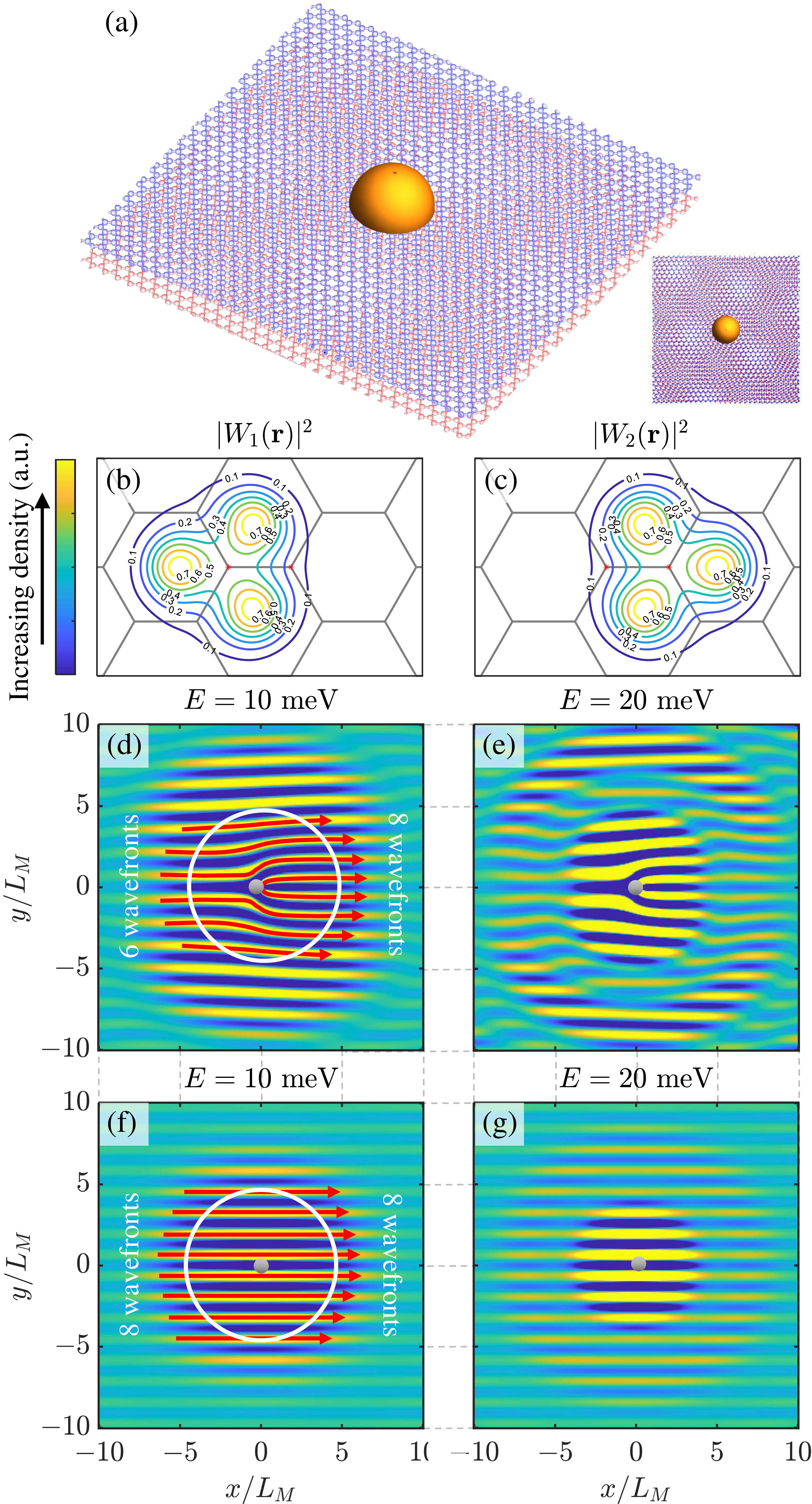}
\caption{\textbf{Wannier orbitals and impurity-induced LDOS in topologically-inequivalent effective models of the flat bands.} (a) Schematic representation of a possible experimental setup in which an impurity is placed atop an $AB$ region. (b)-(c) The spatial distribution of the two Wannier orbitals in a single valley of TBLG. $\ket{W_1}$ and $\ket{W_2}$ are centered at an $AB$ and $BA$ region respectively; however, most of their density is concentrated at the adjacent $AA$ regions, corresponding to the observed density peaks seen in experiments. Because of that, the wavefunctions of these Wannier orbitals are nonlocal in space. (d)-(e) The LDOS simulated from inter-Dirac-cone scatterings at a bias energy $E = 10$ meV  and $E = 20$ meV for the Wannier-representable model where the chirality at the two Dirac cones is opposite. We observe two wavefront dislocations in the interference pattern. In addition, we also observe that as the probing energy increases, the period of radial oscillations decreases. (f)-(g) The LDOS simulated from inter-Dirac-cone scatterings for the Wannier-obstructed model where the chirality at the two Dirac cones is opposite. Unlike before, here, we observe no wavefront dislocations.   }
\label{fig2}
\end{figure}

We now place a spatially-localized impurity atop one of the Wannier orbitals, as shown schematically in Fig.~\ref{fig2}a. Suppose this impurity contains a resonant bound state  with energy $U_0$ that has significant wavefunction overlap with only the Wannier orbital on which it sits, say the $AB$ orbital, then the scattering potential in the basis of equation~\eqref{eq: Eq1} simplifies to
\begin{equation}
\label{eq2}
\mathcal{U}(\mathbf{r},\mathbf{r}') \approx U_0 \begin{pmatrix}
1 & 0 \\
0 & 0
\end{pmatrix} \delta^{(2)}(\mathbf{r}) \delta^{(2)}(\mathbf{r}').
\end{equation}
Using this approximation, we can calculate the induced change in LDOS using the Green's function formalism. This framework is especially convenient when the impurity can be modeled as an delta impurity as in equation~\eqref{eq2}. In our context, we only require that it is localized on the moiré scale rather than the atomic scale.

\begin{figure*}[htbp]
\centering
\includegraphics[width=\textwidth]{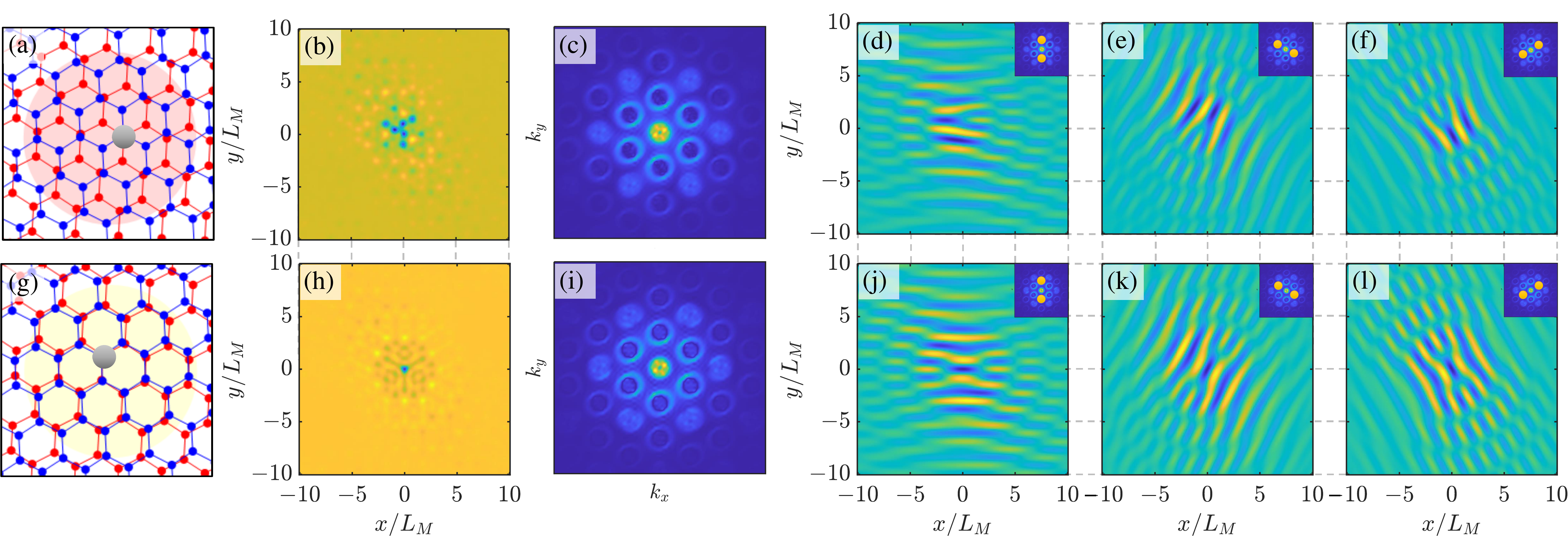}
\caption{\label{fig3}\textbf{The change in LDOS induced by an atomic impurity simulated by the continuum model.} Each row shows the change in LDOS induced by an impurity placed on an $AB$ region, as in (a) for the top row, and on an $AA$ region, as in (g) for the bottom row. (b) and (h) show the full interference pattern, and the magnitudes of their fast Fourier transforms (FFT) are plotted in (c) and (i). From the FFT, we filter out time-reversed pairs of momenta in three independent directions, and then calculate the inverse FFT to observe the density wavefronts. (d)-(f) and (j)-(i) show the wavefront interference pattern after applying corresponding FFT filters indicated on the insets. As can clearly be seen, (d)-(f) feature one wavefront dislocation. On the contrary, (j)-(l) do not have any dislocation. This is consistent with the interpretation that when the impurity is placed on an $AB$ region, it overlaps with the host Wannier orbital as well as the three nearest-neighbor Wannier states, demonstrating the non-negligible spatial spread of the   Wannier states into the adjacent sublattice. }
\end{figure*}

The change in LDOS at energy $E$ is given by $\Delta \rho(E,\mathbf{r}) = - \pi^{-1}\Im \text{Tr } \left[\mathcal{G}^{(0)} \left(E, \mathbf{r}  \right) \mathcal{T}(E) \mathcal{G}^{(0)} \left(E, -\mathbf{r}  \right)  \right],$ where $\mathcal{G}^{(0)} \left(E, \mathbf{r}  \right)$ is the bare Green's function, and $\mathbf{r}$ is measured from the location of the impurity. The trace is taken over the Wannier sublattice degree of freedom, which amounts to performing a unit-cell average. When $E$ is sufficiently close to the energy of the Dirac cones, we use equation~\eqref{eq: Eq1} to calculate the bare Green's function exactly \cite{SM}. We find that the momentum-space phase in the Hamiltonian that defines the chirality is mapped to a real-space phase in the bare Green's function upon integration over all momenta near the Dirac cones. For scattering processes that exchange momentum within the same Dirac cone, this real-space phase cancels out, but for scattering processes that exchange momentum between different Dirac cones within a single valley, this real-space phase instead has a non-trivial signature for the interference pattern of  LDOS. Explicitly, we decompose $\Delta \rho(E,\mathbf{r})$ into two parts
\begin{equation}
\label{eq3}
\begin{split}
\Delta \rho(E,\mathbf{r}) &= \Delta \rho_\text{intra}(E,\mathbf{r}) + \Delta \rho_\text{inter}(E,\mathbf{r}) ,\\
\Delta \rho_\text{intra}(E,\mathbf{r}) &\propto \Im \left[ t(E) \left(K_0^2 \left( r /i\ell  \right)-  K_1^2 \left(  r /i\ell  \right) \right) \right], \\
\Delta \rho_\text{inter}(E,\mathbf{r})  & \propto\Im \left[ t(E) K_1^2 \left( r/i \ell \right)  \right] \cos \left( \Delta \bar{\mathbf{K}} \cdot \mathbf{r} - 2 \phi_\mathbf{r}\right) \\
&- \Im \left[ t(E) K_0^2 \left( r/i \ell \right)  \right] \cos \left( \Delta \bar{\mathbf{K}} \cdot \mathbf{r} \right),
\end{split}
\end{equation}
where $\Delta \bar{\mathbf{K}} = \bar{\mathbf{K}} - \bar{\mathbf{K}}',$ $\ell = \hbar \tilde{v}_F/E,$ $t(E)$ is the non-zero matrix element of $\mathcal{T}(E)$, $\phi_\mathbf{r}$ is the real-space angle, and $K_\alpha(z)$ is the $\alpha^\text{th}$ modified Bessel function of the second kind. From equation~\eqref{eq3}, we observe that $\Delta \rho_\text{intra}(E,\mathbf{r})$ is only a function of the radial direction. $\Delta \rho_\text{inter}(E,\mathbf{r})$ is more interesting; in addition to radial oscillations with period $\ell$ coming from the Bessel functions, there are also wavefronts propagating in the $\Delta \bar{\mathbf{K}}$ direction with period $|\Delta \bar{\mathbf{K}}|^{-1}.$  Importantly, the two oscillatory terms differ by a spatial phase $2 \phi_\mathbf{r}.$ To understand the effect of this phase,  we write  $\cos \left( \Delta \bar{\mathbf{K}} \cdot \mathbf{r} - 2 \phi_\mathbf{r}\right) =\cos \left( \Delta \bar{\mathbf{K}} \cdot \mathbf{r} \right) \cos \left(  2 \phi_\mathbf{r}\right)  +\sin \left( \Delta \bar{\mathbf{K}} \cdot \mathbf{r} \right) \sin \left(  2 \phi_\mathbf{r}\right) ,$ and observe that the wavefronts are modulated by trigonometric functions that change signs \textit{twice} as we go around a simple closed loop, resulting in two wavefront dislocations. If we go around in a circle at a fixed $r$ for which $\Im \left[ t(E) K_0^2 \left( r/i \ell \right)  \right] < \Im \left[ t(E) K_1^2 \left( r/i \ell \right)  \right],$ then the change in  LDOS contains these two wavefront dislocations, as shown in Fig.~\ref{fig2}de. The prefactor of $\phi_\mathbf{r}$ is given by the difference in chirality of the two Dirac cones, and thus the presence of wavefront dislocations in LDOS is a measure of the relative chirality.  The LDOS obtained experimentally will contain all backscattering processes, including those related by reciprocal lattice vectors. Because of that, to observe the density dislocations in practice, we need to filter out only time-reversed pairs of momenta that are near the relevant Dirac cones, as illustrated in Fig.~\ref{fig3}. Then, the dislocations, if present, will be contained in the dark-field reconstruction of the LDOS.

We now consider a different effective model of TBLG in which the flat-band Dirac cones in the same valley have identical chirality. In this case, it is not possible to construct a tight-binding model which  contains only two orbitals and still retains a local representation of the $C_2 T$ symmetry that protects the valley-projected Dirac points. We need to include additional bands to capture the correct topology, as done in \cite{ZP18}. However, these auxiliary bands can be sent to  high energies since they do not correspond to the actual band structure of TBLG. Thus, as long as we are probing at energies close to the Dirac cones, these auxiliary bands can be safely neglected. In this limit, the low-energy effective Hamiltonian is similar to equation~\eqref{eq: Eq1}, but the chirality is identical at the two cones
\begin{equation}
\label{eq: Eq4}
\begin{split}
\mathcal{H} \left( \bar{\mathbf{K}} + \mathbf{q} \right) &= -\hbar \tilde{v}_F \mathbf{q} \cdot \boldsymbol{\sigma}, \\
\mathcal{H} \left( \bar{\mathbf{K}}' + \mathbf{q} \right) &= -\hbar \tilde{v}_F \mathbf{q} \cdot \boldsymbol{\sigma}.
\end{split}
\end{equation}
If we now place an impurity in this system with scattering potential as in equation~\eqref{eq2}, then the induced change in  LDOS will not feature wavefront dislocations previously seen, as shown in Fig.~\ref{fig2}fg. This markedly different interference pattern serves as a diagnostic of the two competing effective models of the valley-projected flat bands in TBLG.

 The Wannier nonlocality can be demonstrated directly from the continuum model without explicit construction of the orbitals. To do so, we place an atomic impurity atop one carbon atom at location $\mathbf{s}$ within a moiré unit cell. The induced change of LDOS is calculated analytically in \cite{SM}. The presence of a dislocation in LDOS depends on $\mathbf{s}.$ When $\mathbf{s}$ is in the $AA$ region, we do not observe a dislocation, but when $\mathbf{s}$ is in a region where the sublattice on which the impurity sits is aligned with another sublattice of the other layer, then we observe one wavefront dislocation. For example, if a hydrogen atom is chemisorbed by an $A$ carbon of the top layer, then we expect to find one phase dislocation if the impurity is in the $AB$ region. This is consistent with the interpretation that an atomic impurity deposited in an $AB$ or $BA$ region must necessarily have non-negligible projection on the host Wannier orbital and the three nearest-neighbor Wannier states. We confirm this with numerical simulation of the LDOS induced by an atomic impurity using the tetrahedron method applied to the full continuum model \cite{SY16}. The results are shown in Fig.~\ref{fig3}.

Our analysis so far relies crucially on the ability to design an impurity that polarizes one Wannier sublattice. This is, however, a nontrivial task because the Wannier orbitals in TBLG are spatially distributed and overlapping \cite{KF18, ZP18, KV18},  as shown in Fig.~\ref{fig2}bc.  It is known that the Wannier orbitals in a single valley have $p_\pm$ symmetry \cite{KF18, ZP18}. So if the resonant bound state carries angular momentum $m \neq  \pm 1,$ then its overlap with the Wannier orbital on which it is centered vanishes. In particular, if we place a $T$-symmetric quantum dot centered on an $AB$ region, then its bound state will only have significant overlap with Wannier orbitals at the {\it three neighboring} $BA$ regions. In this case, the interference patterns induced will be identical to those produced by the scattering potential in equation~\eqref{eq2}. This establishes the relevance of our proposal in experimental settings. Next, we provide some parameter estimates as further motivation, using values from \cite{KF18}. In an STS experiment that probes LDOS, the bias potential must be set at an energy where the linear approximation to the flat bands holds. Approaching the magic angle, this is challenging because the bandwidth there becomes quite small. At slightly larger angles, $\theta = 2^\circ - 1.5^\circ ,$ the bias potential can be several tens of meV, well within experimental capacity. The period of the wavefronts is $|a \Delta \bar{\mathbf{K}}|^{-1} \approx 6 - 8,$ while the decay of the radial oscillation is  $\ell E/a = 500-1000$ meV for $\theta = 2^\circ - 1.5^\circ.$ In order to clearly observe the density dislocations, we must probe at an energy low enough where $\ell >| \Delta \bar{\mathbf{K}}|^{-1} .$ However, it must not be too small that the amplitude of  LDOS is suppressed, $V_\text{cell}\Delta \rho \sim U_0 E^2 / (10^6 \text{ meV}^4)$ for $\theta = 2^\circ - 1.5^\circ,$ where $V_\text{cell}$ is the unit-cell area and $U_0$ is small. The scanning window must span multiple wavefronts in order to observe the dislocations, which is on the order of $100-200$ graphene lattice constants.

We note that this proposal for experimentally  defining the topology of the low-energy bands needs not be restricted to the magic-angle regime. Indeed, there are some advantages to studying other nearby angles where the low-energy bands are predicted to be more dispersive and can be interrogated as the chemical potential is varied. Indeed, the reentrant regime where dispersive bands re-appear at very low rotation angles has not been widely explored and it remains an interesting speculation that the topology of the low energy bands could change through a series of band flattening transitions. Finally,  it will be important to extend this approach to look at the effects of mean field interaction-driven density wave instabilities.

 We thank fruitful conversations with Francisco Guinea, Abhay Pasupathy, and Oskar Vafek. V. T. P. acknowledges financial support from the National Science Foundation through the Graduate Research Fellowship Program and from the P.D. Soros Fellowship for New Americans.  E. J. M.'s work on this problem is supported by the Department of Energy under Grant No. DE-FG02-84ER45118.

\bibliography{arXiv_submission}

\begin{thebibliography}{44}%
\makeatletter
\providecommand \@ifxundefined [1]{%
 \@ifx{#1\undefined}
}%
\providecommand \@ifnum [1]{%
 \ifnum #1\expandafter \@firstoftwo
 \else \expandafter \@secondoftwo
 \fi
}%
\providecommand \@ifx [1]{%
 \ifx #1\expandafter \@firstoftwo
 \else \expandafter \@secondoftwo
 \fi
}%
\providecommand \natexlab [1]{#1}%
\providecommand \enquote  [1]{``#1''}%
\providecommand \bibnamefont  [1]{#1}%
\providecommand \bibfnamefont [1]{#1}%
\providecommand \citenamefont [1]{#1}%
\providecommand \href@noop [0]{\@secondoftwo}%
\providecommand \href [0]{\begingroup \@sanitize@url \@href}%
\providecommand \@href[1]{\@@startlink{#1}\@@href}%
\providecommand \@@href[1]{\endgroup#1\@@endlink}%
\providecommand \@sanitize@url [0]{\catcode `\\12\catcode `\$12\catcode
  `\&12\catcode `\#12\catcode `\^12\catcode `\_12\catcode `\%12\relax}%
\providecommand \@@startlink[1]{}%
\providecommand \@@endlink[0]{}%
\providecommand \url  [0]{\begingroup\@sanitize@url \@url }%
\providecommand \@url [1]{\endgroup\@href {#1}{\urlprefix }}%
\providecommand \urlprefix  [0]{URL }%
\providecommand \Eprint [0]{\href }%
\providecommand \doibase [0]{http://dx.doi.org/}%
\providecommand \selectlanguage [0]{\@gobble}%
\providecommand \bibinfo  [0]{\@secondoftwo}%
\providecommand \bibfield  [0]{\@secondoftwo}%
\providecommand \translation [1]{[#1]}%
\providecommand \BibitemOpen [0]{}%
\providecommand \bibitemStop [0]{}%
\providecommand \bibitemNoStop [0]{.\EOS\space}%
\providecommand \EOS [0]{\spacefactor3000\relax}%
\providecommand \BibitemShut  [1]{\csname bibitem#1\endcsname}%
\let\auto@bib@innerbib\@empty
\bibitem [{\citenamefont {Cao}\ \emph {et~al.}(2018{\natexlab{a}})\citenamefont
  {Cao}, \citenamefont {Fatemi}, \citenamefont {Fang}, \citenamefont
  {Watanabe}, \citenamefont {Taniguchi}, \citenamefont {Kaxiras},\ and\
  \citenamefont {Jarillo-Herrero}}]{CFF18}%
  \BibitemOpen
  \bibfield  {author} {\bibinfo {author} {\bibfnamefont {Yuan}\ \bibnamefont
  {Cao}}, \bibinfo {author} {\bibfnamefont {Valla}\ \bibnamefont {Fatemi}},
  \bibinfo {author} {\bibfnamefont {Shiang}\ \bibnamefont {Fang}}, \bibinfo
  {author} {\bibfnamefont {Kenji}\ \bibnamefont {Watanabe}}, \bibinfo {author}
  {\bibfnamefont {Takashi}\ \bibnamefont {Taniguchi}}, \bibinfo {author}
  {\bibfnamefont {Efthimios}\ \bibnamefont {Kaxiras}}, \ and\ \bibinfo {author}
  {\bibfnamefont {Pablo}\ \bibnamefont {Jarillo-Herrero}},\ }\bibfield  {title}
  {\enquote {\bibinfo {title} {Unconventional superconductivity in magic-angle
  graphene superlattices},}\ }\href@noop {} {\bibfield  {journal} {\bibinfo
  {journal} {Nature}\ }\textbf {\bibinfo {volume} {556}},\ \bibinfo {pages}
  {43--50} (\bibinfo {year} {2018}{\natexlab{a}})}\BibitemShut {NoStop}%
\bibitem [{\citenamefont {Cao}\ \emph {et~al.}(2018{\natexlab{b}})\citenamefont
  {Cao}, \citenamefont {Fatemi}, \citenamefont {Demir}, \citenamefont {Fang},
  \citenamefont {Tomarken}, \citenamefont {Luo}, \citenamefont
  {Sanchez-Yamagishi}, \citenamefont {Watanabe}, \citenamefont {Taniguchi},
  \citenamefont {Kaxiras} \emph {et~al.}}]{CFD18}%
  \BibitemOpen
  \bibfield  {author} {\bibinfo {author} {\bibfnamefont {Yuan}\ \bibnamefont
  {Cao}}, \bibinfo {author} {\bibfnamefont {Valla}\ \bibnamefont {Fatemi}},
  \bibinfo {author} {\bibfnamefont {Ahmet}\ \bibnamefont {Demir}}, \bibinfo
  {author} {\bibfnamefont {Shiang}\ \bibnamefont {Fang}}, \bibinfo {author}
  {\bibfnamefont {Spencer~L.}\ \bibnamefont {Tomarken}}, \bibinfo {author}
  {\bibfnamefont {Jason~Y.}\ \bibnamefont {Luo}}, \bibinfo {author}
  {\bibfnamefont {Javier~D.}\ \bibnamefont {Sanchez-Yamagishi}}, \bibinfo
  {author} {\bibfnamefont {Kenji}\ \bibnamefont {Watanabe}}, \bibinfo {author}
  {\bibfnamefont {Takashi}\ \bibnamefont {Taniguchi}}, \bibinfo {author}
  {\bibfnamefont {Efthimios}\ \bibnamefont {Kaxiras}},  \emph {et~al.},\
  }\bibfield  {title} {\enquote {\bibinfo {title} {Correlated insulator
  behaviour at half-filling in magic-angle graphene superlattices},}\
  }\href@noop {} {\bibfield  {journal} {\bibinfo  {journal} {Nature}\ }\textbf
  {\bibinfo {volume} {556}},\ \bibinfo {pages} {80} (\bibinfo {year}
  {2018}{\natexlab{b}})}\BibitemShut {NoStop}%
\bibitem [{\citenamefont {Yankowitz}\ \emph {et~al.}(2019)\citenamefont
  {Yankowitz}, \citenamefont {Chen}, \citenamefont {Polshyn}, \citenamefont
  {Zhang}, \citenamefont {Watanabe}, \citenamefont {Taniguchi}, \citenamefont
  {Graf}, \citenamefont {Young},\ and\ \citenamefont {Dean}}]{YC19}%
  \BibitemOpen
  \bibfield  {author} {\bibinfo {author} {\bibfnamefont {Matthew}\ \bibnamefont
  {Yankowitz}}, \bibinfo {author} {\bibfnamefont {Shaowen}\ \bibnamefont
  {Chen}}, \bibinfo {author} {\bibfnamefont {Hryhoriy}\ \bibnamefont
  {Polshyn}}, \bibinfo {author} {\bibfnamefont {Yuxuan}\ \bibnamefont {Zhang}},
  \bibinfo {author} {\bibfnamefont {K.}~\bibnamefont {Watanabe}}, \bibinfo
  {author} {\bibfnamefont {T.}~\bibnamefont {Taniguchi}}, \bibinfo {author}
  {\bibfnamefont {David}\ \bibnamefont {Graf}}, \bibinfo {author}
  {\bibfnamefont {Andrea~F.}\ \bibnamefont {Young}}, \ and\ \bibinfo {author}
  {\bibfnamefont {Cory~R.}\ \bibnamefont {Dean}},\ }\bibfield  {title}
  {\enquote {\bibinfo {title} {Tuning superconductivity in twisted bilayer
  graphene},}\ }\href@noop {} {\bibfield  {journal} {\bibinfo  {journal}
  {Science}\ }\textbf {\bibinfo {volume} {363}},\ \bibinfo {pages} {1059--1064}
  (\bibinfo {year} {2019})}\BibitemShut {NoStop}%
\bibitem [{\citenamefont {Lu}\ \emph {et~al.}(2019)\citenamefont {Lu},
  \citenamefont {Stepanov}, \citenamefont {Yang}, \citenamefont {Xie},
  \citenamefont {Aamir}, \citenamefont {Das}, \citenamefont {Urgell},
  \citenamefont {Watanabe}, \citenamefont {Taniguchi}, \citenamefont {Zhang}
  \emph {et~al.}}]{LSY19}%
  \BibitemOpen
  \bibfield  {author} {\bibinfo {author} {\bibfnamefont {Xiaobo}\ \bibnamefont
  {Lu}}, \bibinfo {author} {\bibfnamefont {Petr}\ \bibnamefont {Stepanov}},
  \bibinfo {author} {\bibfnamefont {Wei}\ \bibnamefont {Yang}}, \bibinfo
  {author} {\bibfnamefont {Ming}\ \bibnamefont {Xie}}, \bibinfo {author}
  {\bibfnamefont {Mohammed~Ali}\ \bibnamefont {Aamir}}, \bibinfo {author}
  {\bibfnamefont {Ipsita}\ \bibnamefont {Das}}, \bibinfo {author}
  {\bibfnamefont {Carles}\ \bibnamefont {Urgell}}, \bibinfo {author}
  {\bibfnamefont {Kenji}\ \bibnamefont {Watanabe}}, \bibinfo {author}
  {\bibfnamefont {Takashi}\ \bibnamefont {Taniguchi}}, \bibinfo {author}
  {\bibfnamefont {Guangyu}\ \bibnamefont {Zhang}},  \emph {et~al.},\ }\bibfield
   {title} {\enquote {\bibinfo {title} {Superconductors, orbital magnets and
  correlated states in magic-angle bilayer graphene},}\ }\href@noop {}
  {\bibfield  {journal} {\bibinfo  {journal} {Nature}\ }\textbf {\bibinfo
  {volume} {574}},\ \bibinfo {pages} {653--657} (\bibinfo {year}
  {2019})}\BibitemShut {NoStop}%
\bibitem [{\citenamefont {Sharpe}\ \emph {et~al.}(2019)\citenamefont {Sharpe},
  \citenamefont {Fox}, \citenamefont {Barnard}, \citenamefont {Finney},
  \citenamefont {Watanabe}, \citenamefont {Taniguchi}, \citenamefont
  {Kastner},\ and\ \citenamefont {Goldhaber-Gordon}}]{SS19}%
  \BibitemOpen
  \bibfield  {author} {\bibinfo {author} {\bibfnamefont {Aaron~L.}\
  \bibnamefont {Sharpe}}, \bibinfo {author} {\bibfnamefont {Eli~J.}\
  \bibnamefont {Fox}}, \bibinfo {author} {\bibfnamefont {Arthur~W.}\
  \bibnamefont {Barnard}}, \bibinfo {author} {\bibfnamefont {Joe}\ \bibnamefont
  {Finney}}, \bibinfo {author} {\bibfnamefont {Kenji}\ \bibnamefont
  {Watanabe}}, \bibinfo {author} {\bibfnamefont {Takashi}\ \bibnamefont
  {Taniguchi}}, \bibinfo {author} {\bibfnamefont {M.~A.}\ \bibnamefont
  {Kastner}}, \ and\ \bibinfo {author} {\bibfnamefont {David}\ \bibnamefont
  {Goldhaber-Gordon}},\ }\bibfield  {title} {\enquote {\bibinfo {title}
  {Emergent ferromagnetism near three-quarters filling in twisted bilayer
  graphene},}\ }\href@noop {} {\bibfield  {journal} {\bibinfo  {journal}
  {Science}\ }\textbf {\bibinfo {volume} {365}},\ \bibinfo {pages} {605--608}
  (\bibinfo {year} {2019})}\BibitemShut {NoStop}%
\bibitem [{\citenamefont {Kang}\ and\ \citenamefont {Vafek}(2018)}]{KV18}%
  \BibitemOpen
  \bibfield  {author} {\bibinfo {author} {\bibfnamefont {Jian}\ \bibnamefont
  {Kang}}\ and\ \bibinfo {author} {\bibfnamefont {Oskar}\ \bibnamefont
  {Vafek}},\ }\bibfield  {title} {\enquote {\bibinfo {title} {Symmetry,
  maximally localized wannier states, and a low-energy model for twisted
  bilayer graphene narrow bands},}\ }\href {\doibase 10.1103/PhysRevX.8.031088}
  {\bibfield  {journal} {\bibinfo  {journal} {Phys. Rev. X}\ }\textbf {\bibinfo
  {volume} {8}},\ \bibinfo {pages} {031088} (\bibinfo {year}
  {2018})}\BibitemShut {NoStop}%
\bibitem [{\citenamefont {Koshino}\ \emph {et~al.}(2018)\citenamefont
  {Koshino}, \citenamefont {Yuan}, \citenamefont {Koretsune}, \citenamefont
  {Ochi}, \citenamefont {Kuroki},\ and\ \citenamefont {Fu}}]{KF18}%
  \BibitemOpen
  \bibfield  {author} {\bibinfo {author} {\bibfnamefont {Mikito}\ \bibnamefont
  {Koshino}}, \bibinfo {author} {\bibfnamefont {Noah F.~Q.}\ \bibnamefont
  {Yuan}}, \bibinfo {author} {\bibfnamefont {Takashi}\ \bibnamefont
  {Koretsune}}, \bibinfo {author} {\bibfnamefont {Masayuki}\ \bibnamefont
  {Ochi}}, \bibinfo {author} {\bibfnamefont {Kazuhiko}\ \bibnamefont {Kuroki}},
  \ and\ \bibinfo {author} {\bibfnamefont {Liang}\ \bibnamefont {Fu}},\
  }\bibfield  {title} {\enquote {\bibinfo {title} {Maximally localized wannier
  orbitals and the extended hubbard model for twisted bilayer graphene},}\
  }\href {\doibase 10.1103/PhysRevX.8.031087} {\bibfield  {journal} {\bibinfo
  {journal} {Phys. Rev. X}\ }\textbf {\bibinfo {volume} {8}},\ \bibinfo {pages}
  {031087} (\bibinfo {year} {2018})}\BibitemShut {NoStop}%
\bibitem [{\citenamefont {Zou}\ \emph {et~al.}(2018)\citenamefont {Zou},
  \citenamefont {Po}, \citenamefont {Vishwanath},\ and\ \citenamefont
  {Senthil}}]{ZP18}%
  \BibitemOpen
  \bibfield  {author} {\bibinfo {author} {\bibfnamefont {Liujun}\ \bibnamefont
  {Zou}}, \bibinfo {author} {\bibfnamefont {Hoi~Chun}\ \bibnamefont {Po}},
  \bibinfo {author} {\bibfnamefont {Ashvin}\ \bibnamefont {Vishwanath}}, \ and\
  \bibinfo {author} {\bibfnamefont {T.}~\bibnamefont {Senthil}},\ }\bibfield
  {title} {\enquote {\bibinfo {title} {Band structure of twisted bilayer
  graphene: Emergent symmetries, commensurate approximants, and wannier
  obstructions},}\ }\href {\doibase 10.1103/PhysRevB.98.085435} {\bibfield
  {journal} {\bibinfo  {journal} {Phys. Rev. B}\ }\textbf {\bibinfo {volume}
  {98}},\ \bibinfo {pages} {085435} (\bibinfo {year} {2018})}\BibitemShut
  {NoStop}%
\bibitem [{\citenamefont {Po}\ \emph {et~al.}(2018)\citenamefont {Po},
  \citenamefont {Zou}, \citenamefont {Vishwanath},\ and\ \citenamefont
  {Senthil}}]{PZ18}%
  \BibitemOpen
  \bibfield  {author} {\bibinfo {author} {\bibfnamefont {Hoi~Chun}\
  \bibnamefont {Po}}, \bibinfo {author} {\bibfnamefont {Liujun}\ \bibnamefont
  {Zou}}, \bibinfo {author} {\bibfnamefont {Ashvin}\ \bibnamefont
  {Vishwanath}}, \ and\ \bibinfo {author} {\bibfnamefont {T.}~\bibnamefont
  {Senthil}},\ }\bibfield  {title} {\enquote {\bibinfo {title} {Origin of mott
  insulating behavior and superconductivity in twisted bilayer graphene},}\
  }\href {\doibase 10.1103/PhysRevX.8.031089} {\bibfield  {journal} {\bibinfo
  {journal} {Phys. Rev. X}\ }\textbf {\bibinfo {volume} {8}},\ \bibinfo {pages}
  {031089} (\bibinfo {year} {2018})}\BibitemShut {NoStop}%
\bibitem [{\citenamefont {Song}\ \emph {et~al.}(2019)\citenamefont {Song},
  \citenamefont {Wang}, \citenamefont {Shi}, \citenamefont {Li}, \citenamefont
  {Fang},\ and\ \citenamefont {Bernevig}}]{SW19}%
  \BibitemOpen
  \bibfield  {author} {\bibinfo {author} {\bibfnamefont {Zhida}\ \bibnamefont
  {Song}}, \bibinfo {author} {\bibfnamefont {Zhijun}\ \bibnamefont {Wang}},
  \bibinfo {author} {\bibfnamefont {Wujun}\ \bibnamefont {Shi}}, \bibinfo
  {author} {\bibfnamefont {Gang}\ \bibnamefont {Li}}, \bibinfo {author}
  {\bibfnamefont {Chen}\ \bibnamefont {Fang}}, \ and\ \bibinfo {author}
  {\bibfnamefont {B.~Andrei}\ \bibnamefont {Bernevig}},\ }\bibfield  {title}
  {\enquote {\bibinfo {title} {All magic angles in twisted bilayer graphene are
  topological},}\ }\href {\doibase 10.1103/PhysRevLett.123.036401} {\bibfield
  {journal} {\bibinfo  {journal} {Phys. Rev. Lett.}\ }\textbf {\bibinfo
  {volume} {123}},\ \bibinfo {pages} {036401} (\bibinfo {year}
  {2019})}\BibitemShut {NoStop}%
\bibitem [{\citenamefont {Ahn}\ \emph {et~al.}(2019)\citenamefont {Ahn},
  \citenamefont {Park},\ and\ \citenamefont {Yang}}]{APY19}%
  \BibitemOpen
  \bibfield  {author} {\bibinfo {author} {\bibfnamefont {Junyeong}\
  \bibnamefont {Ahn}}, \bibinfo {author} {\bibfnamefont {Sungjoon}\
  \bibnamefont {Park}}, \ and\ \bibinfo {author} {\bibfnamefont {Bohm-Jung}\
  \bibnamefont {Yang}},\ }\bibfield  {title} {\enquote {\bibinfo {title}
  {Failure of nielsen-ninomiya theorem and fragile topology in two-dimensional
  systems with space-time inversion symmetry: Application to twisted bilayer
  graphene at magic angle},}\ }\href {\doibase 10.1103/PhysRevX.9.021013}
  {\bibfield  {journal} {\bibinfo  {journal} {Phys. Rev. X}\ }\textbf {\bibinfo
  {volume} {9}},\ \bibinfo {pages} {021013} (\bibinfo {year}
  {2019})}\BibitemShut {NoStop}%
\bibitem [{\citenamefont {Dutreix}\ \emph {et~al.}(2019)\citenamefont
  {Dutreix}, \citenamefont {Gonz{\'a}lez-Herrero}, \citenamefont {Brihuega},
  \citenamefont {Katsnelson}, \citenamefont {Chapelier},\ and\ \citenamefont
  {Renard}}]{DG19}%
  \BibitemOpen
  \bibfield  {author} {\bibinfo {author} {\bibfnamefont {C.}~\bibnamefont
  {Dutreix}}, \bibinfo {author} {\bibfnamefont {H.}~\bibnamefont
  {Gonz{\'a}lez-Herrero}}, \bibinfo {author} {\bibfnamefont {I.}~\bibnamefont
  {Brihuega}}, \bibinfo {author} {\bibfnamefont {M.~I.}\ \bibnamefont
  {Katsnelson}}, \bibinfo {author} {\bibfnamefont {C.}~\bibnamefont
  {Chapelier}}, \ and\ \bibinfo {author} {\bibfnamefont {V.~T.}\ \bibnamefont
  {Renard}},\ }\bibfield  {title} {\enquote {\bibinfo {title} {Measuring the
  berry phase of graphene from wavefront dislocations in friedel
  oscillations},}\ }\href@noop {} {\bibfield  {journal} {\bibinfo  {journal}
  {Nature}\ }\textbf {\bibinfo {volume} {574}},\ \bibinfo {pages} {219--222}
  (\bibinfo {year} {2019})}\BibitemShut {NoStop}%
\bibitem [{\citenamefont {Lopes~dos Santos}\ \emph
  {et~al.}(2007{\natexlab{a}})\citenamefont {Lopes~dos Santos}, \citenamefont
  {Peres},\ and\ \citenamefont {Castro~Neto}}]{dSPN07}%
  \BibitemOpen
  \bibfield  {author} {\bibinfo {author} {\bibfnamefont {J.~M.~B.}\
  \bibnamefont {Lopes~dos Santos}}, \bibinfo {author} {\bibfnamefont
  {N.~M.~R.}\ \bibnamefont {Peres}}, \ and\ \bibinfo {author} {\bibfnamefont
  {A.~H.}\ \bibnamefont {Castro~Neto}},\ }\bibfield  {title} {\enquote
  {\bibinfo {title} {Graphene bilayer with a twist: Electronic structure},}\
  }\href {\doibase 10.1103/PhysRevLett.99.256802} {\bibfield  {journal}
  {\bibinfo  {journal} {Phys. Rev. Lett.}\ }\textbf {\bibinfo {volume} {99}},\
  \bibinfo {pages} {256802} (\bibinfo {year} {2007}{\natexlab{a}})}\BibitemShut
  {NoStop}%
\bibitem [{\citenamefont {Su\'arez~Morell}\ \emph {et~al.}(2010)\citenamefont
  {Su\'arez~Morell}, \citenamefont {Correa}, \citenamefont {Vargas},
  \citenamefont {Pacheco},\ and\ \citenamefont {Barticevic}}]{SCV10}%
  \BibitemOpen
  \bibfield  {author} {\bibinfo {author} {\bibfnamefont {E.}~\bibnamefont
  {Su\'arez~Morell}}, \bibinfo {author} {\bibfnamefont {J.~D.}\ \bibnamefont
  {Correa}}, \bibinfo {author} {\bibfnamefont {P.}~\bibnamefont {Vargas}},
  \bibinfo {author} {\bibfnamefont {M.}~\bibnamefont {Pacheco}}, \ and\
  \bibinfo {author} {\bibfnamefont {Z.}~\bibnamefont {Barticevic}},\ }\bibfield
   {title} {\enquote {\bibinfo {title} {Flat bands in slightly twisted bilayer
  graphene: Tight-binding calculations},}\ }\href {\doibase
  10.1103/PhysRevB.82.121407} {\bibfield  {journal} {\bibinfo  {journal} {Phys.
  Rev. B}\ }\textbf {\bibinfo {volume} {82}},\ \bibinfo {pages} {121407}
  (\bibinfo {year} {2010})}\BibitemShut {NoStop}%
\bibitem [{\citenamefont {Mele}(2010)}]{M10}%
  \BibitemOpen
  \bibfield  {author} {\bibinfo {author} {\bibfnamefont {E.~J.}\ \bibnamefont
  {Mele}},\ }\bibfield  {title} {\enquote {\bibinfo {title} {Commensuration and
  interlayer coherence in twisted bilayer graphene},}\ }\href {\doibase
  10.1103/PhysRevB.81.161405} {\bibfield  {journal} {\bibinfo  {journal} {Phys.
  Rev. B}\ }\textbf {\bibinfo {volume} {81}},\ \bibinfo {pages} {161405}
  (\bibinfo {year} {2010})}\BibitemShut {NoStop}%
\bibitem [{\citenamefont {Bistritzer}\ and\ \citenamefont
  {MacDonald}(2011)}]{BM11}%
  \BibitemOpen
  \bibfield  {author} {\bibinfo {author} {\bibfnamefont {Rafi}\ \bibnamefont
  {Bistritzer}}\ and\ \bibinfo {author} {\bibfnamefont {Allan~H.}\ \bibnamefont
  {MacDonald}},\ }\bibfield  {title} {\enquote {\bibinfo {title} {Moir{\'e}
  bands in twisted double-layer graphene},}\ }\href@noop {} {\bibfield
  {journal} {\bibinfo  {journal} {PNAS}\ }\textbf {\bibinfo {volume} {108}},\
  \bibinfo {pages} {12233--12237} (\bibinfo {year} {2011})}\BibitemShut
  {NoStop}%
\bibitem [{\citenamefont {Guinea}\ and\ \citenamefont {Walet}(2019)}]{GW19}%
  \BibitemOpen
  \bibfield  {author} {\bibinfo {author} {\bibfnamefont {Francisco}\
  \bibnamefont {Guinea}}\ and\ \bibinfo {author} {\bibfnamefont {Niels~R.}\
  \bibnamefont {Walet}},\ }\bibfield  {title} {\enquote {\bibinfo {title}
  {Continuum models for twisted bilayer graphene: Effect of lattice deformation
  and hopping parameters},}\ }\href {\doibase 10.1103/PhysRevB.99.205134}
  {\bibfield  {journal} {\bibinfo  {journal} {Phys. Rev. B}\ }\textbf {\bibinfo
  {volume} {99}},\ \bibinfo {pages} {205134} (\bibinfo {year}
  {2019})}\BibitemShut {NoStop}%
\bibitem [{\citenamefont {Angeli}\ \emph {et~al.}(2018)\citenamefont {Angeli},
  \citenamefont {Mandelli}, \citenamefont {Valli}, \citenamefont {Amaricci},
  \citenamefont {Capone}, \citenamefont {Tosatti},\ and\ \citenamefont
  {Fabrizio}}]{AMV18}%
  \BibitemOpen
  \bibfield  {author} {\bibinfo {author} {\bibfnamefont {M.}~\bibnamefont
  {Angeli}}, \bibinfo {author} {\bibfnamefont {D.}~\bibnamefont {Mandelli}},
  \bibinfo {author} {\bibfnamefont {A.}~\bibnamefont {Valli}}, \bibinfo
  {author} {\bibfnamefont {A.}~\bibnamefont {Amaricci}}, \bibinfo {author}
  {\bibfnamefont {M.}~\bibnamefont {Capone}}, \bibinfo {author} {\bibfnamefont
  {E.}~\bibnamefont {Tosatti}}, \ and\ \bibinfo {author} {\bibfnamefont
  {M.}~\bibnamefont {Fabrizio}},\ }\bibfield  {title} {\enquote {\bibinfo
  {title} {Emergent ${D}_{6}$ symmetry in fully relaxed magic-angle twisted
  bilayer graphene},}\ }\href {\doibase 10.1103/PhysRevB.98.235137} {\bibfield
  {journal} {\bibinfo  {journal} {Phys. Rev. B}\ }\textbf {\bibinfo {volume}
  {98}},\ \bibinfo {pages} {235137} (\bibinfo {year} {2018})}\BibitemShut
  {NoStop}%
\bibitem [{\citenamefont {Po}\ \emph {et~al.}(2019)\citenamefont {Po},
  \citenamefont {Zou}, \citenamefont {Senthil},\ and\ \citenamefont
  {Vishwanath}}]{PZSV19}%
  \BibitemOpen
  \bibfield  {author} {\bibinfo {author} {\bibfnamefont {Hoi~Chun}\
  \bibnamefont {Po}}, \bibinfo {author} {\bibfnamefont {Liujun}\ \bibnamefont
  {Zou}}, \bibinfo {author} {\bibfnamefont {T.}~\bibnamefont {Senthil}}, \ and\
  \bibinfo {author} {\bibfnamefont {Ashvin}\ \bibnamefont {Vishwanath}},\
  }\bibfield  {title} {\enquote {\bibinfo {title} {Faithful tight-binding
  models and fragile topology of magic-angle bilayer graphene},}\ }\href
  {\doibase 10.1103/PhysRevB.99.195455} {\bibfield  {journal} {\bibinfo
  {journal} {Phys. Rev. B}\ }\textbf {\bibinfo {volume} {99}},\ \bibinfo
  {pages} {195455} (\bibinfo {year} {2019})}\BibitemShut {NoStop}%
\bibitem [{\citenamefont {Fidrysiak}\ \emph {et~al.}(2018)\citenamefont
  {Fidrysiak}, \citenamefont {Zegrodnik},\ and\ \citenamefont
  {Spa\l{}ek}}]{FZS18}%
  \BibitemOpen
  \bibfield  {author} {\bibinfo {author} {\bibfnamefont {M.}~\bibnamefont
  {Fidrysiak}}, \bibinfo {author} {\bibfnamefont {M.}~\bibnamefont
  {Zegrodnik}}, \ and\ \bibinfo {author} {\bibfnamefont {J.}~\bibnamefont
  {Spa\l{}ek}},\ }\bibfield  {title} {\enquote {\bibinfo {title}
  {Unconventional topological superconductivity and phase diagram for an
  effective two-orbital model as applied to twisted bilayer graphene},}\ }\href
  {\doibase 10.1103/PhysRevB.98.085436} {\bibfield  {journal} {\bibinfo
  {journal} {Phys. Rev. B}\ }\textbf {\bibinfo {volume} {98}},\ \bibinfo
  {pages} {085436} (\bibinfo {year} {2018})}\BibitemShut {NoStop}%
\bibitem [{\citenamefont {Venderbos}\ and\ \citenamefont
  {Fernandes}(2018)}]{VF18}%
  \BibitemOpen
  \bibfield  {author} {\bibinfo {author} {\bibfnamefont {J\"orn W.~F.}\
  \bibnamefont {Venderbos}}\ and\ \bibinfo {author} {\bibfnamefont {Rafael~M.}\
  \bibnamefont {Fernandes}},\ }\bibfield  {title} {\enquote {\bibinfo {title}
  {Correlations and electronic order in a two-orbital honeycomb lattice model
  for twisted bilayer graphene},}\ }\href {\doibase 10.1103/PhysRevB.98.245103}
  {\bibfield  {journal} {\bibinfo  {journal} {Phys. Rev. B}\ }\textbf {\bibinfo
  {volume} {98}},\ \bibinfo {pages} {245103} (\bibinfo {year}
  {2018})}\BibitemShut {NoStop}%
\bibitem [{\citenamefont {Lin}\ and\ \citenamefont {Nandkishore}(2018)}]{LN18}%
  \BibitemOpen
  \bibfield  {author} {\bibinfo {author} {\bibfnamefont {Yu-Ping}\ \bibnamefont
  {Lin}}\ and\ \bibinfo {author} {\bibfnamefont {Rahul~M.}\ \bibnamefont
  {Nandkishore}},\ }\bibfield  {title} {\enquote {\bibinfo {title}
  {Kohn-luttinger superconductivity on two orbital honeycomb lattice},}\ }\href
  {\doibase 10.1103/PhysRevB.98.214521} {\bibfield  {journal} {\bibinfo
  {journal} {Phys. Rev. B}\ }\textbf {\bibinfo {volume} {98}},\ \bibinfo
  {pages} {214521} (\bibinfo {year} {2018})}\BibitemShut {NoStop}%
\bibitem [{\citenamefont {Thomson}\ \emph {et~al.}(2018)\citenamefont
  {Thomson}, \citenamefont {Chatterjee}, \citenamefont {Sachdev},\ and\
  \citenamefont {Scheurer}}]{TCS18}%
  \BibitemOpen
  \bibfield  {author} {\bibinfo {author} {\bibfnamefont {Alex}\ \bibnamefont
  {Thomson}}, \bibinfo {author} {\bibfnamefont {Shubhayu}\ \bibnamefont
  {Chatterjee}}, \bibinfo {author} {\bibfnamefont {Subir}\ \bibnamefont
  {Sachdev}}, \ and\ \bibinfo {author} {\bibfnamefont {Mathias~S.}\
  \bibnamefont {Scheurer}},\ }\bibfield  {title} {\enquote {\bibinfo {title}
  {Triangular antiferromagnetism on the honeycomb lattice of twisted bilayer
  graphene},}\ }\href {\doibase 10.1103/PhysRevB.98.075109} {\bibfield
  {journal} {\bibinfo  {journal} {Phys. Rev. B}\ }\textbf {\bibinfo {volume}
  {98}},\ \bibinfo {pages} {075109} (\bibinfo {year} {2018})}\BibitemShut
  {NoStop}%
\bibitem [{\citenamefont {Zhu}\ \emph {et~al.}(2019)\citenamefont {Zhu},
  \citenamefont {Sheng},\ and\ \citenamefont {Fu}}]{ZSF19}%
  \BibitemOpen
  \bibfield  {author} {\bibinfo {author} {\bibfnamefont {Zheng}\ \bibnamefont
  {Zhu}}, \bibinfo {author} {\bibfnamefont {D.~N.}\ \bibnamefont {Sheng}}, \
  and\ \bibinfo {author} {\bibfnamefont {Liang}\ \bibnamefont {Fu}},\
  }\bibfield  {title} {\enquote {\bibinfo {title} {Spin-orbital density wave
  and a mott insulator in a two-orbital hubbard model on a honeycomb
  lattice},}\ }\href {\doibase 10.1103/PhysRevLett.123.087602} {\bibfield
  {journal} {\bibinfo  {journal} {Phys. Rev. Lett.}\ }\textbf {\bibinfo
  {volume} {123}},\ \bibinfo {pages} {087602} (\bibinfo {year}
  {2019})}\BibitemShut {NoStop}%
\bibitem [{\citenamefont {Da~Liao}\ \emph {et~al.}(2019)\citenamefont
  {Da~Liao}, \citenamefont {Meng},\ and\ \citenamefont {Xu}}]{DMX19}%
  \BibitemOpen
  \bibfield  {author} {\bibinfo {author} {\bibfnamefont {Yuan}\ \bibnamefont
  {Da~Liao}}, \bibinfo {author} {\bibfnamefont {Zi~Yang}\ \bibnamefont {Meng}},
  \ and\ \bibinfo {author} {\bibfnamefont {Xiao~Yan}\ \bibnamefont {Xu}},\
  }\bibfield  {title} {\enquote {\bibinfo {title} {Valence bond orders at
  charge neutrality in a possible two-orbital extended hubbard model for
  twisted bilayer graphene},}\ }\href {\doibase 10.1103/PhysRevLett.123.157601}
  {\bibfield  {journal} {\bibinfo  {journal} {Phys. Rev. Lett.}\ }\textbf
  {\bibinfo {volume} {123}},\ \bibinfo {pages} {157601} (\bibinfo {year}
  {2019})}\BibitemShut {NoStop}%
\bibitem [{\citenamefont {Crommie}\ \emph {et~al.}(1993)\citenamefont
  {Crommie}, \citenamefont {Lutz},\ and\ \citenamefont {Eigler}}]{CLE93}%
  \BibitemOpen
  \bibfield  {author} {\bibinfo {author} {\bibfnamefont {M.~F.}\ \bibnamefont
  {Crommie}}, \bibinfo {author} {\bibfnamefont {Ch.~P.}\ \bibnamefont {Lutz}},
  \ and\ \bibinfo {author} {\bibfnamefont {D.~M.}\ \bibnamefont {Eigler}},\
  }\bibfield  {title} {\enquote {\bibinfo {title} {Imaging standing waves in a
  two-dimensional electron gas},}\ }\href@noop {} {\bibfield  {journal}
  {\bibinfo  {journal} {Nature}\ }\textbf {\bibinfo {volume} {363}},\ \bibinfo
  {pages} {524--527} (\bibinfo {year} {1993})}\BibitemShut {NoStop}%
\bibitem [{\citenamefont {Dutreix}\ and\ \citenamefont
  {Delplace}(2017)}]{DD17}%
  \BibitemOpen
  \bibfield  {author} {\bibinfo {author} {\bibfnamefont {C.}~\bibnamefont
  {Dutreix}}\ and\ \bibinfo {author} {\bibfnamefont {P.}~\bibnamefont
  {Delplace}},\ }\bibfield  {title} {\enquote {\bibinfo {title} {Geometrical
  phase shift in friedel oscillations},}\ }\href {\doibase
  10.1103/PhysRevB.96.195207} {\bibfield  {journal} {\bibinfo  {journal} {Phys.
  Rev. B}\ }\textbf {\bibinfo {volume} {96}},\ \bibinfo {pages} {195207}
  (\bibinfo {year} {2017})}\BibitemShut {NoStop}%
\bibitem [{SM()}]{SM}%
  \BibitemOpen
  \href@noop {} {}\bibinfo {note} {See Supplementary Material}\BibitemShut
  {NoStop}%
\bibitem [{\citenamefont {Seki}\ and\ \citenamefont {Yunoki}(2016)}]{SY16}%
  \BibitemOpen
  \bibfield  {author} {\bibinfo {author} {\bibfnamefont {K.}~\bibnamefont
  {Seki}}\ and\ \bibinfo {author} {\bibfnamefont {S.}~\bibnamefont {Yunoki}},\
  }\bibfield  {title} {\enquote {\bibinfo {title} {Brillouin-zone integration
  scheme for many-body density of states: Tetrahedron method combined with
  cluster perturbation theory},}\ }\href {\doibase 10.1103/PhysRevB.93.245115}
  {\bibfield  {journal} {\bibinfo  {journal} {Phys. Rev. B}\ }\textbf {\bibinfo
  {volume} {93}},\ \bibinfo {pages} {245115} (\bibinfo {year}
  {2016})}\BibitemShut {NoStop}%
\bibitem [{\citenamefont {Lopes~dos Santos}\ \emph
  {et~al.}(2007{\natexlab{b}})\citenamefont {Lopes~dos Santos}, \citenamefont
  {Peres},\ and\ \citenamefont {Castro~Neto}}]{LS07}%
  \BibitemOpen
  \bibfield  {author} {\bibinfo {author} {\bibfnamefont {J.~M.~B.}\
  \bibnamefont {Lopes~dos Santos}}, \bibinfo {author} {\bibfnamefont
  {N.~M.~R.}\ \bibnamefont {Peres}}, \ and\ \bibinfo {author} {\bibfnamefont
  {A.~H.}\ \bibnamefont {Castro~Neto}},\ }\bibfield  {title} {\enquote
  {\bibinfo {title} {Graphene bilayer with a twist: Electronic structure},}\
  }\href {\doibase 10.1103/PhysRevLett.99.256802} {\bibfield  {journal}
  {\bibinfo  {journal} {Phys. Rev. Lett.}\ }\textbf {\bibinfo {volume} {99}},\
  \bibinfo {pages} {256802} (\bibinfo {year} {2007}{\natexlab{b}})}\BibitemShut
  {NoStop}%
\bibitem [{\citenamefont {Lopes~dos Santos}\ \emph {et~al.}(2012)\citenamefont
  {Lopes~dos Santos}, \citenamefont {Peres},\ and\ \citenamefont
  {Castro~Neto}}]{LS12}%
  \BibitemOpen
  \bibfield  {author} {\bibinfo {author} {\bibfnamefont {J.~M.~B.}\
  \bibnamefont {Lopes~dos Santos}}, \bibinfo {author} {\bibfnamefont
  {N.~M.~R.}\ \bibnamefont {Peres}}, \ and\ \bibinfo {author} {\bibfnamefont
  {A.~H.}\ \bibnamefont {Castro~Neto}},\ }\bibfield  {title} {\enquote
  {\bibinfo {title} {Continuum model of the twisted graphene bilayer},}\ }\href
  {\doibase 10.1103/PhysRevB.86.155449} {\bibfield  {journal} {\bibinfo
  {journal} {Phys. Rev. B}\ }\textbf {\bibinfo {volume} {86}},\ \bibinfo
  {pages} {155449} (\bibinfo {year} {2012})}\BibitemShut {NoStop}%
\bibitem [{\citenamefont {Nam}\ and\ \citenamefont {Koshino}(2017)}]{NK17}%
  \BibitemOpen
  \bibfield  {author} {\bibinfo {author} {\bibfnamefont {Nguyen N.~T.}\
  \bibnamefont {Nam}}\ and\ \bibinfo {author} {\bibfnamefont {Mikito}\
  \bibnamefont {Koshino}},\ }\bibfield  {title} {\enquote {\bibinfo {title}
  {Lattice relaxation and energy band modulation in twisted bilayer
  graphene},}\ }\href {\doibase 10.1103/PhysRevB.96.075311} {\bibfield
  {journal} {\bibinfo  {journal} {Phys. Rev. B}\ }\textbf {\bibinfo {volume}
  {96}},\ \bibinfo {pages} {075311} (\bibinfo {year} {2017})}\BibitemShut
  {NoStop}%
\bibitem [{\citenamefont {Koshino}\ and\ \citenamefont {Nam}(2019)}]{KN19}%
  \BibitemOpen
  \bibfield  {author} {\bibinfo {author} {\bibfnamefont {Mikito}\ \bibnamefont
  {Koshino}}\ and\ \bibinfo {author} {\bibfnamefont {Nguyen N.~T.}\
  \bibnamefont {Nam}},\ }\bibfield  {title} {\enquote {\bibinfo {title}
  {Continuum model for relaxed twisted bilayer graphenes and moir{\'e}
  electron-phonon interaction},}\ }\href@noop {} {\bibfield  {journal}
  {\bibinfo  {journal} {arXiv preprint arXiv:1909.10786}\ } (\bibinfo {year}
  {2019})}\BibitemShut {NoStop}%
\bibitem [{\citenamefont {Hejazi}\ \emph {et~al.}(2019)\citenamefont {Hejazi},
  \citenamefont {Liu}, \citenamefont {Shapourian}, \citenamefont {Chen},\ and\
  \citenamefont {Balents}}]{HC19}%
  \BibitemOpen
  \bibfield  {author} {\bibinfo {author} {\bibfnamefont {Kasra}\ \bibnamefont
  {Hejazi}}, \bibinfo {author} {\bibfnamefont {Chunxiao}\ \bibnamefont {Liu}},
  \bibinfo {author} {\bibfnamefont {Hassan}\ \bibnamefont {Shapourian}},
  \bibinfo {author} {\bibfnamefont {Xiao}\ \bibnamefont {Chen}}, \ and\
  \bibinfo {author} {\bibfnamefont {Leon}\ \bibnamefont {Balents}},\ }\bibfield
   {title} {\enquote {\bibinfo {title} {Multiple topological transitions in
  twisted bilayer graphene near the first magic angle},}\ }\href {\doibase
  10.1103/PhysRevB.99.035111} {\bibfield  {journal} {\bibinfo  {journal} {Phys.
  Rev. B}\ }\textbf {\bibinfo {volume} {99}},\ \bibinfo {pages} {035111}
  (\bibinfo {year} {2019})}\BibitemShut {NoStop}%
\bibitem [{\citenamefont {San-Jose}\ and\ \citenamefont {Prada}(2013)}]{SJP13}%
  \BibitemOpen
  \bibfield  {author} {\bibinfo {author} {\bibfnamefont {Pablo}\ \bibnamefont
  {San-Jose}}\ and\ \bibinfo {author} {\bibfnamefont {Elsa}\ \bibnamefont
  {Prada}},\ }\bibfield  {title} {\enquote {\bibinfo {title} {Helical networks
  in twisted bilayer graphene under interlayer bias},}\ }\href {\doibase
  10.1103/PhysRevB.88.121408} {\bibfield  {journal} {\bibinfo  {journal} {Phys.
  Rev. B}\ }\textbf {\bibinfo {volume} {88}},\ \bibinfo {pages} {121408}
  (\bibinfo {year} {2013})}\BibitemShut {NoStop}%
\bibitem [{\citenamefont {Efimkin}\ and\ \citenamefont
  {MacDonald}(2018)}]{EMA18}%
  \BibitemOpen
  \bibfield  {author} {\bibinfo {author} {\bibfnamefont {Dmitry~K.}\
  \bibnamefont {Efimkin}}\ and\ \bibinfo {author} {\bibfnamefont {Allan~H.}\
  \bibnamefont {MacDonald}},\ }\bibfield  {title} {\enquote {\bibinfo {title}
  {Helical network model for twisted bilayer graphene},}\ }\href {\doibase
  10.1103/PhysRevB.98.035404} {\bibfield  {journal} {\bibinfo  {journal} {Phys.
  Rev. B}\ }\textbf {\bibinfo {volume} {98}},\ \bibinfo {pages} {035404}
  (\bibinfo {year} {2018})}\BibitemShut {NoStop}%
\bibitem [{\citenamefont {Tsim}\ \emph {et~al.}(2020)\citenamefont {Tsim},
  \citenamefont {Nam},\ and\ \citenamefont {Koshino}}]{TNK20}%
  \BibitemOpen
  \bibfield  {author} {\bibinfo {author} {\bibfnamefont {Bonnie}\ \bibnamefont
  {Tsim}}, \bibinfo {author} {\bibfnamefont {Nguyen N.~T.}\ \bibnamefont
  {Nam}}, \ and\ \bibinfo {author} {\bibfnamefont {Mikito}\ \bibnamefont
  {Koshino}},\ }\bibfield  {title} {\enquote {\bibinfo {title} {Perfect
  one-dimensional chiral states in biased twisted bilayer graphene},}\
  }\href@noop {} {\bibfield  {journal} {\bibinfo  {journal} {arXiv preprint
  arXiv:2001.06257}\ } (\bibinfo {year} {2020})}\BibitemShut {NoStop}%
\bibitem [{\citenamefont {Gonzalez-Arraga}\ \emph {et~al.}(2017)\citenamefont
  {Gonzalez-Arraga}, \citenamefont {Lado}, \citenamefont {Guinea},\ and\
  \citenamefont {San-Jose}}]{GAGSJ17}%
  \BibitemOpen
  \bibfield  {author} {\bibinfo {author} {\bibfnamefont {Luis~A.}\ \bibnamefont
  {Gonzalez-Arraga}}, \bibinfo {author} {\bibfnamefont {J.~L.}\ \bibnamefont
  {Lado}}, \bibinfo {author} {\bibfnamefont {Francisco}\ \bibnamefont
  {Guinea}}, \ and\ \bibinfo {author} {\bibfnamefont {Pablo}\ \bibnamefont
  {San-Jose}},\ }\bibfield  {title} {\enquote {\bibinfo {title} {Electrically
  controllable magnetism in twisted bilayer graphene},}\ }\href {\doibase
  10.1103/PhysRevLett.119.107201} {\bibfield  {journal} {\bibinfo  {journal}
  {Phys. Rev. Lett.}\ }\textbf {\bibinfo {volume} {119}},\ \bibinfo {pages}
  {107201} (\bibinfo {year} {2017})}\BibitemShut {NoStop}%
\bibitem [{\citenamefont {Tarnopolsky}\ \emph {et~al.}(2019)\citenamefont
  {Tarnopolsky}, \citenamefont {Kruchkov},\ and\ \citenamefont
  {Vishwanath}}]{TKV19}%
  \BibitemOpen
  \bibfield  {author} {\bibinfo {author} {\bibfnamefont {Grigory}\ \bibnamefont
  {Tarnopolsky}}, \bibinfo {author} {\bibfnamefont {Alex~Jura}\ \bibnamefont
  {Kruchkov}}, \ and\ \bibinfo {author} {\bibfnamefont {Ashvin}\ \bibnamefont
  {Vishwanath}},\ }\bibfield  {title} {\enquote {\bibinfo {title} {Origin of
  magic angles in twisted bilayer graphene},}\ }\href {\doibase
  10.1103/PhysRevLett.122.106405} {\bibfield  {journal} {\bibinfo  {journal}
  {Phys. Rev. Lett.}\ }\textbf {\bibinfo {volume} {122}},\ \bibinfo {pages}
  {106405} (\bibinfo {year} {2019})}\BibitemShut {NoStop}%
\bibitem [{\citenamefont {Economou}(2006)}]{EE06}%
  \BibitemOpen
  \bibfield  {author} {\bibinfo {author} {\bibfnamefont {Eleftherios~N.}\
  \bibnamefont {Economou}},\ }\href@noop {} {\emph {\bibinfo {title} {Green's
  functions in quantum physics}}},\ Vol.~\bibinfo {volume} {7}\ (\bibinfo
  {publisher} {Springer Science \& Business Media},\ \bibinfo {year}
  {2006})\BibitemShut {NoStop}%
\bibitem [{\citenamefont {Katsnelson}(2012)}]{K12}%
  \BibitemOpen
  \bibfield  {author} {\bibinfo {author} {\bibfnamefont {Mikhail~I.}\
  \bibnamefont {Katsnelson}},\ }\href@noop {} {\emph {\bibinfo {title}
  {Graphene: carbon in two dimensions}}}\ (\bibinfo  {publisher} {Cambridge
  University Press},\ \bibinfo {year} {2012})\BibitemShut {NoStop}%
\bibitem [{\citenamefont {Vanderbilt}(2018)}]{V18}%
  \BibitemOpen
  \bibfield  {author} {\bibinfo {author} {\bibfnamefont {David}\ \bibnamefont
  {Vanderbilt}},\ }\href@noop {} {\emph {\bibinfo {title} {Berry Phases in
  Electronic Structure Theory: Electric Polarization, Orbital Magnetization and
  Topological Insulators}}}\ (\bibinfo  {publisher} {Cambridge University
  Press},\ \bibinfo {year} {2018})\BibitemShut {NoStop}%
\bibitem [{\citenamefont {Rath}\ and\ \citenamefont {Freeman}(1975)}]{RF75}%
  \BibitemOpen
  \bibfield  {author} {\bibinfo {author} {\bibfnamefont {J.}~\bibnamefont
  {Rath}}\ and\ \bibinfo {author} {\bibfnamefont {A.~J.}\ \bibnamefont
  {Freeman}},\ }\bibfield  {title} {\enquote {\bibinfo {title} {Generalized
  magnetic susceptibilities in metals: Application of the analytic tetrahedron
  linear energy method to sc},}\ }\href {\doibase 10.1103/PhysRevB.11.2109}
  {\bibfield  {journal} {\bibinfo  {journal} {Phys. Rev. B}\ }\textbf {\bibinfo
  {volume} {11}},\ \bibinfo {pages} {2109--2117} (\bibinfo {year}
  {1975})}\BibitemShut {NoStop}%
\bibitem [{\citenamefont {Bl\"ochl}\ \emph {et~al.}(1994)\citenamefont
  {Bl\"ochl}, \citenamefont {Jepsen},\ and\ \citenamefont {Andersen}}]{BJA94}%
  \BibitemOpen
  \bibfield  {author} {\bibinfo {author} {\bibfnamefont {Peter~E.}\
  \bibnamefont {Bl\"ochl}}, \bibinfo {author} {\bibfnamefont {O.}~\bibnamefont
  {Jepsen}}, \ and\ \bibinfo {author} {\bibfnamefont {O.~K.}\ \bibnamefont
  {Andersen}},\ }\bibfield  {title} {\enquote {\bibinfo {title} {Improved
  tetrahedron method for brillouin-zone integrations},}\ }\href {\doibase
  10.1103/PhysRevB.49.16223} {\bibfield  {journal} {\bibinfo  {journal} {Phys.
  Rev. B}\ }\textbf {\bibinfo {volume} {49}},\ \bibinfo {pages} {16223--16233}
  (\bibinfo {year} {1994})}\BibitemShut {NoStop}%
\end{thebibliography}%

\pagebreak
\appendix

\onecolumngrid

\setcounter{equation}{0}
\setcounter{figure}{0}

\renewcommand{\theequation}{S\arabic{equation}}
\renewcommand{\thefigure}{S\arabic{figure}}

\renewcommand{\bibnumfmt}[1]{[#1]}
\renewcommand{\citenumfont}[1]{#1}

\pagebreak

\begin{center}
\begin{Large}
Supplementary Material
\end{Large}
\end{center}

\section{Continuum Theory}

\begin{figure}
\includegraphics[scale=0.25]{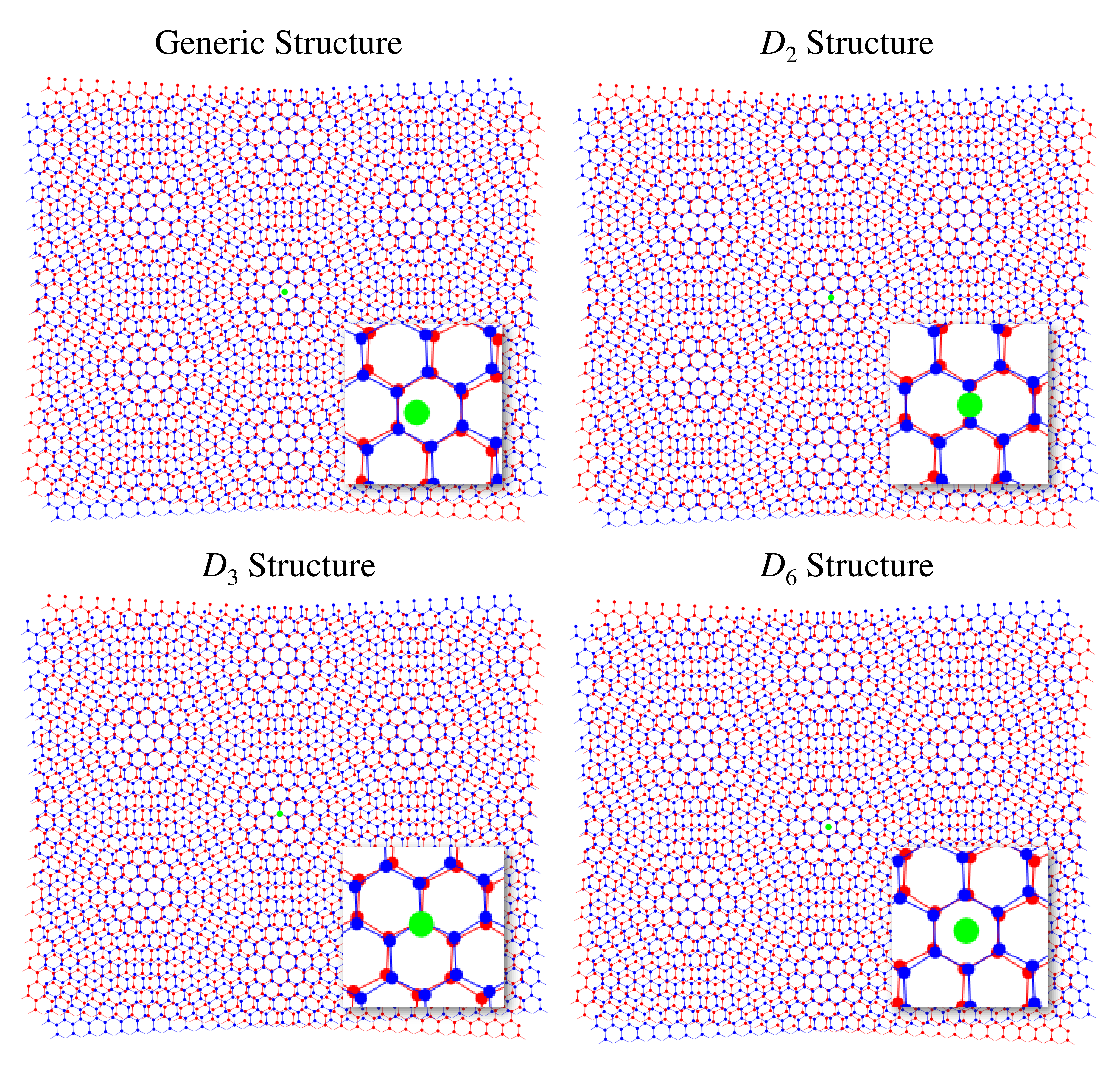}
\caption{Moiré structure of TBLG with different point symmetries at  rotation angle $\theta = 6^{\circ}.$ For each structure, two sheets of monolayer graphene are first stacked one on top of the other in perfect atomic alignment, then one sheet is rotated relative to the other by an angle $\theta$ about some chosen point, shown in green in the insets. As can be clearly be seen from the large-scale patterns, the four structures look indistinguishable on the moiré length scale. Only when we zoom in to the atomic length scale, shown in the insets, do we notice a difference between the structures. }
\label{fig. pointsymmetries}
\end{figure}

In the small-angle limit, twisted bilayer graphene (TBG) forms a long-wavelength two-dimensional moiré solid whose unit cells typically consist of thousands of relevent orbitals each. Because of that, calculating the band structure of TBLG from microscopic tight-binding models is prohibitively expensive computationally. To work around this limitation, we instead approximate the interlayer coupling as a smooth function of coodinate space \cite{M10, LS07,BM11, LS12, NK17, KN19}. In doing so, we must forgo our ability to resolve the microscopic origin of the interlayer hoppings. Thus, a \textit{continuum theory} constructed in this spirit is unable to differentiate between lattice structures with different exact point symmetries. For example, starting with two sheets of monolayer graphene stacked one on top of the other with perfect atomic alignment, a rotation of one layer relative to the other by a small angle about a registered hexagon center yields a structure with $D_6$ lattice symmetry; meanwhile, a  rotation done about a registered pair of carbon atoms gives a $D_3$ lattice structure; and similarly, a rotation about a registered bond center only has $D_2$ lattice symmetry. In practice, the rotation center is not easily controlled; as such, the exact symmetry group of a structure in experiments is not typically known. However, if we are only interested in the physics that operates at the moiré length scale, fine details of the exact symmetries should not qualitatively matter. Instead, the continuum theory accounts only for emergent symmetries that are approximately preserved on the moiré length scale, as shown in Fig.~\ref{fig. pointsymmetries}. In this sense, the continuum theory, in the absence of interlayer bias, has emergent $D_6$ symmetries \cite{PZ18, ZP18}.

\begin{figure}
\includegraphics[scale=0.35]{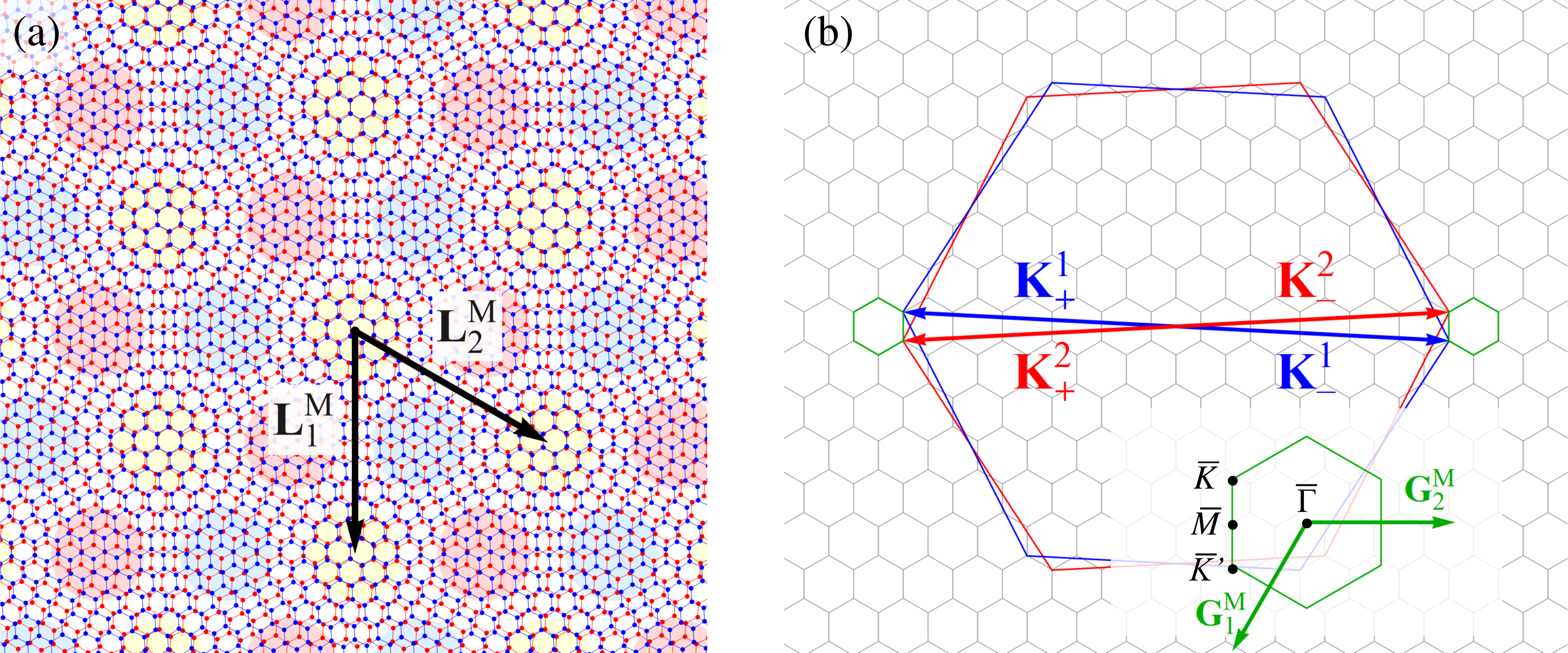}
\caption{(a) Emergent crystal structure of TBLG with $\theta = 6^{\circ}.$ The yellow-shaded regions are $AA$-aligned, while the red-shaded and blue-shaded regions correspond to $AB$ and $BA$ stacking respectively. The emergent lattice is spanned by $\mathbf{L}_1^\text{M}$ and $\mathbf{L}_2^\text{M}.$ (b) Reciprocal-space representation of TBLG. The large red and blue hexagons represent the rotated Brillouin zones of the individual layers. The mismatch in the two monolayer Brillouin zones generates a smaller moiré Brillouin zone shown in green. The moiré reciprocal space is spanned by $\mathbf{G}_1^\text{M}$ and $\mathbf{G}_2^\text{M}.$ The special high-symmetry points in the moiré zone are indicated in the inset. }
\label{fig. crystalstructure}
\end{figure}

In addition to approximating point symmetries, the continuum theory also assumes translation symmetry even though it is not generally exact. At a generic rotation angle, the structure does not usually form a perioidic solid for which, strictly speaking, Bloch's theorem can be applied. Only at certain commemsurate angles do we obtain spatially-periodic structures. However, the emergent moiré-scale system is usually close enough to being periodic that we can approximate it as such. Following this line of reasoning, we can write the interlayer coupling as a periodic function with moiré periodicity. An approximately periodic moiré structure of TBLG forms an emergent hexagonal lattice that consists of three regions with different atomic alignments. An $AA$ region is where the two layers are approximately aligned with one another; an $AB$ region is where the $A$ sublattice of layer 1, denoted $A_1,$ is aligned with the $B$ sublattice of layer 2, denoted $B_2;$ and likewise, a $BA$ region is where the $B$ sublattice of layer 1, denoted $B_1,$ is aligned with the $A$ sublattice of layer 2, denoted $A_2,$ as shown in Fig.~\ref{fig. crystalstructure}. To describe the emergent lattice structure of TBLG, we begin with two planes of monolayer graphene stacked with all the carbon atoms from one layer aligned with those of the other layer. The in-plane primitive lattice vectors for both layers are \cite{KF18}
\begin{equation}
\mathbf{l}_1 = a \left(1,0  \right) \quad \text{and} \quad \mathbf{l}_2 = a \left( \frac{1}{2}, \frac{\sqrt{3}}{2} \right),
\end{equation}
where $a$ is the graphene lattice constant. The common primitive reciprocal lattice constants for both layers are 
\begin{equation}
\mathbf{g}_1 = \frac{2\pi}{a} \left( 1, - \frac{1}{\sqrt{3}} \right) \quad \text{and} \quad \mathbf{g}_2 = \frac{2 \pi}{a} \left(  0 , \frac{2}{\sqrt{3}} \right).
\end{equation}
We then rotate layer $\ell = 1$ by an angle $-\theta/2$ and layer $\ell = 2$ by an angle $+\theta/2.$ The lattice vectors and reciprocal lattice vectors transform as $\mathbf{l}_i^{\ell} = R\left(  \mp \theta/2 \right) \mathbf{l}_i $ and $\mathbf{g}_i^{\ell} = R\left(  \mp \theta/2 \right) \mathbf{g}_i ,$ where $R \left(  \theta \right)$ is the operator that performs rotation about the $z$-axis by an angle $\theta.$ The mismatch in the two monolayer Brillouin zones generates a much smaller moiré Brillouin zone defined by the vectors
\begin{equation}
\mathbf{G}_1^\text{M} = \mathbf{g}_1^1 - \mathbf{g}_1^2 = \frac{4\pi \sin \left( \theta /2 \right)}{a} \left( - \frac{1}{\sqrt{3}}, -1 \right) \quad \text{and} \quad \mathbf{G}_2^\text{M} = \mathbf{g}_2^1 - \mathbf{g}_2^2 = \frac{4\pi \sin \left( \theta /2 \right)}{a} \left( \frac{2}{\sqrt{3}}, 0 \right).
\end{equation}
From this, we can determine the primitive lattice vectors of the emergent moiré lattice
\begin{equation}
\mathbf{L}_1^\mathbf{M} = L_\mathbf{M} \left( 0, -1 \right) \quad \text{and} \quad \mathbf{L}_2^\mathbf{M} = L_\mathbf{M} \left( \frac{\sqrt{3}}{2}, -\frac{1}{2} \right),
\end{equation}
where $L_\text{M} = a/2\sin \left( \theta /2 \right)$ is the moiré lattice constant. We note that we have not defined a basis within a unit cell of each layer. Thus, this procedure is independent of the origin of rotation, and is devoid of any information of the microscopic symmetries.

Using this emergent lattice structure, we can construct an effective theory at low energy where the monolayer dispersion is well-approximated by the Dirac equation with definite chirality near the two valleys. We denote the valley index by $\nu = \pm 1.$ The interlayer coupling is written as a superposition of long-wavelength plane waves where the wave-vectors are integral multiples of $\mathbf{G}_1^\text{M}$ and $\mathbf{G}_2^\text{M}.$ Keeping only the first few dominant harmonics in the interlayer coupling, the low-energy Hamiltonian can be factored into two effectively independent valleys. Terms which couple the two valleys are expected to be exponentially small when the wavevectors which connect the two valleys are large compared to the wavevectors belonging to the moiré Brillouin zone (mBZ), and are typically completely neglected. Though not fundamentally required by symmetries, this valley polarization is typically assumed in many continuum models and is critical in the present work \cite{BM11}. With this additional valley symmetry, the continuum Hamiltonian can be written as \cite{KF18}
\begin{equation}
\mathcal{H}(\mathbf{k}) = \begin{pmatrix}
\mathcal{H}_+(\mathbf{k}) & 0 \\
0 & \mathcal{H}_-(\mathbf{k})
\end{pmatrix}  ,
\end{equation}
where $\mathbf{k}$ is measured from $\bar{\Gamma},$ and $\mathcal{H}_\pm$ are the valley-polarized Hamiltonians given by, in the basis of $\ket{A_1},$ $\ket{B_1},$ $\ket{A_2},$ $\ket{B_2},$
\begin{equation}
\label{eq: valley Hamiltonian}
\mathcal{H}_\nu(\mathbf{k}) = \begin{pmatrix}
\mathcal{H}_\nu^1(\mathbf{k}) & U_\nu^\dagger \\
U_\nu & \mathcal{H}_\nu^2(\mathbf{k})
\end{pmatrix},
\end{equation}
with $\mathcal{H}_\nu^\ell(\mathbf{k})$ is the monolayer Hamiltonian of layer $\ell$ at valley $\nu,$ and $U_\nu$ is the interlayer coupling matrix. In the sublattice representation in momentum space, the monolayer Hamiltonians are 
\begin{equation}
\label{eq: layer Hamiltonian}
\mathcal{H}_\nu^\ell(\mathbf{k}) = - \hbar v_F \left( R\left( \pm \theta /2 \right) \left( \mathbf{k} - \mathbf{K}_\nu^\ell \right)    \right) \cdot \left( \nu \sigma_x, \sigma_y \right),
\end{equation}
where $v_F$ is Dirac velocity of monolayer graphene, $\mathbf{K}_\nu^\ell$ is defined in Fig.~\ref{fig. crystalstructure}b,  the $\sigma$ Pauli matrices act on the sublattice space, and the sign $+$ in the rotation operator is for layer $\ell = 1,$  $-$ for layer $\ell = 2.$ In real space, the interlayer hopping matrix is
\begin{equation}
\begin{split}
U_\nu &= \begin{pmatrix}
w_\text{AA} & w_\text{AB} \\
w_\text{AB} & w_\text{AA}
\end{pmatrix} + \begin{pmatrix}
w_\text{AA} & w_\text{AB} e^{-i \nu 2 \pi /3}\\
w_\text{AB}e^{i \nu 2 \pi /3} & w_\text{AA}
\end{pmatrix} e^{i \nu \mathbf{G}_1^\text{M} \cdot \mathbf{r}} + \begin{pmatrix}
w_\text{AA} & w_\text{AB} e^{i \nu 2 \pi /3}\\
w_\text{AB}e^{-i \nu 2 \pi /3} & w_\text{AA}
\end{pmatrix} e^{i \nu \left(\mathbf{G}_1^\text{M} + \mathbf{G}_2^\text{M} \right)\cdot \mathbf{r}} \\
&= U^0+ U_\nu^1 e^{i \nu \mathbf{G}_1^\text{M} \cdot \mathbf{r}} + U_\nu^2e^{i \nu \left(\mathbf{G}_1^\text{M} + \mathbf{G}_2^\text{M} \right)\cdot \mathbf{r}},
\end{split}
\end{equation}
where $w_\text{AA}$ and $w_\text{AB}$ are coupling constants at an $AA$ and $AB$ regions. To diagonalize equation~\eqref{eq: valley Hamiltonian}, we write the wavefunction at a wavevector $\mathbf{k}$ and band $n$ as 
\begin{equation}
\label{eq: wavefunction}
\ket{\psi_\nu^{n \mathbf{k}}(\mathbf{r})} = e^{i \mathbf{k} \cdot \mathbf{r}} \sum_{\mathbf{G}} e^{i \mathbf{G} \cdot \mathbf{r}} \begin{pmatrix}
A_{1,\nu}^{n \mathbf{k}}\left( \mathbf{G} \right) \\
B_{1,\nu}^{n \mathbf{k}}\left( \mathbf{G} \right) \\
A_{2,\nu}^{n \mathbf{k}}\left( \mathbf{G} \right) \\
B_{2,\nu}^{n \mathbf{k}}\left( \mathbf{G} \right) 
\end{pmatrix} = e^{i \mathbf{k} \cdot \mathbf{r}} \sum_{\mathbf{G}} e^{i \mathbf{G} \cdot \mathbf{r}} \left(  A_{1,\nu}^{n \mathbf{k}}\left( \mathbf{G} \right) \ket{A_1} + B_{1,\nu}^{n \mathbf{k}}\left( \mathbf{G} \right)\ket{B_1} +  A_{2,\nu}^{n \mathbf{k}}\left( \mathbf{G} \right) \ket{A_2}  + B_{2,\nu}^{n \mathbf{k}}\left( \mathbf{G} \right) \ket{B_2} \right),
\end{equation}
where $\mathbf{G} = n_1 \mathbf{G}_1^\text{M} +  n_2 \mathbf{G}_2^\text{M} $ and $n_1$ and $n_2$ are integers. For the Hamiltonian at a particular valley, we expect the physics at wavevectors near the original monolayer Dirac cones to dominate; so we expand the Hamitlonian about a locus of $\mathbf{G}$ points in reciprocal space near to the original Dirac cones. We then diagonalize the resulting Hamiltonian numerically to find the eigenspectrum.

\begin{figure}
\includegraphics[scale=0.5]{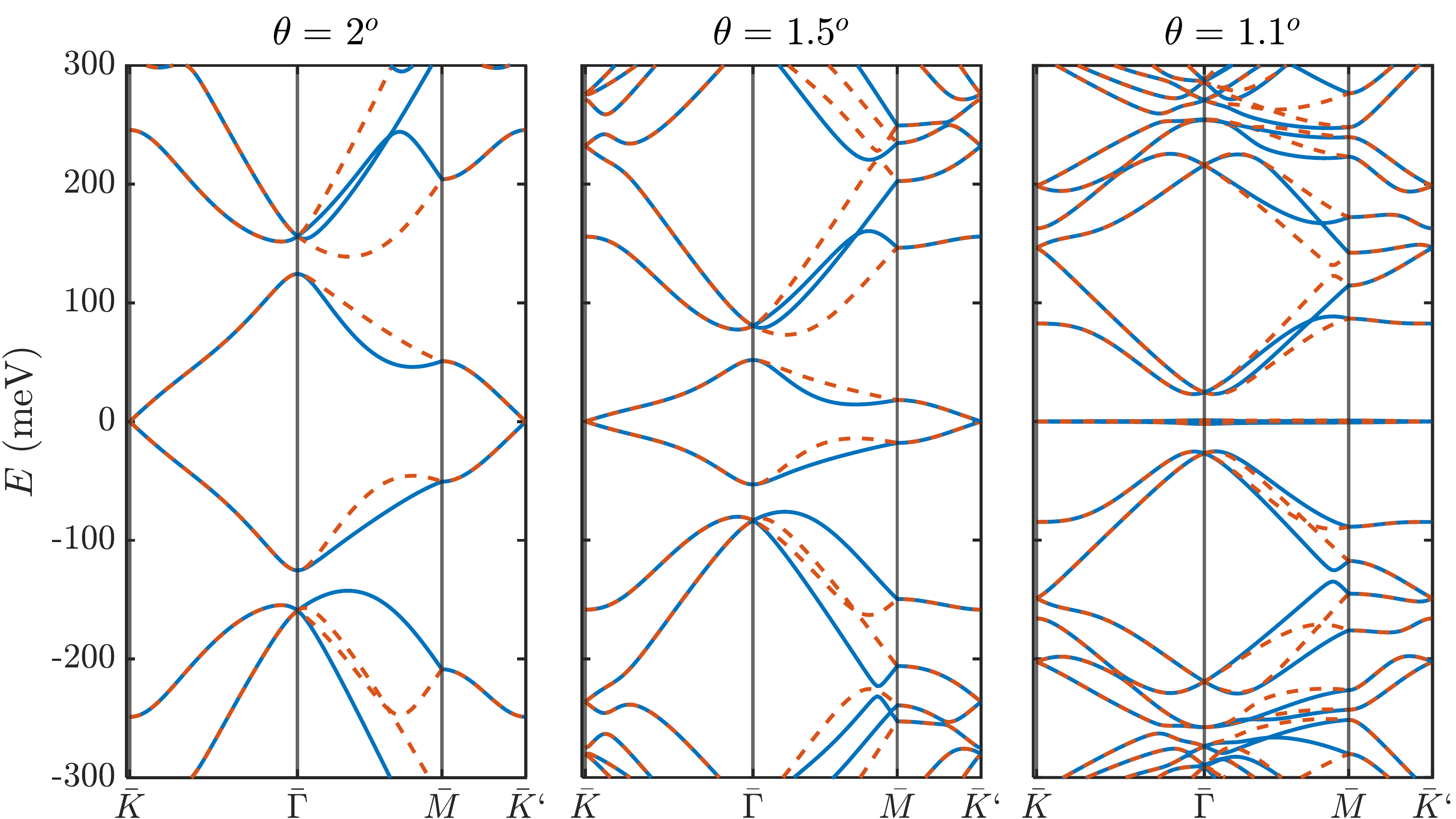}
\caption{Band structure of TBLG at small angles for $\theta = 2^\circ{},$ $\theta = 1.5^\circ{},$ and $\theta = 1.1^\circ{}$ along high-symmetry lines of the mBZ with $\hbar v_F/a = 2135.4$ meV, $w_\text{AA} = 79.7$ meV, and $w_\text{AB} = 97.5$ meV . The solid lines are energy bands from $\nu = +1,$ while the dashed lines are energy bands from $\nu = -1.$ At a sufficiently large angle $( \theta = 2^\circ{}),$ the Dirac cones at $\bar{K}$ and $\bar{K}'$ resemble the original monolayer cones from which they descend. Near $\bar{\Gamma},$ because of interlayer hybridization, the Dirac-like energy bands are separated from the higher- and lower-energy bands by small energy gaps. As we decrease the angle, the Fermi velocity  at $\bar{K}$ and $\bar{K}'$ is renormalized downward, and the bandwidth of these bands near neutrality is significantly reduced. As we approached a magic angle near $\theta = 1.1^\circ{},$ the bandwidth is much smaller than the energy gaps separating these flat bands from the remote higher- and lower-energy bands.  }
\label{fig. spectrum}
\end{figure}

\begin{figure}
\includegraphics[scale=0.5]{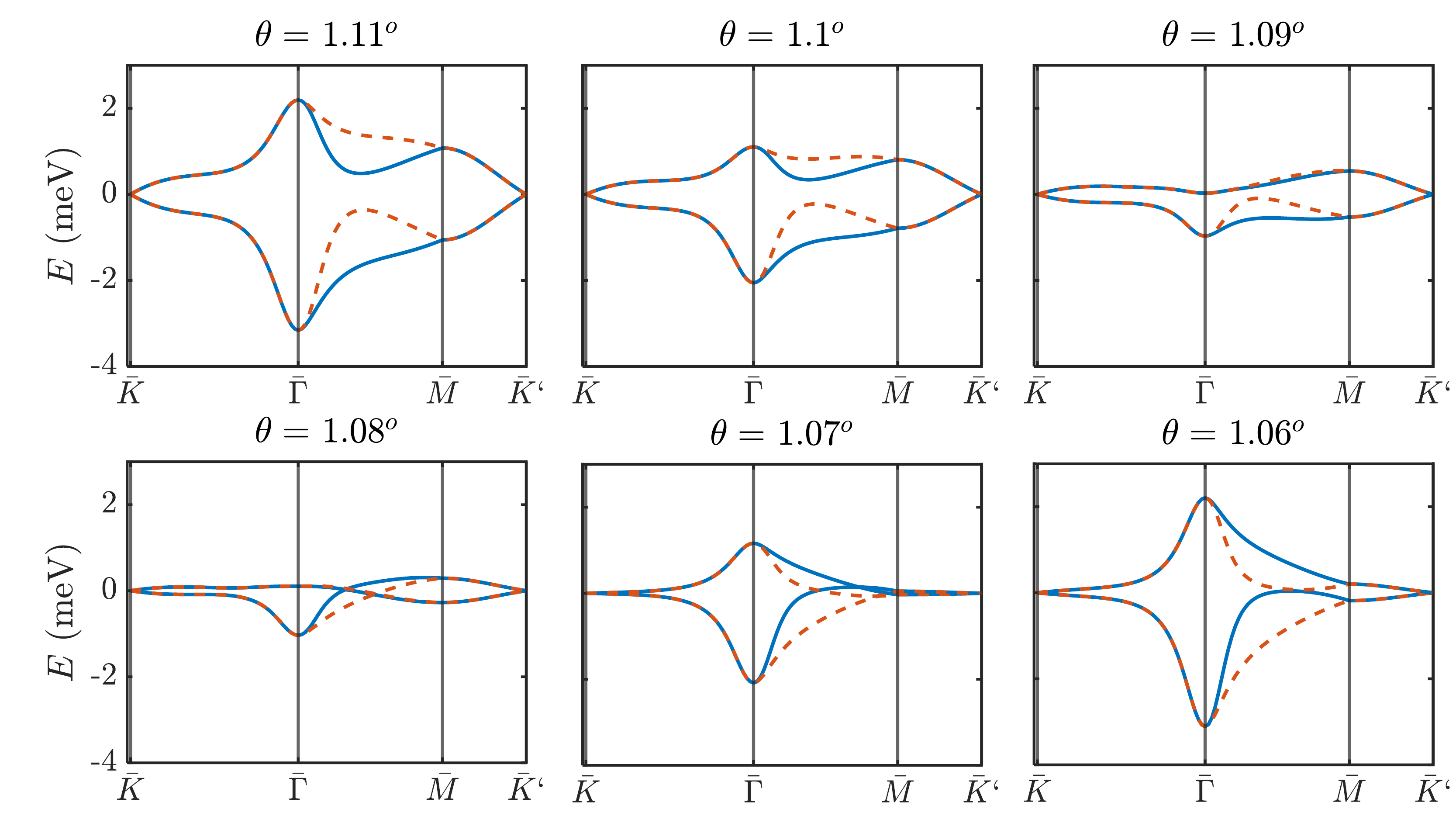}
\caption{ Band structure of TBLG near the first magic angle for $\theta = 1.11^\circ{}-1.06^\circ{}.$ The parameters used here are the same as those used for Fig.~\ref{fig. spectrum}.  As we vary the angle in this range, the Fermi velocity at the $\bar{K}$ and $\bar{K}'$ is reduced until it vanishes at some special angle, typically dubbed the magic angle. As we approach the magic angle from above, pairs of secondary Dirac points first emerge at $\bar{\Gamma},$ propagate along the reflection symmetric lines from $\bar{\Gamma}$ to $\bar{M}$, and vanish once they reach the $\bar{M}.$ These secondary Dirac points can be seen at $\theta = 1.08^\circ{}$ and $\theta = 1.07^\circ{}.$ }
\label{fig. spectrum2}
\end{figure}

The band structures obtained from numerical diagonalization for various small angles are shown for both valleys in Figs.~\ref{fig. spectrum} and~\ref{fig. spectrum2}. We use the parameters $\hbar v_F/a = 2135.4$ meV, $w_\text{AA} = 79.7$ meV, and $w_\text{AB} = 97.5$ meV for simulation \cite{KF18}. The difference in the interlayer coupling constants $w_\text{AA}$ and $w_\text{AB}$ accounts for lattice relaxation in the interlayer distance at the $AA$ and $AB$ regions \cite{NK17, KN19, GW19}. At small, but not too small, angles, e.g. $\theta \approx 1.5^\circ{}-2^\circ{},$ the Dirac cones near neutrality at $\bar{K}$ and $\bar{K}'$ look similar to the original Dirac cones of monolayer graphene from which they descend. As we move away from the zone corners and approach $\bar{\Gamma},$ these linear dispersions are significantly modified by interlayer coupling. The energy bands close to neutrality are isolated from the other bands which form a continuum of states at lower and higher energies, called the remote bands. Much of the interesting physics of TBLG arises from the bands near neutrality; hence, these are called the active bands. As we decrease the angle, the Fermi velocity of the Dirac cones is renormalized downward, and the bandwidth of the active bands is significantly decreased. Near a special angle, the Fermi velocity of the Dirac cones is entirely quenched, resulting in a very flat band well-isolated from the rest of the band structure. In Fig.~\ref{fig. spectrum2}, we zoom into the flat bands in the small-angle range where the magic angle is obtained. For our choice of parameters, this range of angles is about from $\theta \approx 1.1^\circ{}$ to $\theta \approx 1.05^\circ.$ In this range, as we decrease the angle, pairs of secondary Dirac points are generated, first at $\bar{\Gamma}$, then propagate along the $\bar{\Gamma}$-$\bar{M}$ lines, and annihilate at $\bar{M}$ \cite{HC19}. Because we are interested in the chirality of the Dirac cones at the zone corners near charge neutrality, we will concentrate on the region of angles where the active bands only host two Dirac cones that are descendant from the original monolayer cones. The generation of secondary Dirac cones complicates the count of winding numbers in the mBZ. Thus, we will work primarily with angles $\theta > 1.1^\circ{}$ to avoid the topological transitions at occur at small angles near the magic angle. 

\begin{figure}
\includegraphics[scale=0.5]{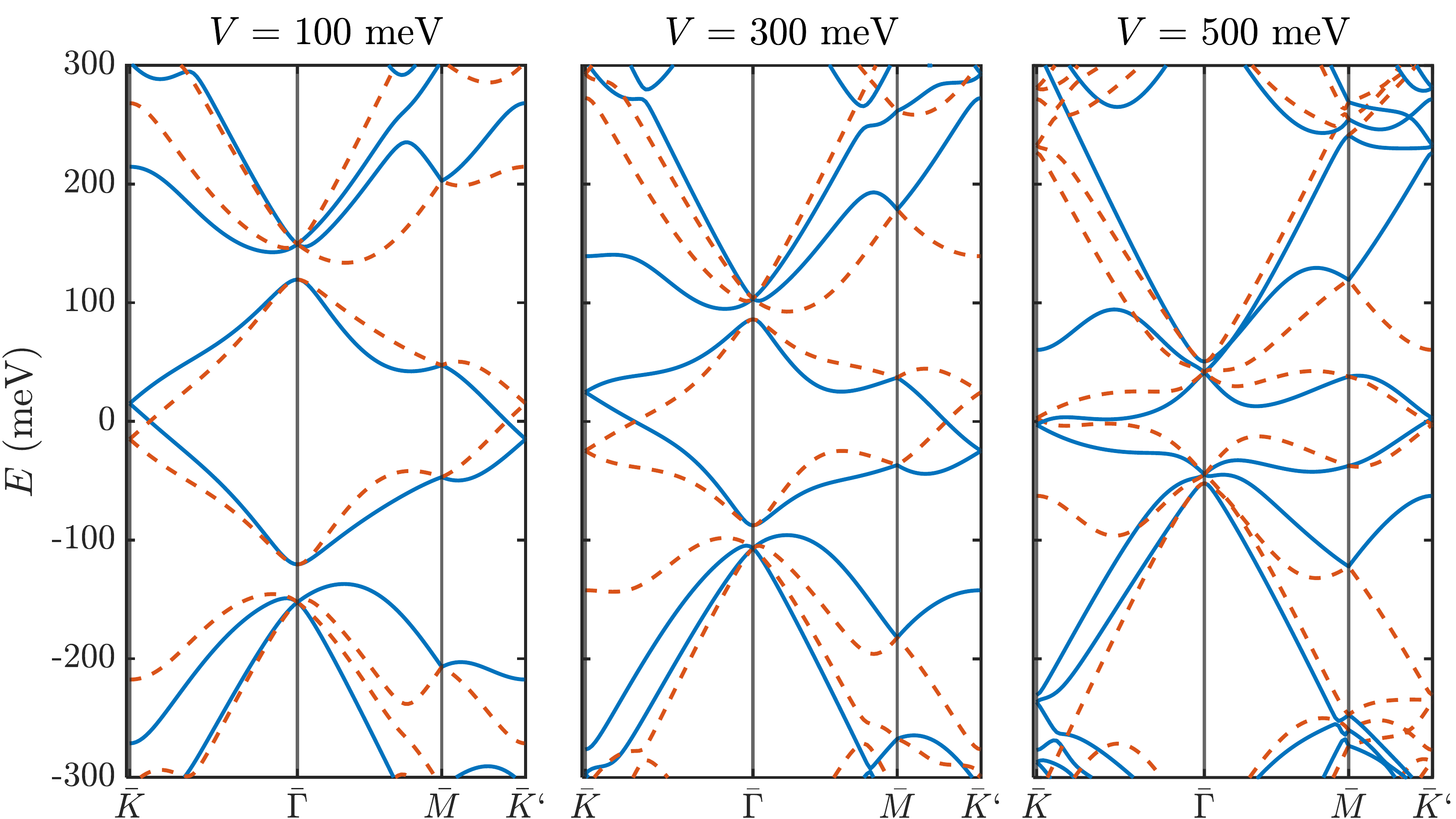}
\caption{ Band structure of TBLG at $\theta = 2^\circ{}$ for various values of the interlayer bias $V.$ The parameters used here are the same as those used for Fig.~\ref{fig. spectrum}.  As we increase $V,$ the two Dirac cones in the same valley are shifted in energy in opposite direction relative to zero energy.  The Dirac cones at a single zone corner in different valleys are also shifted in opposite direction. For $V \lesssim 450$ meV, the active bands remain isolated from the remote bands. However, for larger values of $V,$ the active bands are no longer spectrally separated from the rest of the band structure.   }
\label{fig. spectrum3}
\end{figure}

If we break the symmetry between layers by applying a perpendicular displacement field, then the Dirac points at the zone corners in a single valley are not longer pinned to the same energy \cite{SJP13, EMA18, TNK20}. In this case, equation~\eqref{eq: layer Hamiltonian} is modified to 
\begin{equation}
\label{eq: layer Hamiltonian 2}
\mathcal{H}_\nu^\ell(\mathbf{k}) = - \hbar v_F \left( R\left( \pm \theta /2 \right) \left( \mathbf{k} - \mathbf{K}_\nu^\ell \right)    \right) \cdot \left( \nu \sigma_x, \sigma_y \right) \pm V,
\end{equation}
where $V$ is the interlayer bias and the $\pm$ signs correspond to $\ell = 1$ and $\ell = 2$ respectively. For a sufficiently small bias, the topology of the active bands remains unchanged as no gap closing is induced. As shown in Fig.~\ref{fig. spectrum3}, the effect of a small interlayer bias is to induce spectral shifts of the Dirac cones in a single valley in opposite direction relative to zero energy. Related by time-reversal symmetry, Dirac cones at the same zone corner in different valleys also shift in opposite direction in energy. As we increase the displacement field, the gaps between the active bands and the remote bands narrow until the active bands join the continuum at some parameter-dependent critical field value.

\section{Effective Low-Energy Dirac Theory}

As shown in Figs.~\ref{fig. spectrum} and~\ref{fig. spectrum3}, the band structure near neutrality, at the zone corners, is well-described by linear dispersions. For larger values of $\theta,$ the linear-band approximation becomes valid for a larger range of energies. In this approximation, we can obtain analytic expressions to the wavefunctions of the continuum theory for the four Dirac cones located at 
\begin{equation}
\begin{split}
\mathbf{K}_{-}^1 &= \frac{4 \pi}{3a} \left( \cos \left( \theta/2 \right),- \sin \left( \theta/2 \right) \right), \\
\mathbf{K}_{-}^2 &= \frac{4 \pi}{3a} \left( \cos \left( \theta/2 \right), \sin \left( \theta/2 \right) \right) \\
\mathbf{K}_{+}^1 &= -\frac{4 \pi}{3a} \left( \cos \left( \theta/2 \right),- \sin \left( \theta/2 \right) \right), \\
\mathbf{K}_{+}^2 &= -\frac{4 \pi}{3a} \left( \cos \left( \theta/2 \right), \sin \left( \theta/2 \right) \right).
\end{split}
\end{equation}
To obtain states near these points, we truncate the continuum Hamiltonian to include just the first star of reciprocal lattice vectors about these points \cite{BM11}. In doing so, the effective Hamiltonian for states near $\mathbf{K}_\nu^1$ is 
\begin{equation}
\label{eq: truncated Hamitlonian near K1}
\mathcal{H}_{\mathbf{K}_\nu^1}(\mathbf{k}) =  \begin{pmatrix}
\mathcal{H}_\nu^1 \left( \mathbf{k}  + \mathbf{K}_\nu^1 \right) & U^0 & U_\nu^{1} & U_\nu^{2} \\
U^0 &  \mathcal{H}_\nu^2 \left( \mathbf{k}  + \mathbf{K}_\nu^1 \right) & 0 & 0 \\ 
U^1_\nu & 0 &  \mathcal{H}_\nu^2 \left( \mathbf{k}  + \mathbf{K}_\nu^1  + \nu \mathbf{G}_1^\text{M}\right) & 0 \\
U^2_\nu & 0 & 0 & \mathcal{H}_\nu^2 \left( \mathbf{k}  + \mathbf{K}_\nu^1  + \nu \mathbf{G}_1^\text{M} + \nu \mathbf{G}_2^\text{M}\right) 
\end{pmatrix}.
\end{equation}
The effective Hamiltonian for  states near $\mathbf{K}_\nu^2$ is
\begin{equation}
\label{eq: truncated Hamiltonian near K2}
\mathcal{H}_{\mathbf{K}_\nu^2}(\mathbf{k}) = \begin{pmatrix}
\mathcal{H}_\nu^2 \left( \mathbf{k}  + \mathbf{K}_\nu^2 \right) & U^0 & U_{\nu}^{1} & U_{\nu}^{2} \\ 
U^0 & \mathcal{H}_\nu^1 \left( \mathbf{k}  + \mathbf{K}_\nu^2 \right) & 0 & 0 \\
U_{\nu}^{1} & 0 & \mathcal{H}_\nu^1 \left( \mathbf{k}  + \mathbf{K}_\nu^2 - \nu \mathbf{G}_1^\text{M} \right) & 0 \\
U_{\nu}^{2} & 0 & 0 & \mathcal{H}_\nu^1 \left( \mathbf{k}  + \mathbf{K}_\nu^2 - \nu \mathbf{G}_1^\text{M}- \nu \mathbf{G}_2^\text{M} \right)
\end{pmatrix}.
\end{equation}
For small angles,  we can neglect the rotation of the Pauli matrices to simplify equations~\eqref{eq: truncated Hamitlonian near K1} and~\eqref{eq: truncated Hamiltonian near K2} to, in the absence of interlayer bias, 
\begin{equation}
\label{eq: simplified truncated Hamiltonian near K1}
\mathcal{H}_{\mathbf{K}_\nu^1}(\mathbf{k}) = \begin{pmatrix}
h_\nu \left( \mathbf{k} \right)  & U^0 & U_\nu^1 & U^2_\nu \\
U^0 & h_\nu \left( \mathbf{k} + \nu \boldsymbol{\mathcal{G}}_0  \right)  & 0 & 0 \\
U^1_\nu & 0 & h_\nu \left( \mathbf{k} + \nu \boldsymbol{\mathcal{G}}_1  \right)  & 0 \\
U^2_\nu & 0 & 0 & h_\nu \left( \mathbf{k} + \nu \boldsymbol{\mathcal{G}}_2  \right) 
\end{pmatrix},
\end{equation}
\begin{equation}
\label{eq: simplified truncated Hamiltonian near K2}
\mathcal{H}_{\mathbf{K}_\nu^2}(\mathbf{k}) = \begin{pmatrix}
h_\nu \left( \mathbf{k} \right) & U^0 & U_\nu^1 & U^2_\nu \\
U^0 & h_\nu \left( \mathbf{k} - \nu \boldsymbol{\mathcal{G}}_0  \right)  & 0 & 0 \\
U^1_\nu & 0 & h_\nu \left( \mathbf{k} - \nu \boldsymbol{\mathcal{G}}_1  \right) & 0 \\
U^2_\nu & 0 & 0 & h_\nu \left( \mathbf{k} - \nu \boldsymbol{\mathcal{G}}_2  \right) 
\end{pmatrix},
\end{equation}
where $h_\nu \left(  \mathbf{k} \right) = - \hbar v_F \mathbf{k} \cdot \left( \nu \sigma_x, \sigma_y \right),$ $\boldsymbol{\mathcal{G}}_0 = \mathbf{K}_+
^1 - \mathbf{K}_+^2 = G \left( 0 , 1 \right),$ $ \boldsymbol{\mathcal{G}}_1 = \boldsymbol{\mathcal{G}}_0 +\mathbf{G}_1^\text{M} = G \left( - \frac{\sqrt{3}}{2},-\frac{1}{2} \right), $ $ \boldsymbol{\mathcal{G}}_2 = \boldsymbol{\mathcal{G}}_0 +\mathbf{G}_1^\text{M} +\mathbf{G}_2^\text{M} = G \left(  \frac{\sqrt{3}}{2},-\frac{1}{2} \right),$ and $G = \frac{8 \pi \sin \theta/2}{3a}.$ Even though the $8 \times 8$ matrices in equations~\eqref{eq: simplified truncated Hamiltonian near K1} and~\eqref{eq: simplified truncated Hamiltonian near K2} are not too complicated, they cannot be diagonalized exactly for a general value of $\mathbf{k}.$ Thus, we have to resort to perturbation theory in order to find approximate eigenstates. Since we are most interested in the active bands near neutrality, we first find the eigenstates at $\mathbf{k} = \mathbf{0}$ for zero energy. Using the notation in equation~\eqref{eq: wavefunction}, writing a two-component spinor in layer $\ell$ and wavevector $\mathbf{G}$ as $\psi_{\ell , \nu}^{n \mathbf{k}}(\mathbf{G}) = \begin{pmatrix}
A_{\ell , \nu}^{n \mathbf{k}}(\mathbf{G}) & B_{\ell , \nu}^{n \mathbf{k}}(\mathbf{G})
\end{pmatrix}^T,$ the eigenstates for equations~\eqref{eq: simplified truncated Hamiltonian near K1} and~\eqref{eq: simplified truncated Hamiltonian near K2} can be written as 
\begin{equation}
\ket{\psi_{\mathbf{K}_\nu^1}^{n \mathbf{k}}} = \begin{pmatrix}
\psi_{1 , \nu}^{n \mathbf{k}}\left( \mathbf{0} \right) \\
\psi_{2 , \nu}^{n \mathbf{k}}\left( \mathbf{0} \right) \\
\psi_{2 , \nu}^{n \mathbf{k}}\left( \nu\mathbf{G}_1^\text{M} \right) \\
\psi_{2 , \nu}^{n \mathbf{k}}\left( \nu\mathbf{G}_1^\text{M}+ \nu\mathbf{G}_2^\text{M} \right)
\end{pmatrix} \quad \text{and} \quad 
\ket{\psi_{\mathbf{K}_\nu^2}^{n \mathbf{k}}} = \begin{pmatrix}
\psi_{2 , \nu}^{n \mathbf{k}}\left( \mathbf{0} \right) \\
\psi_{1 , \nu}^{n \mathbf{k}}\left( \mathbf{0} \right) \\
\psi_{1 , \nu}^{n \mathbf{k}}\left( -\nu\mathbf{G}_1^\text{M} \right) \\
\psi_{1 , \nu}^{n \mathbf{k}}\left( -\nu\mathbf{G}_1^\text{M}- \nu\mathbf{G}_2^\text{M} \right)
\end{pmatrix}.
\end{equation}
Let us  solve for states at zero energy at $\mathbf{k} = \mathbf{0}.$ Solving for $\mathcal{H}_{\mathbf{K}_\nu^1}(\mathbf{0}) \ket{\psi_{\mathbf{K}_\nu^1}^{n \mathbf{k}}}  = 0,$ we find 
\begin{equation}
\begin{split}
 \sum_{m = 0}^2 U^m_\nu \psi_{2,\nu}^{n,m}  &= 0, \\
U^m_\nu \psi_{1,\nu}^n +  h_\nu \left( \nu \boldsymbol{\mathcal{G}}_m \right) \psi_{2,\nu}^{n,m} &= 0,
\end{split}
\end{equation}
where we have used a condensed notation $\psi_{2,\nu}^{n,0} = \psi_{2 , \nu}^{n \mathbf{0}}\left( \mathbf{0} \right),$  $\psi_{2,\nu}^{n,1} = \psi_{2 , \nu}^{n \mathbf{0}}\left( \nu\mathbf{G}_1^\text{M} \right),$ and $\psi_{2,\nu}^{n,2} = \psi_{2 , \nu}^{n \mathbf{0}}\left( \nu\mathbf{G}_1^\text{M} + \nu\mathbf{G}_2^\text{M}\right).$ Using substitution, we find that $\psi_{1,\nu}^n$ must satisfy the equation 
\begin{equation}
\left(     \sum_{m = 0}^2 U^m_\nu \left[ h_\nu \left( \nu \boldsymbol{\mathcal{G}}_m \right) \right]^{-1}U^m_\nu \right)\psi_{1,\nu}^n = 0,
\end{equation}
which has two degenerate solutions
\begin{equation}
\psi_{1,\nu}^1 = \begin{pmatrix}
1 \\
0
\end{pmatrix} \quad \text{and} \quad \psi_{1,\nu}^2 = \begin{pmatrix}
0 \\
1
\end{pmatrix}.
\end{equation}
The other components are obtained consistently via $\psi_{2,\nu}^{n,m} = - \left[ h_\nu \left( \nu \boldsymbol{\mathcal{G}}_m \right) \right]^{-1} U^m_\nu \psi_{1,\nu}^n$
\begin{equation}
\begin{aligned}
\psi_{2,\nu}^{1,0} &= \frac{1}{\hbar v_F G}\begin{pmatrix}
-i \nu w_\text{AB} \\
i \nu w_\text{AA}
\end{pmatrix},  \quad &&\psi_{2,\nu}^{2,0} = \frac{1}{\hbar v_F G}\begin{pmatrix}
-i \nu w_\text{AA} \\
i \nu w_\text{AB}
\end{pmatrix}, \\
\psi_{2,\nu}^{1,1} &= \frac{1}{\hbar v_F G}\begin{pmatrix}
-i \nu  w_\text{AB}\\
i \nu w_\text{AA} e^{2 \pi \nu  i /3}
\end{pmatrix}, \quad 
&&\psi_{2,\nu}^{2,1} = \frac{1}{\hbar v_F G}\begin{pmatrix}
-i \nu w_\text{AA} e^{-2 \pi \nu i /3} \\
i \nu  w_\text{AB} 
\end{pmatrix}, \\
\psi_{2,\nu}^{1,2} &= \frac{1}{\hbar v_F G}\begin{pmatrix}
-i \nu w_\text{AB} \\
i \nu w_\text{AA} e^{-2 \pi \nu i /3}
\end{pmatrix}, \quad 
&&\psi_{2,\nu}^{2,2} = \frac{1}{\hbar v_F G}\begin{pmatrix}
 -i \nu w_\text{AA}e^{2\pi \nu i /3} \\
i\nu  w_\text{AB} 
\end{pmatrix}. 
\end{aligned}
\end{equation}
The eigenstates for $\mathcal{H}_{\mathbf{K}_\nu^2}(\mathbf{0})$ can be found similarly by exchanging the layer and valley indices. They are
\begin{equation}
\begin{aligned}
\psi_{2,\nu}^1 &= \begin{pmatrix}
1 \\
0
\end{pmatrix} ,&&\psi_{2,\nu}^2 = \begin{pmatrix}
0 \\
1
\end{pmatrix},\\
\psi_{1,\nu}^{1,0} &= \frac{1}{\hbar v_F G}\begin{pmatrix}
i \nu w_\text{AB} \\
-i \nu w_\text{AA}
\end{pmatrix},  \quad &&\psi_{1,\nu}^{2,0} = \frac{1}{\hbar v_F G}\begin{pmatrix}
i \nu w_\text{AA} \\
-i \nu w_\text{AB}
\end{pmatrix}, \\
\psi_{1,\nu}^{1,1} &= \frac{1}{\hbar v_F G}\begin{pmatrix}
i \nu  w_\text{AB}\\
-i\nu w_\text{AA} e^{2 \pi \nu  i /3}
\end{pmatrix}, \quad 
&&\psi_{1,\nu}^{2,1} = \frac{1}{\hbar v_F G}\begin{pmatrix}
i \nu w_\text{AA} e^{-2 \pi \nu i /3} \\
-i \nu  w_\text{AB} 
\end{pmatrix}, \\
\psi_{1,\nu}^{1,2} &= \frac{1}{\hbar v_F G}\begin{pmatrix}
i \nu w_\text{AB} \\
-i \nu w_\text{AA} e^{ -2\pi \nu i /3}
\end{pmatrix}, \quad 
&&\psi_{1,\nu}^{2,2} = \frac{1}{\hbar v_F G}\begin{pmatrix}
i\nu w_\text{AA}e^{2\pi \nu i /3} \\
-i\nu  w_\text{AB} 
\end{pmatrix}. 
\end{aligned}
\end{equation}

\begin{SCfigure}
 \centering
  \caption{Contour plots of the overlap of the wavefunctions of the active bands, indexed by $\chi,$ calculated analytically from the truncated Hamiltonian and those calculated numerically from the full continuum model, $ \bigg\vert \bra{\psi_{\mathbf{K}_-^1,\text{truncated}}^{\chi \mathbf{k}}}\ket{\psi_{\mathbf{K}_-^1,\text{continuum}}^{\chi \mathbf{k}}} \bigg \vert^2,$ for different angles. For $\theta$ well above the magic angle shown in (a)-(d), the overlap is near unity for momentum close to the zone corners. As we approach the magic angle near $\theta = 1.08^\circ,$ shown in (e)-(h), the eigenstates of the truncated Hamiltonians are no longer a good approximation of the eigenstates of the continuum model. We plot only the results for $\mathbf{k}$ near to $\mathbf{K}_-^1,$ but the results for $\mathbf{k}$ near the other three Dirac cones are similar. The parameters used to simulate the continuum model here are the same to those used to simulate Fig.~\ref{fig. spectrum}.}
\includegraphics[scale=0.55]{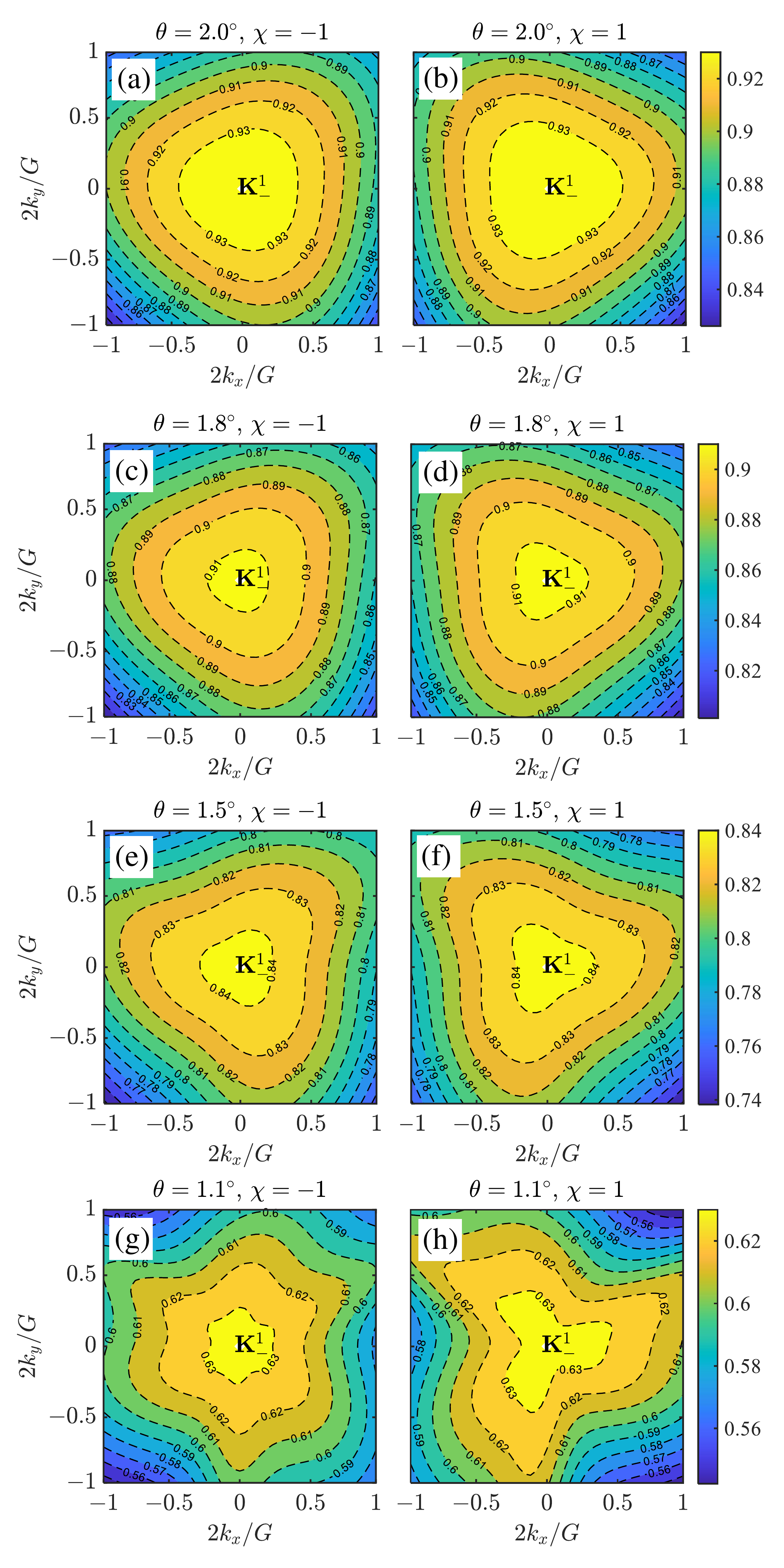}
\label{fig: overlap}
\end{SCfigure}

We can use the states at zero energy as the basis states to perform a perturbation expansion for finite $\mathbf{k}.$ The effective Hamiltonians expanded to linear order in $\mathbf{k}$  are
\begin{equation}
\label{eq: effective Hamiltonian}
\mathcal{H}^\text{eff}_{\mathbf{K}_\nu^1}(\mathbf{k})  = \begin{pmatrix}
\bra{\psi_{\mathbf{K}_\nu^1}^{1, \mathbf{0}}} \mathcal{H}_{\mathbf{K}_\nu^1}(\mathbf{k})\ket{\psi_{\mathbf{K}_\nu^1}^{1, \mathbf{0}}} & \bra{\psi_{\mathbf{K}_\nu^1}^{1, \mathbf{0}}} \mathcal{H}_{\mathbf{K}_\nu^1}(\mathbf{k})\ket{\psi_{\mathbf{K}_\nu^1}^{2, \mathbf{0}}} \\
\bra{\psi_{\mathbf{K}_\nu^1}^{2, \mathbf{0}}} \mathcal{H}_{\mathbf{K}_\nu^1}(\mathbf{k})\ket{\psi_{\mathbf{K}_\nu^1}^{1, \mathbf{0}}} & \bra{\psi_{\mathbf{K}_\nu^1}^{2, \mathbf{0}}} \mathcal{H}_{\mathbf{K}_\nu^1}(\mathbf{k})\ket{\psi_{\mathbf{K}_\nu^1}^{2, \mathbf{0}}}
\end{pmatrix} =   - \hbar \tilde{v}_F \mathbf{k} \cdot \left(\nu  \sigma_x, \sigma_y \right),
\end{equation} 
where the renormalized velocity is 
\begin{equation}
\tilde{v}_F = \frac{v_F \left(\hbar^2 v_F^2 G^2-3 w_\text{AB}^2\right)}{3w^2_\text{AA}+3w^2_\text{AB}+ \hbar^2 v_F^2 G^2}.
\end{equation}
In this approximation, while the magnitude of the renormalized velocity is in general dependent on both interlayer couplings $w_\text{AA}$ and $w_\text{AB},$ the vanishing of the Dirac velocity, and hence the location of the magic angle, is only determined by $w_\text{AB}$ through the relation $\hbar v_F G = \sqrt{3} w_\text{AB}$  \cite{GAGSJ17, TKV19}. The effective Hamiltonian at $\mathbf{K}_\nu^2$ is of the same form as equation~\eqref{eq: effective Hamiltonian}. The eigenvalues and eigenvectors for equation~\eqref{eq: effective Hamiltonian} are 
\begin{equation}
\varepsilon_{\chi,\mathbf{k} } = \chi \hbar \tilde{v}_F k, \quad \Phi^\nu_\chi(\mathbf{k}) = \frac{1}{\sqrt{2}}\begin{pmatrix}
e^{-i \nu \phi_\mathbf{k}}\\
-\chi \nu
\end{pmatrix},
\end{equation}
where $\chi = \pm$ indicates the valence and conduction bands respectively, $k = \sqrt{k_x^2+k_y^2},$ and $\phi_\mathbf{k} = \text{angle} \left( k_x+ik_y \right).$ The real-space representation in the composite layer and sublattice basis of the Bloch states near $\mathbf{K}_\nu^1$ is 
\begin{equation}
\label{eq: wavefunction at K1}
\begin{split}
\ket{\psi_{\mathbf{K}_\nu^1}^{\chi \mathbf{k}} \left( \mathbf{r} \right) } &= \frac{e^{i \left(\mathbf{K}_\nu^1 + \mathbf{k}   \right) \cdot \mathbf{r}}}{\sqrt{2}\sqrt{3 w_\text{AA}^2 + 3 w_\text{AB}^2 + \hbar^2 v_F^2 G^2}}   \begin{pmatrix}
\hbar v_F G e^{- i \nu \phi_\mathbf{k}} \\
-\hbar v_F G \chi \nu \\
-i \nu w_\text{AB} e^{-i\nu \phi_\mathbf{k}} g_\nu^{(1)}(\mathbf{r}) + i\chi w_\text{AA} g_\nu^{(2)}(\mathbf{r}) \\
-i \chi w_\text{AB}  g_\nu^{(1)}(\mathbf{r}) + i \nu w_\text{AA} e^{-i\nu \phi_\mathbf{k}}g_\nu^{(3)}(\mathbf{r})
\end{pmatrix},
\end{split}
\end{equation}
where 
\begin{equation}
\begin{split}
g_\nu^{(1)}(\mathbf{r}) &= 1 + e^{i \nu \mathbf{G}_1^\text{M} \cdot \mathbf{r}} + e^{i \nu \left( \mathbf{G}_1^\text{M} +\mathbf{G}_2^\text{M} \right) \cdot \mathbf{r}} =  1+2 e^{ - \frac{3}{2} i \nu G y } \cos \left( \frac{\sqrt{3}}{2} G x \right), \\
g_\nu^{(2)}(\mathbf{r}) &=  1 + e^{-2\pi \nu i /3}e^{i \nu \mathbf{G}_1^\text{M} \cdot \mathbf{r}} + e^{2\pi \nu i /3}e^{i \nu \left( \mathbf{G}_1^\text{M} +\mathbf{G}_2^\text{M} \right) \cdot \mathbf{r}} = 1-e^{ - \frac{3}{2} i \nu G y }\left[\sqrt{3}\sin \left( \frac{\sqrt{3}}{2} G x  \right)+ \cos \left( \frac{\sqrt{3}}{2} G x  \right)\right],\\
g_\nu^{(3)}(\mathbf{r}) &= 1 + e^{2\pi \nu i /3}e^{i \nu \mathbf{G}_1^\text{M} \cdot \mathbf{r}} + e^{-2\pi \nu i /3}e^{i \nu \left( \mathbf{G}_1^\text{M} +\mathbf{G}_2^\text{M} \right) \cdot \mathbf{r}} = 1+e^{ - \frac{3}{2} i \nu G y }\left[\sqrt{3}\sin \left( \frac{\sqrt{3}}{2} G x  \right)- \cos \left( \frac{\sqrt{3}}{2} G x  \right)\right].
\end{split}
\end{equation}
Likewise, the real-space representation for the Bloch states near $\mathbf{K}_\nu^2$ is 
\begin{equation}
\label{eq: wavefunction at K2}
\begin{split}
\ket{\psi_{\mathbf{K}_\nu^2}^{\chi \mathbf{k}} \left( \mathbf{r} \right) } &= \frac{e^{i \left(\mathbf{K}_\nu^2 + \mathbf{k}   \right) \cdot \mathbf{r}}}{\sqrt{2}\sqrt{3 w_\text{AA}^2 + 3 w_\text{AB}^2 + \hbar^2 v_F^2 G^2}}   \begin{pmatrix}
 i \nu w_\text{AB} e^{-i\nu \phi_\mathbf{k}} g_{-\nu}^{(1)}(\mathbf{r}) - i\chi w_\text{AA} g_{-\nu}^{(3)}(\mathbf{r}) \\
 i \chi w_\text{AB}  g_{-\nu}^{(1)}(\mathbf{r}) - i \nu w_\text{AA} e^{-i\nu \phi_\mathbf{k}}g_{-\nu}^{(2)}(\mathbf{r})\\
 \hbar v_F G e^{- i \nu \phi_\mathbf{k}} \\
 -\hbar v_F G \chi \nu
\end{pmatrix}.
\end{split}
\end{equation}
As a first check of consistency, we see that if interlayer coupling is made to vanish, the low-energy modes at $\mathbf{K}^1_\nu$  only have non-zero components on layer 1 because these are just the eigenstates of the decoupled monolayer Dirac cone of layer 1; similarly, the low-energy modes at $\mathbf{K}^2_\nu$  only have non-zero components on layer 2. To further determine the validity of this $G$-truncated approximation at finite interlayer hopping, we calculate the overlap of the wavefunctions from equations~\eqref{eq: wavefunction at K1} and~\eqref{eq: wavefunction at K2} and those numerically obtained from the full continuum model. The results are shown in Fig.~\ref{fig: overlap}. As seen in Fig.~\ref{fig: overlap}a-d, for angles well above the magic angle, the eigenstates obtained from the truncated effective Hamiltonians agree well with the eigenstates of the active bands calculated numerically from the full continuum model, especially for wavevectors near the zone corners. As we approach the magic angle near $\theta \approx 1.08^\circ,$ the eigenstates of the continuum model become much more complicated. Thus, including just the first star of reciprocal lattice vectors in the effective Hamiltonians is no longer sufficient, resulting in poor overlap with the numerical eigenstates, as shown in Fig.~\ref{fig: overlap}e-h. For our purpose, it is not necessary to work strictly at the magic angle. In fact, it is more practically relevant to work at larger angles where the bands are not too flat and the moiré unit cell is not too large. In this regime, the eigenstates in equations~\eqref{eq: wavefunction at K1} and~\eqref{eq: wavefunction at K2} are a good approximation to the continuum eigenstates at low energies. The behavior of the charge distribution is characteristically different at the three different regions. At $\mathbf{r} = \mathbf{0},$ corresponding to an $AA$ region, $g_\nu^{(1)}(\mathbf{0}) = 3$ and $g_\nu^{(2)}(\mathbf{0})= g_\nu^{(3)}(\mathbf{0}) = 0.$ Thus, all elements of the density matrix for $\mathbf{k}$ near both $\mathbf{K}_\nu^{1}$ and $\mathbf{K}_\nu^{2}$ are generically all non-zero. At an AB region, $\mathbf{r} = \mathbf{r}_\text{AB} = L_\text{M} \left( -1/2, \sqrt{3}/2 \right)/\sqrt{3},$  $g_\nu^{(3)}(\mathbf{r}_\text{AB}) = 3$ and $g_\nu^{(2)}(\mathbf{r}_\text{AB})= g_\nu^{(1)}(\mathbf{r}_\text{AB}) = 0.$ States near $\mathbf{K}_\nu^{1}$ have no density in the $A_2$ sublattice; while states near $\mathbf{K}_\nu^{2}$ have no density in the $B_1$ sublattice. Likewise, at an BA region, $\mathbf{r} = \mathbf{r}_\text{BA} = L_\text{M} \left( 1/2, \sqrt{3}/2 \right)/\sqrt{3},$  $g_\nu^{(2)}(\mathbf{r}_\text{BA}) = 3$ and $g_\nu^{(1)}(\mathbf{r}_\text{BA})= g_\nu^{(3)}(\mathbf{r}_\text{BA}) = 0.$ States near $\mathbf{K}_\nu^{1}$ have no density in the $B_2$ sublattice; while states near $\mathbf{K}_\nu^{2}$ have no density in the $A_1$ sublattice.

\section{$\mathcal{T}$ Matrix Formalism}

The Green's function formalism is a standard technique to study impurity scattering \cite{EE06, K12}. Here, we briefly review this general approach to clarify our notation.  The Green's function operator is defined as the resolvant of the Hamiltonian operator $\mathcal{H}$
\begin{equation}
\mathcal{G}(z) = \frac{1}{z- \mathcal{H}},
\end{equation}
where $z$ is a complex variable. By construction, the resolvant solves the operator equation $\left(z - \mathcal{H} \right)\mathcal{G}\left( z \right) = 1,$ justifying the name Green's function. Let $\mathcal{H}$ be diagonalized by eigenstates $\lbrace \ket{\Psi_n} \rbrace$ with energies $\lbrace \varepsilon_n \rbrace$ labeled by quantum numbers $n.$ The states are normalized by $\bra{\Psi_n} \ket{\Psi_m} = N \delta_{nm},$ where $N$ is a positive normalization constant. With this normalization convention, the spectral resolution of the identity is
\begin{equation}
\mathbb{I} = \frac{1}{N} \sum_n \ket{\Psi_n}\bra{\Psi_n}.
\end{equation}
The Green's function has the following spectral representation
\begin{equation}
\label{eq: spectral representation of Green's function}
\mathcal{G}\left( z \right) = \frac{1}{N^2} \sum_{n,m} \ket{\Psi_n}\bra{\Psi_n} \frac{1}{z - \mathcal{H}} \ket{\Psi_m}\bra{\Psi_m} =\frac{1}{N}\sum_n \frac{ \ket{\Psi_n}\bra{\Psi_n}}{z- \varepsilon_n}.
\end{equation}
From equation~\eqref{eq: spectral representation of Green's function}, we see that the poles of the Green's function on the real axis are precisely the locations of the energies. The residues contain information about the wavefunctions. From equation~\eqref{eq: spectral representation of Green's function}, we can project the Green's function onto any desired representation. For example, we can write the resolvant in the momentum representation 
\begin{equation}
\mathcal{G}\left( z, \mathbf{p}, \mathbf{p}' \right) = \frac{1}{N} \sum_n \frac{\ket{\Psi_{n, \mathbf{p}}}\bra{\Psi_{n, \mathbf{p}'}}}{z - \varepsilon_n},
\end{equation}
where $\bra{\mathbf{p}}\ket{\Psi_n} = \ket{\Psi_{n,\mathbf{p}}}.$ Likewise, we can write the Green's function in the real-space representation 
\begin{equation}
\label{eq: real space representation of Green's function}
\begin{split}
\mathcal{G}\left( z, \mathbf{r}, \mathbf{r}' \right) &= \frac{1}{N}\sum_n \frac{\ket{\Psi_n (\mathbf{r})}\bra{\Psi_n(\mathbf{r}')}}{z - \varepsilon_n} = \frac{1}{N} \sum_n \int \frac{d^d \mathbf{p}d^d \mathbf{p}'}{(2 \pi)^{2d}} \frac{\bra{\mathbf{r}}\ket{\mathbf{p}}\bra{\mathbf{p}}\ket{\Psi_n}\bra{\Psi_n}\ket{\mathbf{p}'}\bra{\mathbf{p}'}\ket{\mathbf{r}'}  }{z-\varepsilon_n} \\
&= \frac{1}{N} \sum_n \int \frac{d^d \mathbf{p}d^d \mathbf{p}'}{(2 \pi)^{2d}} e^{i \mathbf{p} \cdot \mathbf{r}-i \mathbf{p}' \cdot \mathbf{r}'}\frac{\ket{\Psi_{n, \mathbf{p}}}\bra{\Psi_{n, \mathbf{p}'}} }{z - \varepsilon_n} = \int \frac{d^d \mathbf{p}d^d \mathbf{p}'}{(2 \pi)^{2d}} e^{i \mathbf{p} \cdot \mathbf{r}-i \mathbf{p}' \cdot \mathbf{r}'} \mathcal{G}\left( z, \mathbf{p}, \mathbf{p}' \right),
\end{split}
\end{equation}
where $d$ is the dimension. If the Hamiltonian has continuous translation symmetry, then the energy eigenstates can be chosen to be simultaneous momentum eigenstates. In that case, the Green's function is diagonal in the momentum representation 
\begin{equation}
\mathcal{G}(z, \mathbf{p}, \mathbf{p}') = \frac{1}{N} \sum_n \frac{\ket{\Psi_{n, \mathbf{p}}}\bra{\Psi_{n, \mathbf{p}'}}}{z - \varepsilon_n} \left( 2 \pi \right)^d \delta^d \left( \mathbf{p} - \mathbf{p}' \right),
\end{equation}
and thus, we can simplify
\begin{equation}
\mathcal{G}(z, \mathbf{p}) = \frac{1}{N} \sum_{n} \frac{\ket{\Psi_{n, \mathbf{p}}}\bra{\Psi_{n, \mathbf{p}}}}{z-\varepsilon_{n,\mathbf{p}}},
\end{equation}
where the energies are now also indexed by $\mathbf{p}$ in addition to other degrees of freedom indexed by $n.$ In real space, the Green function only depends on $\mathbf{r}-\mathbf{r}',$ a restatement of translation symmetry. We then can write 
\begin{equation}
\mathcal{G}(z, \mathbf{r}) = \int \frac{d^d\mathbf{p}}{\left( 2 \pi \right)^d} \mathcal{G}\left( z,\mathbf{p} \right) e^{i \mathbf{p} \cdot \mathbf{r}}.
\end{equation}
Having continuous translational symmetry greatly simplifies the Green's function formalism. However, condensed matter systems typically only have discrete translational symmetry.  In this case, Bloch's theorem shows that energy eigenstates can be labeled by a band index $n$ and a crystal momentum $\mathbf{k}$ in the first Brillouin zone, possibly in addition to labels of other internal symmetries. A Bloch state $\ket{\Psi_{n, \mathbf{k}}}$ with energy $\varepsilon_{n, \mathbf{k}}$ satisfying $\mathcal{H}\ket{\Psi_{n, \mathbf{k}}} = \varepsilon_{n, \mathbf{k}}\ket{\Psi_{n, \mathbf{k}}}$ is normalized to unity within one unit cell $\Omega_\text{cell}.$ Due to Bloch's theorem, energy eigenstates can be decomposed into a plane wave and a cell-periodic part $\ket{\Psi_{n,\mathbf{k}}} = e^{i \mathbf{k} \cdot \mathbf{r}} \ket{u_{n,\mathbf{k}}}.$ In the Bloch representation, equation~\eqref{eq: real space representation of Green's function} takes the form 
\begin{equation}
\label{eq: Bloch Green's function}
\begin{split}
\mathcal{G}\left(z, \mathbf{r}+\mathbf{R}, \mathbf{r}'   \right) &= \frac{V_\text{cell}}{\left( 2 \pi \right)^d} \sum_n\int_\text{BZ} d^d\mathbf{k} \frac{\ket{\Psi_{n,\mathbf{k}}\left( \mathbf{r}+\mathbf{R}\right)}\bra{\Psi_{n,\mathbf{k}}\left( \mathbf{r}' \right)}}{z-\varepsilon_{n,\mathbf{k}}} \\
&= \frac{V_\text{cell}}{\left(2 \pi  \right)^d} \sum_n \int_\text{BZ} d^d\mathbf{k} e^{i \mathbf{k} \cdot \left( \mathbf{R}+\mathbf{r}-\mathbf{r}' \right)} \frac{\ket{u_{n,\mathbf{k}}\left( \mathbf{r}\right)}\bra{u_{n,\mathbf{k}}\left( \mathbf{r}' \right)}}{z-\varepsilon_{n,\mathbf{k}}} = \frac{V_\text{cell}}{\left(2 \pi  \right)^d}  \int_\text{BZ} d^d\mathbf{k} e^{i \mathbf{k} \cdot \left( \mathbf{R}+\mathbf{r}-\mathbf{r}' \right)} \mathcal{G} \left( z, \mathbf{r},\mathbf{r}',\mathbf{k} \right),
\end{split}
\end{equation}
where $\mathbf{R}$ is a reciprocal lattice vector, $\mathbf{r}$ and $\mathbf{r}'$ are defined within a single unit cell, and $\mathcal{G}\left(z, \mathbf{r},\mathbf{r}',\mathbf{k}\right)$ is defined only by the periodic parts of the Bloch wavefunctions. Computing $\ket{u_{n,\mathbf{k}}\left( \mathbf{r} \right)}$ for a general $\mathbf{r}$ within a unit cell is  a difficult task because one needs to diagonalize the periodic Hamiltonian exactly. Fortunately, knowledge of the exact wavefunction is often not necessary when we are interested in long-wavelength phenomena. In that approximation, an especially useful representation for periodic Hamiltonians is the tight-binding basis where we approximate the energy eigenstates as linear combinations of a finite (usually small) set of active Wannier orbitals within each unit cell.

In the tight-binding representation, let the Wannier orbitals be located at $\boldsymbol{\tau}_\alpha$ within a unit cell, where $\alpha = 1,2,3,...,m,$ and $m$ is the number of orbitals relevant to the physics that we wish to describe. We denote the Wannier state at $\mathbf{R} + \boldsymbol{\tau}_\alpha$ as $\ket{W_{\alpha,\mathbf{R}}}.$ Then, we can form basis states using linear combinations of the following form \cite{V18}
\begin{equation}
\label{eq: Bloch states}
\ket{\phi_{\alpha, \mathbf{k}}} = \frac{1}{\sqrt{N}}\sum_{\mathbf{R}} e^{i \mathbf{k} \cdot \left(\mathbf{R} + \boldsymbol{\tau}_\alpha \right)} \ket{W_{\alpha,\mathbf{R}}}
\end{equation}
The matrix elements of the Hamiltonian are
\begin{equation}
\label{eq: tight-binding Hamiltonian}
\mathcal{H}_{\alpha \beta}(\mathbf{k}) = \bra{\phi_{\alpha, \mathbf{k}}} \mathcal{H} \ket{\phi_{\beta, \mathbf{k}}} = \sum_\mathbf{R} e^{i \mathbf{k} \cdot \left( \mathbf{R}+\boldsymbol{\tau}_\alpha - \boldsymbol{\tau}_\beta \right)} \mathcal{H}_{\alpha \beta} \left( \mathbf{R} \right),
\end{equation}
where $\mathcal{H}_{\alpha \beta} \left(\mathbf{R}\right) = \bra{W_{\alpha,\mathbf{0}}}\mathcal{H}\ket{W_{\beta ,\mathbf{R}}}$ are the on-site and hopping constants. The Hamiltonian in equation~\eqref{eq: tight-binding Hamiltonian} is now a matrix for each $\mathbf{k}$ that can be diagonalized 
\begin{equation}
\mathcal{H}(\mathbf{k}) \Phi_{n,\mathbf{k}} = \varepsilon_{n,\mathbf{k}}\Phi_{n,\mathbf{k}},
\end{equation} 
where $\Phi_{n,\mathbf{k}}$ is an eigenvector of $\mathcal{H}(\mathbf{k}).$ The corresponding Bloch wavefunction is
\begin{equation}
\ket{\Psi_{n,\mathbf{k}}} = \sum_{\alpha}\Phi_{n,\mathbf{k}}^\alpha \ket{\phi_{\alpha,\mathbf{k}}}.
\end{equation}
The Green's function in the tight-binding basis in momentum space is 
\begin{equation}
\mathcal{G}(z,\mathbf{k}) = \frac{1}{z-\mathcal{H}\left( \mathbf{k} \right)} =\sum_n \frac{\Phi_{n,\mathbf{k}} \Phi_{n,\mathbf{k}}^\dagger}{z-\varepsilon_{n,\mathbf{k}}},
\end{equation}
and in real space is 
\begin{equation}
\label{eq: tight-binding Green's function}
\mathcal{G}_{\alpha\beta}(z,\mathbf{R}) = \frac{1}{\left( 2\pi \right)^d} \int_\text{BZ} d^d \mathbf{k} e^{i \mathbf{k} \cdot \left( \mathbf{R}+ \boldsymbol{\tau}_\alpha  -\boldsymbol{\tau}_\beta \right)}\mathcal{G}_{\alpha \beta}(z,\mathbf{k}).
\end{equation}
We note that equation~\eqref{eq: tight-binding Green's function} is just equation~\eqref{eq: Bloch Green's function} with the continuous variables $\mathbf{r}$ and $\mathbf{r}'$ replaced by discrete variables $\alpha$ and $\beta.$ Green's functions as defined above, in any representation, are manifestly independent of gauge transformations of the energy eigenstates. So, we can obtain physical observables directly from them. One such observable is the local density of states (LDOS), which is given by 
\begin{equation}
\label{eq: LDOS}
\rho \left(E,\mathbf{r}  \right) = - \frac{1}{\pi} \lim_{\lambda \searrow 0} \Im \text{Tr } \mathcal{G} \left(z =E+i\lambda, \mathbf{r},\mathbf{r} \right),
\end{equation}
where the trace is taken over internal degrees of freedom. The LDOS defined in equation~\eqref{eq: LDOS} has units of inverse energy and area. 

The formalism outlined so far is only useful if the eigenstates and spectrum of the Hamiltonian, which are typically not easy to calculate, are known exactly. Thus, we often must resort to perturbation theory expanded about some known basis. Suppose the Hamiltonian $\mathcal{H}$ can be partitioned into a simple part $\mathcal{H}_0$ which can be diagonalized exactly and the remainder $\mathcal{U}$ that may be quite complicated
\begin{equation}
\mathcal{H} = \mathcal{H}_0+\mathcal{U}.
\end{equation}
With $\mathcal{G}^{(0)}(z) = \left( z-\mathcal{H}_0 \right)^{-1}$ denoting the unperturbed bare Green's function, the full Green's function is
\begin{equation}
\label{eq: transfer matrix def}
\begin{split}
\mathcal{G}(z) &= \left(z-\mathcal{H}_0-\mathcal{U}  \right)^{-1} = \left(\left(z-\mathcal{H}_0  \right) \left( 1-\mathcal{G}^{(0)}(z) \mathcal{U} \right)  \right)^{-1} = \left(1-\mathcal{G}^{(0)}(z) \mathcal{U}  \right)^{-1} \mathcal{G}^{(0)}(z) = \sum_{n = 0}^\infty \left( \mathcal{G}^{(0)}(z) \mathcal{U}  \right)^n \mathcal{G}^{(0)}(z) \\\
&= \mathcal{G}^{(0)}(z) + \mathcal{G}^{(0)}(z) \mathcal{U} \left( 1- \mathcal{G}^{(0)}(z)\mathcal{U} \right)^{-1}  \mathcal{G}^{(0)}(z)  = \mathcal{G}^{(0)}(z) + \Delta \mathcal{G}(z) = \mathcal{G}^{(0)}(z) + \mathcal{G}^{(0)}(z) \mathcal{T}(z)  \mathcal{G}^{(0)}(z) ,
\end{split}
\end{equation}
where $\mathcal{T}(z) = \mathcal{U} \left( 1- \mathcal{G}^{(0)}(z)\mathcal{U} \right)^{-1}$ is the $\cal{T}$ matrix that encodes all information about the perturbation potential $\mathcal{U},$ and $\Delta \mathcal{G}(z) = \mathcal{G}^{(0)}(z) \mathcal{T}(z)  \mathcal{G}^{(0)}(z)$ is the difference between the full Green's function and the bare Green's function. If $\mathcal{U}$ is sufficiently complicated, then only the first few dominant terms in the power series expansion of the $\mathcal{T}$-matrix are usually kept. In the special case that $\mathcal{U}$ is a localized function, we can calculate the Green's function exactly. For our purpose, we are interested in the situation where $\mathcal{U}$ is localized in real space. We can write $\mathcal{U}(\mathbf{r}, \mathbf{r}') = U \delta^d \left( \mathbf{r} - \mathbf{d}\right) \delta^d \left( \mathbf{r}' - \mathbf{d}\right), $ where $U$ is, in general, a matrix in the space of internal degrees of freedom and $\mathbf{d}$ is the center of localization. The change in the Green's function in real space is 
\begin{equation}
\Delta \mathcal{G}(z,\mathbf{r},\mathbf{r}') = \mathcal{G}^{(0)}(z, \mathbf{r},\mathbf{d}) U \left(1-  \mathcal{G}^{(0)}(z, \mathbf{d},\mathbf{d})U  \right)^{-1} \mathcal{G}^{(0)}(z, \mathbf{d},\mathbf{r}') =\mathcal{G}^{(0)}(z, \mathbf{r},\mathbf{d}) \mathcal{T}(z, \mathbf{d}, \mathbf{d}) \mathcal{G}^{(0)}(z, \mathbf{d},\mathbf{r}').
\end{equation}
The change in LDOS is 
\begin{equation}
\Delta \rho(E, \mathbf{r}) = - \frac{1}{\pi} \lim_{\lambda \searrow 0} \Im \text{Tr } \Delta\mathcal{G} \left(z =E+i\lambda, \mathbf{r},\mathbf{r} \right).
\end{equation}

\section{Friedel Oscillations in Toy Models}

We now apply the Green's function formalism to study Friedel oscillations in some toy models which incorporate the emergent symmetries of the continuum model. Because of approximate valley conservation, we can impose a $U_\nu(1)$ gauge symmetry and work with only one valley at a time. The Hamiltonian in each valley  respects $C_3$ symmetry, $M_y$ symmetry, and composite $C_2T$ symmetry. The remaining symmetries like time-reversal symmetry $T$ and mirror symmetry $M_x$ map one valley onto the other, and are thus not preserved in a single valley. Because the valley-polarized low-energy physics of TBLG contains two spectrally-isolated bands, it is to tempting to find a two-orbital tight-binding representation. To capture the energetics of the continuum model, this two-band approximation must contain Dirac band crossings at $\bar{K}$ and $\bar{K}'$ and has no degeneracies anywhere else in the mBZ. This arrangement of band touching points precludes a tight-binding model defined on a triangular lattice. Thus, we instead work with a tight-binding model defined on a honeycomb lattice with Wannier centers located on the $AB$ and $BA$ regions. However, these cannot be ordinary Wannier orbitals whose density is concentrated at the Wannier centers because the charge density is known from the continuum model and from experiments to be concentrated at the $AA$ regions. Though this approach is able to reproduce the spectral features of the isolated bands of TBLG, it fails to capture the topology of these bands. Because the net winding number of any tight-binding model must be zero, the Dirac cones at the zone corners must carry opposite winding numbers. However, the Dirac cones from the continuum model carry the same winding number due to mirror symmetry. Thus, if one trusts the continuum model, then it is not possible to construct an equivalent two-orbital tight-binding representation that respects all the emergent symmetries, reproduces the spectral features well, and  has the correct topology \cite{ZP18, PZ18}. 

Suppose that the valley-polarized low-energy physics of TBLG could in fact be represented by a two-orbital tight-binding model.  We analyze the pattern of Friedel oscillations in this case. Let $\ket{W_{1,\mathbf{R}}}$ and $\ket{W_{2,\mathbf{R}}}$ be exponentially-localized Wannier orbitals centered at the $AB$ and $BA$ regions respectively in the $\mathbf{R}$ unit cell. Then, we can construct a tight-binding model based on these Wannier orbitals. The minimal model includes just nearest-neighbor hoppings, and is topologically equivalent to the band structure of monolayer graphene for spinless electrons. In the Bloch basis of equation~\eqref{eq: Bloch states}, the Hamiltonian can be written as 
\begin{equation}
\label{eq: graphene Hamiltonian}
\mathcal{H}(\mathbf{k}) = \begin{pmatrix}
0 & h(\mathbf{k}) \\
h(\mathbf{k})^* & 0
\end{pmatrix},
\end{equation}
where $h(\mathbf{k}) = -  \sum_{i = 1}^3t_i \exp \left( i \mathbf{k} \cdot \boldsymbol{\delta}_i \right),$ $t_i$ are the nearest-neighbor hopping parameters, and $\boldsymbol{\delta}_i$ are the nearest-neighbor vectors. A $\sigma_z$ term is prohibited by $C_2T$ symmetry.  For $\mathbf{k}$ near the zone corners, the Hamiltonian in a single valley can be expanded to linear order in $\mathbf{q}$ measured from the respective zone corner
\begin{equation}
\label{eq: Dirac cone at K}
\mathcal{H}\left(  \bar{\mathbf{K}}+ \mathbf{q} \right) = \begin{pmatrix}
0 & -\hbar\tilde{v}_F q  e^{i \phi_\mathbf{q}} \\
-\hbar \tilde{v}_F q e^{-i \phi_\mathbf{q}}& 0
\end{pmatrix},
\end{equation}
\begin{equation}
\label{eq: Dirac cone at Kprime}
\mathcal{H}\left(  \bar{\mathbf{K}}'+ \mathbf{q} \right) =  \begin{pmatrix}
0 & -\hbar\tilde{v}_F  q e^{-i \phi_\mathbf{q}} \\
-\hbar\tilde{v}_F  q e^{i \phi_\mathbf{q}} & 0
\end{pmatrix},
\end{equation}
where $\phi_\mathbf{q}$ is the angle that the vector $\mathbf{q}$ makes with the $x$-axis. The eigenvalues and eigenvectors of equation~\eqref{eq: graphene Hamiltonian} are 
\begin{equation}
\varepsilon_{\chi,\mathbf{k}} = \chi |h(\mathbf{k})|, \quad \Phi_\chi(\mathbf{k}) = \frac{1}{\sqrt{2}}\begin{pmatrix}
\exp \left(i \text{arg}\left(h \right) \right)\\
\chi
\end{pmatrix},
\end{equation}
where $h\left(\mathbf{k} \right) = |h\left(\mathbf{k} \right)|\exp \left(i \text{arg}\left(h \right) \right)$ and $\chi = \pm 1$ is the band index.  While the overall phase of the eigenvector is not fixed, the relative phase between the two orbitals in a single unit cell is dictated by the chirality of the Hamiltonian. The Hamiltonian at the other valley is related by time reversal which simultaneously exchanges the Dirac cones, $\bar{\mathbf{K}} \leftrightarrow \bar{\mathbf{K}}'$, and inverts the chirality. So Hamiltonian near the Dirac cones in both valleys has the same form as in  equations~\eqref{eq: Dirac cone at K} and \eqref{eq: Dirac cone at Kprime}. In the long-wavelength approximation, the Green's function in real space in a single valley is 
\begin{equation}
\label{eq: Green function for general k}
\mathcal{G}^{(0)}(z, \mathbf{r}) = \frac{1}{\left(  2 \pi \right)^2} \int d^2 \mathbf{k} e^{i \mathbf{k} \cdot \mathbf{r}} \mathcal{G}^{(0)} \left( z, \mathbf{k} \right),
\end{equation}
where the Green's function in momentum space is 
\begin{equation}
\mathcal{G}^{(0)} \left( z, \mathbf{k} \right) = \frac{1}{z^2- |h(\mathbf{k})|^2} \begin{pmatrix}
z & h(\mathbf{k}) \\
h(\mathbf{k})^* & z
\end{pmatrix}.
\end{equation}
At this point, there is no natural origin from which to measure $\mathbf{r};$ once we impose an impurity, $\mathbf{r}$ will be the displacement vector from the impurity. now, for small $|z|$ near neutrality, only states near $\bar{\mathbf{K}}$ and $\bar{\mathbf{K}}'$ contribute significantly to the momentum integral; so we approximate equation~\eqref{eq: Green function for general k} as 
\begin{equation}
\label{eq: Green function approximation}
\mathcal{G}^{(0)}(z, \mathbf{r}) \approx \mathcal{G}^{(0)}(z, \bar{\mathbf{K}}, \mathbf{r}) + \mathcal{G}^{(0)}(z, \bar{\mathbf{K}}', \mathbf{r}),
\end{equation}
where 
\begin{equation}
\label{eq: Green function at K}
\mathcal{G}^{(0)} \left( z, \bar{\mathbf{K}}, \mathbf{r} \right) = \frac{e^{i \bar{\mathbf{K}} \cdot \mathbf{r}}}{\left( 2 \pi \right)^2}  \int d^2 \mathbf{q} e^{i \mathbf{q} \cdot \mathbf{r}} \frac{1}{z^2 - \hbar^2 \tilde{v}_F^2 q^2} \begin{pmatrix}
z & -\hbar\tilde{v}_F  q e^{i \phi_\mathbf{q}}\\
 -\hbar\tilde{v}_F  q e^{-i \phi_\mathbf{q}} & z 
\end{pmatrix},
\end{equation}
\begin{equation}
\label{eq: Green function at Kp}
\mathcal{G}^{(0)} \left( z, \bar{\mathbf{K}}', \mathbf{r} \right) = \frac{e^{i \bar{\mathbf{K}}' \cdot \mathbf{r}}}{\left( 2 \pi \right)^2}  \int d^2 \mathbf{q} e^{i \mathbf{q} \cdot \mathbf{r}} \frac{1}{z^2 - \hbar^2 \tilde{v}_F^2 q^2} \begin{pmatrix}
z & -\hbar\tilde{v}_F  q e^{-i \phi_\mathbf{q}}\\
 -\hbar\tilde{v}_F  q e^{i \phi_\mathbf{q}} & z 
\end{pmatrix}.
\end{equation}
The integrals can be done exactly. For example, the diagonal elements contain the following integral 
\begin{equation}
\begin{split}
\int d^2 \mathbf{q} e^{i \mathbf{q} \cdot \mathbf{r}} \frac{z}{z^2 - \hbar^2 \tilde{v}_F^2 q^2} &= \int_0^{2 \pi}\int_0^\infty q dq d\phi_\mathbf{q}  \frac{ze^{i q r \cos \left(\phi_\mathbf{q}-\phi_\mathbf{r} \right)}}{z^2 - \hbar^2 \tilde{v}_F^2 q^2} = 2 \pi z \int_0^\infty  dq \frac{qJ_0 \left( qr \right)}{z^2-\hbar^2 \tilde{v}_F^2 q^2} = - \frac{2 \pi z}{\hbar^2 \tilde{v}_F^2} K_0\left( \frac{\sqrt{z^2}r}{i \hbar \tilde{v}_F}  \right),
\end{split}
\end{equation}
where $J_\alpha(z)$ is the $\alpha^\text{th}$ Bessel function of the first kind and $K_\alpha(z)$ is the $\alpha^\text{th}$ modified Bessel function of the second kind. The off-diagonal elements have the integral
\begin{equation}
\label{eq: mapping phi_q to phi_r}
\begin{split}
\int d^2 \mathbf{q} e^{i \mathbf{q} \cdot \mathbf{r}} \frac{qe^{\pm i \phi_\mathbf{q}}}{z^2 - \hbar^2 \tilde{v}_F^2 q^2} &= \int_0^{2 \pi} \int_0^\infty dq d\phi_\mathbf{q} \frac{q^2 e^{i q r \cos \left( \phi_\mathbf{q}- \phi_\mathbf{r} \right)}e^{ \pm i \phi_\mathbf{q}}}{z^2-\hbar^2 \tilde{v}_F^2 q^2} \\
&= 2 \pi i e^{\pm i \phi_\mathbf{r}} \int_0^\infty dq \frac{q^2 J_1(qr)}{z^2-\hbar^2 \tilde{v}_F^2 q^2} = -\frac{2 \pi  e^{\pm i \phi_\mathbf{r}}\sqrt{z^2}}{\hbar^3 \tilde{v}_F^3}K_1\left( \frac{\sqrt{z^2}r}{i \hbar \tilde{v}_F}  \right).
\end{split}
\end{equation}
In equation~\eqref{eq: mapping phi_q to phi_r}, we see that the integral maps the momentum-space phase $\phi_\mathbf{q}$ to a real-space phase $\phi_\mathbf{r},$ which is the origin of the LDOS phase slips we will encounter later.  Explicitly, equations~\eqref{eq: Green function at K} and \eqref{eq: Green function at Kp} once integrated are
\begin{equation}
\mathcal{G}^{(0)} \left( E, \bar{\mathbf{K}}, \mathbf{r} \right) = \frac{ e^{i \bar{\mathbf{K}}\cdot \mathbf{r}}}{2 \pi \hbar \tilde{v}_F \ell } \begin{pmatrix}
- \text{sign}\left( E\right)  K_0 \left(-i  r / \ell   \right) & e^{i \phi_\mathbf{r}} K_1 \left( -i  r / \ell  \right)\\ e^{-i \phi_\mathbf{r}} 
K_1 \left( -i  r /\ell  \right) & - \text{sign}\left( E\right) K_0 \left(-i  r/\ell  \right)
\end{pmatrix},
\end{equation}
\begin{equation}
\mathcal{G}^{(0)} \left( E, \bar{\mathbf{K}}', \mathbf{r} \right) = \frac{e^{i \bar{\mathbf{K}}'\cdot \mathbf{r}}}{2 \pi \hbar \tilde{v}_F \ell }  \begin{pmatrix}
- \text{sign}\left( E\right)  K_0 \left(-i  r / \ell   \right) & e^{-i \phi_\mathbf{r}} K_1 \left( -i  r / \ell  \right)\\ e^{i \phi_\mathbf{r}} 
K_1 \left( -i  r /\ell  \right) & - \text{sign}\left( E\right) K_0 \left(-i  r/\ell  \right)
\end{pmatrix},
\end{equation}
where we have defined a new length scale $\ell = \hbar \tilde{v}_F/E$ with $z = E + i\lambda$ as $\lambda \rightarrow 0$ from above.

\begin{figure}
\includegraphics[scale=0.6]{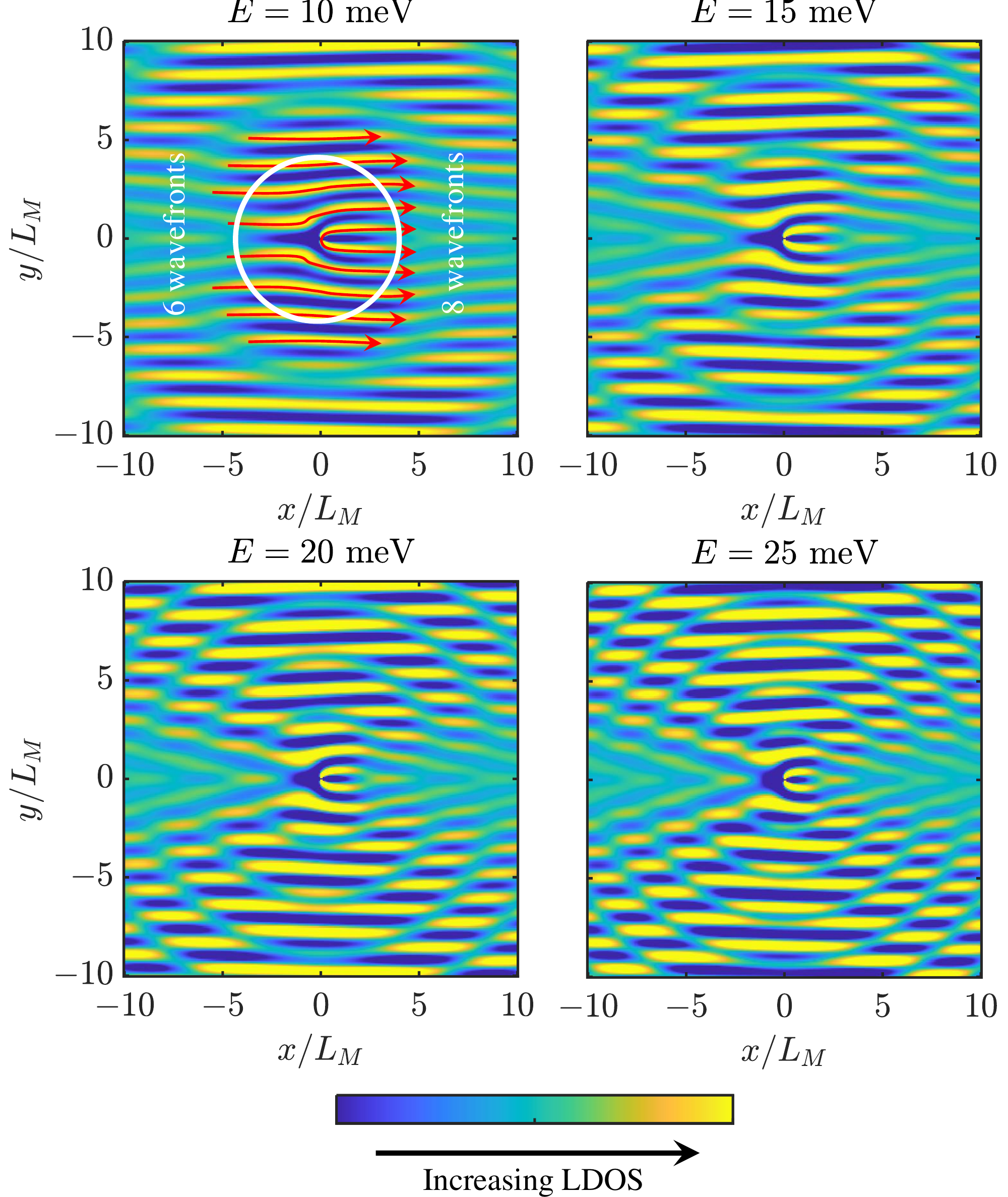}
\caption{Density plots of $r\rho_\text{inter}$ for various energies $E$ with $U_0 = 1000$ meV for the model with opposite chirality. We multiply LDOS by $r$ to amplify the faint signals at large distances away from the impurity. As we increase the energy $E,$ the period of radial oscillations decreases. For certain values of $r$, we see the number of wavefronts to the left of the impurity differs from the number of wavefronts to the right of the impurity by two. }
\label{fig: toy model Friedel oscillations}
\end{figure}

We now describe the scattering potential as a localized impurity in real space. This impurity fixes the origin of our coordinate system. We write the impurity potential as $\mathcal{U} = U_0 \ket{\text{imp}}\bra{\text{imp}},$ where $\ket{\text{imp}}$ is the state of the impurity. We first consider the case where the impurity wavefunction has significant overlap with only one of the two Wanner orbitals. In this case, we can write the scattering potential as 
\begin{equation}
\label{eq: impurity potential}
\mathcal{U}(\mathbf{r}, \mathbf{r}') = U_0 \begin{pmatrix}
1 & 0 \\ 
0 & 0
\end{pmatrix} \delta^{(2)}(\mathbf{r})\delta^{(2)}(\mathbf{r}').
\end{equation}
In TGB, because the Wannier orbitals are non-local states, designing an impurity for which equation~\eqref{eq: impurity potential} is a good approximation for the potential is a challenging task. First, any impurity centered at an $AA$ region will necessarily project onto multiple Wannier orbitals. Thus, the impurity must be placed near an $AB$ or $BA$ region. One might suspect that an atomic impurity placed at the center of an $AB$ or $BA$ region is enough, but this is not true because an atomic impurity could have non-negligible overlap with both Wannier orbitals in a single moiré unit cell \cite{KF18}. In order to increase the overlap with one orbital relative to the other, we need to make the impurity large on the atomic scale, but small on the moiré scale, centered on an $AB$ or $BA$ region. Under these assumptions, we can use equation~\eqref{eq: impurity potential} to calculate the $\mathcal{T}$ matrix 
\begin{equation}
\label{eq: transfer matrix}
\mathcal{T}\left( E \right) = \begin{pmatrix}
t\left( E \right) & 0 \\
 0 & 0
\end{pmatrix},
\end{equation}
where $t\left( E \right) = U_0 \left( 1- U_0 \mathcal{G}^{(0)}_{11}\left(E, \mathbf{0}\right) \right)^{-1}.$ If the bare Green's function is known everywhere in the mBZ, then the $\cal{T}$ matrix can be calculated exactly. Otherwise, we can use resort some approximation. If $U_0$ is sufficiently small, we use the Born approximation where $\mathcal{T} \approx \mathcal{U}.$ Alternatively, we can use the approximation in equation~\eqref{eq: Green function approximation} to calculate $\mathcal{G}_{11}^{(0)}\left(E, \mathbf{0}\right)$
\begin{equation}
\frac{ 1}{\left( 2 \pi \right)^2} \int  d^2 \mathbf{q}\frac{z}{z^2- \hbar^2 \tilde{v}_F^2 q^2} = \frac{1}{2 \pi} \int_0^\Gamma \frac{zqdq}{z^2-\hbar^2\tilde{v}_F^2 q^2} = \frac{z}{4 \pi \hbar^2 \tilde{v}_F^2} \log \left( \frac{z^2}{z^2-\hbar^2 \tilde{v}_F^2 \Gamma^2} \right),
\end{equation}
where $\Gamma$ is some large momentum cutoff. The change in  LDOS, when averaged over a unit cell, is given by 
\begin{equation}
\begin{split}
\Delta \rho \left( E, \mathbf{r} \right) &= - \frac{1}{\pi} \Im \text{Tr} \left[ \left( \mathcal{G}^{(0)}(z, \bar{\mathbf{K}}, \mathbf{r}) + \mathcal{G}^{(0)}(z, \bar{\mathbf{K}}', \mathbf{r}) \right) \mathcal{T} \left( E \right) \left( \mathcal{G}^{(0)}(z, \bar{\mathbf{K}}, -\mathbf{r}) + \mathcal{G}^{(0)}(z, \bar{\mathbf{K}}', -\mathbf{r}) \right) \right] \\
&=  -\frac{1}{\pi} \Im \text{Tr} \left[ \mathcal{G}^{(0)}(z, \bar{\mathbf{K}}, \mathbf{r}) \mathcal{T} \left( E \right)\mathcal{G}^{(0)}(z, \bar{\mathbf{K}}, -\mathbf{r}) + \mathcal{G}^{(0)}(z, \bar{\mathbf{K}}', \mathbf{r}) \mathcal{T} \left( E \right)\mathcal{G}^{(0)}(z, \bar{\mathbf{K}}', -\mathbf{r}) \right] \\
&\quad- \frac{1}{\pi} \Im \text{Tr} \left[ \mathcal{G}^{(0)}(z, \bar{\mathbf{K}}, \mathbf{r}) \mathcal{T} \left( E \right)\mathcal{G}^{(0)}(z, \bar{\mathbf{K}}', -\mathbf{r}) + \mathcal{G}^{(0)}(z, \bar{\mathbf{K}}', \mathbf{r}) \mathcal{T} \left( E \right)\mathcal{G}^{(0)}(z, \bar{\mathbf{K}}, -\mathbf{r}) \right],
\end{split} 
\end{equation}
which consists of two contributions: an inter-Dirac-cone contribution that consists of scattering between $\bar{\mathbf{K}}$ and $\bar{\mathbf{K}}'$ and an intra-Dirac-cone contribution. Here, the trace is taken over the Wannier sublattice degree of freedom, which is equivalent to performing an average over a unit cell. The intra-Dirac-cone contribution is
\begin{equation}
\begin{split}
\Delta \rho_\text{intra}\left(E, \mathbf{r}\right) &=  -\frac{1}{\pi} \Im \text{Tr} \left[ \mathcal{G}^{(0)}(z, \bar{\mathbf{K}}, \mathbf{r}) \mathcal{T} \left( E \right)\mathcal{G}^{(0)}(z, \bar{\mathbf{K}}, -\mathbf{r}) + \mathcal{G}^{(0)}(z, \bar{\mathbf{K}}', \mathbf{r}) \mathcal{T} \left( E \right)\mathcal{G}^{(0)}(z, \bar{\mathbf{K}}', -\mathbf{r}) \right] \\
&= - \frac{1}{2 \pi^3 \hbar^2 \tilde{v}_F^2 \ell^2}\Im \left[t \left( E \right) K_0^2 \left( -i r / \ell \right) -t \left( E \right) K_1^2 \left( -i r / \ell \right) \right],
\end{split}
\end{equation}
which, in real space, is simply a function of the radial coordinate $r.$ The inter-Dirac-cone contribution is more interesting
\begin{equation}
\label{eq: intercone LDOS}
\begin{split}
\Delta \rho_\text{inter} \left( E, \mathbf{r} \right) &= - \frac{1}{\pi} \Im \text{Tr} \left[ \mathcal{G}^{(0)}(z, \bar{\mathbf{K}}, \mathbf{r}) \mathcal{T} \left( E \right)\mathcal{G}^{(0)}(z, \bar{\mathbf{K}}', -\mathbf{r}) + \mathcal{G}^{(0)}(z, \bar{\mathbf{K}}', \mathbf{r}) \mathcal{T} \left( E \right)\mathcal{G}^{(0)}(z, \bar{\mathbf{K}}, -\mathbf{r}) \right] \\
&= - \frac{1}{2 \pi^3 \hbar^2 \tilde{v}_F^2 \ell^2} \left(\Im \left[t(E) K_0^2 \left(-i r/\ell  \right)     \right] \cos \left(\Delta \bar{\mathbf{K}} \cdot \mathbf{r} \right)-\Im \left[t(E) K_1^2 \left(-i r/\ell  \right)     \right] \cos \left(\Delta \bar{\mathbf{K}} \cdot \mathbf{r}-2\phi_\mathbf{r} \right) \right),
\end{split}
\end{equation}
where $\Delta \bar{\mathbf{K}} = \bar{\mathbf{K}}-\bar{\mathbf{K}}'.$ The relative phase $\phi_\mathbf{r}$ determines the wavefront dislocation in  LDOS.  We simulate  LDOS calculated in equation~\eqref{eq: intercone LDOS}, and show in the results in Fig.~\ref{fig: toy model Friedel oscillations}. To enhance the amplitude of LDOS at large distances, we multiply LDOS by $r.$ We observe wavefronts propagating in the direction of $\Delta \bar{\mathbf{K}},$ and the period is given by $|\Delta \bar{\mathbf{K}}|^{-1}.$ There are also oscillations in the radial direction with a period given by $\ell.$ Importantly, we observe wavefront dislocations in the pattern of Friedel oscillations. We establish the wavefront dislocations by drawing a closed loop around the impurity, and counting the number of wavefronts ``entering" the enclosed area and the number of wavefronts ``exiting" the enclosed area. The number of dislocations is the difference between the number of ``entering" and ``exiting" wavefronts. In this model where the two Dirac cones have opposite chirality, we observe a two dislocations.

Let us provide some parameter estimates. First, the radial oscillation length scale is $\ell = \hbar \tilde{v}_F/E,$ which decreases with increasing bias energy $E$ and decreasing angles (when we are above the magic angle). For example, using the parameters for the interlayer hopping as before, we find for $\theta = 2^\circ,$ $\ell/ a \approx \left(1000 \text{ meV}\right)/E,$
and for $\theta = 1.5^\circ,$ $\ell/ a \approx \left(500 \text{ meV}\right)/E.$ On the other hand, the wavelength of the Friedel oscillations is $|a\Delta \bar{\mathbf{K}}|^{-1} \approx 6$ for $\theta = 2^\circ$ and $|a\Delta \bar{\mathbf{K}}|^{-1} \approx 8$ for $\theta = 1.5^\circ.$ In order to clearly observe the wavefront dislocations in LDOS, we must be in a regime where $\ell > |\Delta \bar{\mathbf{K}}|^{-1},$ which implies that $E < 100$ meV. Thus, we must probe LDOS at an energy sufficiently small in order to observe the wavefront dislocations. We must also stay at small energies in order to approximate the band structure as a linear dispersion. However, at low energies, the amplitude of the oscillations is suppressed quadratically in $E$
\begin{equation}
\Delta \rho \sim \frac{U_0 E^2 V_\text{cell} }{\hbar^4 \tilde{v}_F^4 }.
\end{equation}
Again, using the interlayer hopping parameters as before, we obtain $V_\text{cell}\Delta \rho \sim  \frac{U_0E^2}{2 \times 10^6 \text{ meV}^4}$ for  $\theta = 2^\circ$ and  $V_\text{cell}\Delta \rho \sim  \frac{U_0E^2}{6 \times 10^4 \text{ meV}^4}$ for  $\theta = 1.5^\circ.$ At large distances $r\gg\ell,$ this amplitude is further suppressed by the decay of the modified Bessel function as
\begin{equation}
K_\nu^2 \left( - \frac{i r}{\ell} \right) \sim \frac{\pi \ell}{2 r} \exp \left( \frac{ 2ir}{\ell}+ \frac{i\pi}{2} \right)  .
\end{equation}
Therefore, spatial window of observation should only be a couple times $\ell$ before the oscillations are damped out by the $r^{-1}$ decay beyond observable limits. 

\begin{figure}
\includegraphics[scale=0.6]{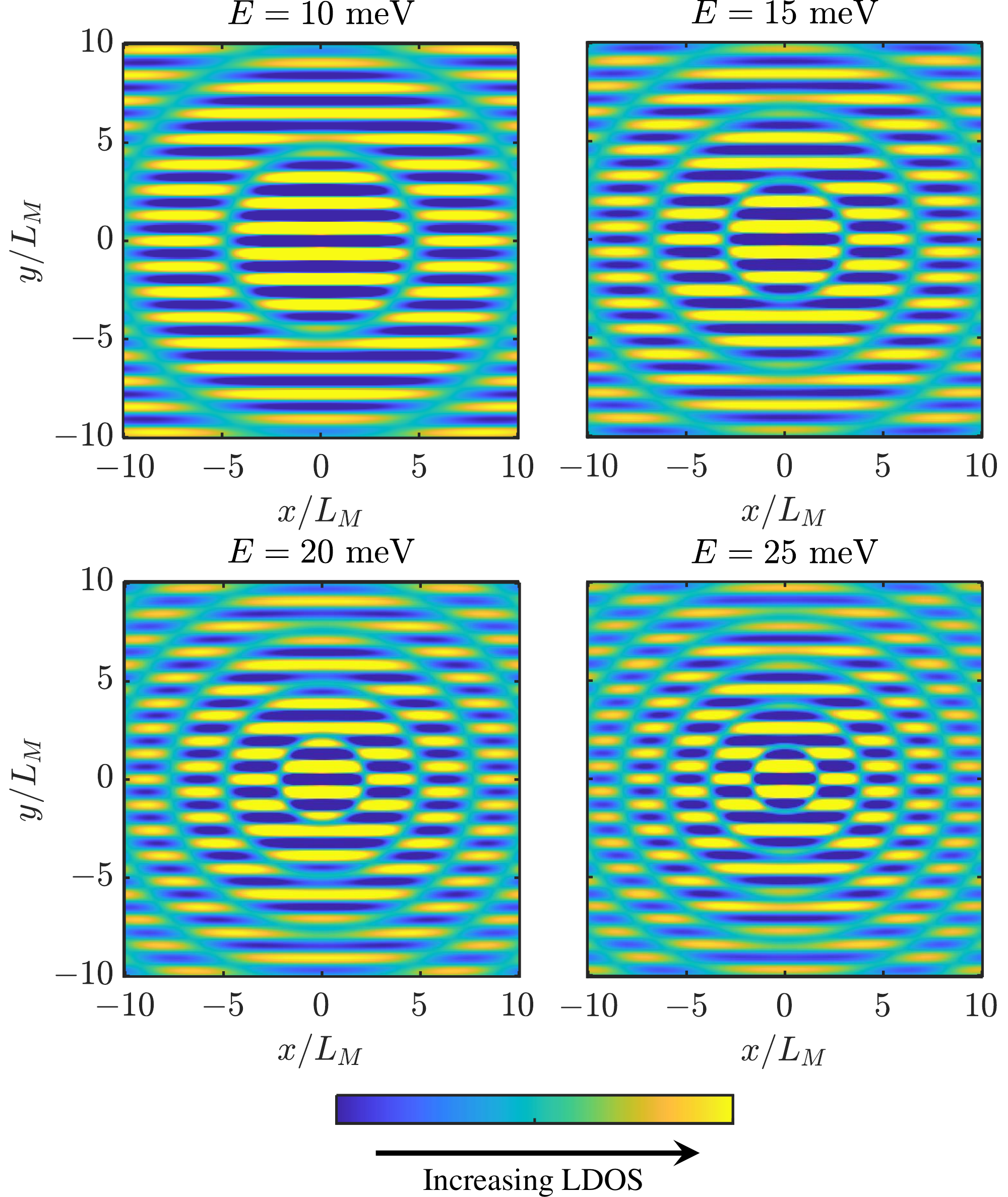}
\caption{Density plots of $r\rho_\text{inter}$ for various energies $E$ with $U_0 = 1000$ meV for the model with identical chirality. We multiply LDOS by $r$ to amplify the faint signals at large distances away from the impurity. As we increase the energy $E,$ the period of radial oscillations decreases. Here, we observe no wavefront dislocations. }
\label{fig: toy model Friedel oscillations 2}
\end{figure}

We now study a different effective model where the chirality at two inequivalent Dirac cones in a single valley is the same. This effective model is a long-wavefunction approximation to the continuum model. In the same valley, the two Dirac cones are related by $M_y$ symmetry. 
\begin{equation}
\label{eq: second model Dirac cone at K}
\mathcal{H}\left(  \bar{\mathbf{K}}+ \mathbf{q} \right) = \begin{pmatrix}
0 & -\hbar\tilde{v}_F q  e^{i \phi_\mathbf{q}} \\
-\hbar \tilde{v}_F q e^{-i \phi_\mathbf{q}}& 0
\end{pmatrix},
\end{equation}
\begin{equation}
\label{eq: second model Dirac cone at Kprime}
\mathcal{H}\left(  \bar{\mathbf{K}}'+ \mathbf{q} \right) =  \begin{pmatrix}
0 & -\hbar\tilde{v}_F  q e^{i \phi_\mathbf{q}} \\
-\hbar\tilde{v}_F  q e^{-i \phi_\mathbf{q}} & 0
\end{pmatrix},
\end{equation}
The Hamiltonian at the other valley is obtained by taking the complex conjugate of equations~\eqref{eq: second model Dirac cone at K} and \eqref{eq: second model Dirac cone at Kprime}. By using this representation of the Hamiltonian, the bare Green's function can be calculated straightforwardly as before. We obtain 
\begin{equation}
\mathcal{G}^{(0)} \left( E, \bar{\mathbf{K}}, \mathbf{r} \right) = \frac{ e^{i \bar{\mathbf{K}}\cdot \mathbf{r}}}{2 \pi \hbar \tilde{v}_F \ell } \begin{pmatrix}
- \text{sign}\left( E\right)  K_0 \left(-i  r / \ell   \right) & e^{i \phi_\mathbf{r}} K_1 \left( -i  r / \ell  \right)\\ e^{-i \phi_\mathbf{r}} 
K_1 \left( -i  r /\ell  \right) & - \text{sign}\left( E\right) K_0 \left(-i  r/\ell  \right)
\end{pmatrix},
\end{equation}
\begin{equation}
\mathcal{G}^{(0)} \left( E, \bar{\mathbf{K}}', \mathbf{r} \right) = \frac{e^{i \bar{\mathbf{K}}'\cdot \mathbf{r}}}{2 \pi \hbar \tilde{v}_F \ell }  \begin{pmatrix}
- \text{sign}\left( E\right)  K_0 \left(-i  r / \ell   \right) & e^{i \phi_\mathbf{r}} K_1 \left( -i  r / \ell  \right)\\ e^{-i \phi_\mathbf{r}} 
K_1 \left( -i  r /\ell  \right) & - \text{sign}\left( E\right) K_0 \left(-i  r/\ell  \right)
\end{pmatrix}.
\end{equation}
Now, we again assume that we have an impurity that can be represented by equation~\eqref{eq: impurity potential}. The inter-Dirac-cone contribution to LDOS is now 
\begin{equation}
\Delta \rho_\text{inter} \left(E, \mathbf{r} \right) = - \frac{1}{2 \pi^3 \hbar^2 \tilde{v}_F^2 \ell^2} \Im \left[t(E) K_0^2 \left(-i r/\ell  \right)     -t(E) K_1^2 \left(-i r/\ell  \right)     \right]  \cos \left(\Delta \bar{\mathbf{K}} \cdot \mathbf{r} \right).
\end{equation}
Importantly here, there is no phase mismatch in $\phi_\mathbf{r}$ between the two Bessel functions. Because of that, there are no dislocations, as shown in Fig.~\ref{fig: toy model Friedel oscillations 2}.

\section{Friedel Oscillations in Effective Low-Energy Dirac Theory}

So far, we have only calculated Friedel oscillations in models that  capture features on the moiré length scale. In particular, the charge distribution within a moiré unit cell has been completely neglected in previous calculations. This prevented us from dealing effectively with impurities that are localized on the atomic scale. Now, to account for the intra-cell density distribution, we use the wavefunctions in equations~\eqref{eq: wavefunction at K1} and \eqref{eq: wavefunction at K2} to calculate the Green's function. Like before, we use valley polarization to simplify the calculation. For a single valley, the bare Green's function near $\mathbf{K}_\nu^1$ is 
\begin{equation}
\mathcal{G}^{(0)}\left( z, \mathbf{K}_\nu^1, \mathbf{r}, \mathbf{r}' \right) = \frac{1}{\left( 2 \pi \right)^2} \sum_{\chi = \pm 1}\int d^2 \mathbf{q} \frac{\ket{\psi^{\chi \mathbf{q}}_{\mathbf{K}_\nu^1}(\mathbf{r})}\bra{\psi^{\chi \mathbf{q}}_{\mathbf{K}_\nu^1}(\mathbf{r}')}}{z - \chi \hbar \tilde{v}_F q}.
\end{equation}
Upon integration, the matrix elements  explicitly are
\begin{equation}
\begin{split}
\mathcal{G}_{A_1A_1}^{(0)} \left(E>0, \mathbf{K}_\nu^1, \mathbf{r} , \mathbf{r}' \right) &= -\frac{e^{i \mathbf{K}_\nu^1 \cdot \mathbf{d}}}{\omega} \frac{ \hbar^2 v_F^2 G^2 }{\pi \hbar^2 \tilde{v}_F^2} K_0 \left( \frac{d}{i \ell} \right)  ,\\
\mathcal{G}_{A_1B_1}^{(0)} \left(E>0, \mathbf{K}_\nu^1, \mathbf{r} , \mathbf{r}' \right) &= \frac{e^{i \mathbf{K}_\nu^1 \cdot \mathbf{d}}}{\omega} \frac{ \nu\hbar^2 v_F^2 G^2 }{\pi \hbar^2 \tilde{v}_F^2 } e^{-i\nu \phi_\mathbf{d}} K_1 \left( \frac{d}{i \ell} \right) ,\\
\mathcal{G}_{A_1A_2}^{(0)} \left(E>0, \mathbf{K}_\nu^1, \mathbf{r} , \mathbf{r}' \right) &= \frac{e^{i \mathbf{K}_\nu^1 \cdot \mathbf{d}}}{\omega} \frac{i \hbar v_F G}{\pi \hbar^2 \tilde{v}_F^2} \left[ g_{-\nu}^{(2)} \left(  \mathbf{r}'\right) w_\text{AA} e^{-i \nu \phi_\mathbf{d}} K_1 \left( \frac{d}{i \ell} \right) - \nu g_{-\nu}^{(1)} \left(  \mathbf{r}'\right) w_\text{AB} K_0 \left( \frac{d}{i \ell} \right) \right] ,\\
\mathcal{G}_{A_1B_2}^{(0)} \left(E>0, \mathbf{K}_\nu^1, \mathbf{r} , \mathbf{r}' \right) &= \frac{e^{i \mathbf{K}_\nu^1 \cdot \mathbf{d}}}{\omega} \frac{i \hbar v_F G}{\pi \hbar^2 \tilde{v}_F^2} \left[\nu  g_{-\nu}^{(3)} \left(  \mathbf{r}'\right) w_\text{AA}  K_0 \left( \frac{d}{i \ell} \right) - g_{-\nu}^{(1)} \left(  \mathbf{r}'\right) w_\text{AB}e^{-i \nu \phi_\mathbf{d}} K_1 \left( \frac{d}{i \ell} \right) \right]  ,\\
\mathcal{G}_{B_1A_1}^{(0)} \left(E>0, \mathbf{K}_\nu^1, \mathbf{r} , \mathbf{r}' \right) &= \frac{e^{i \mathbf{K}_\nu^1 \cdot \mathbf{d}}}{\omega} \frac{\nu\hbar^2  v_F^2G^2}{\pi \hbar^2 \tilde{v}_F^2}  e^{i \nu \phi_\mathbf{d}} K_1 \left( \frac{d}{i \ell} \right) ,\\
\mathcal{G}_{B_1B_1}^{(0)} \left(E>0, \mathbf{K}_\nu^1, \mathbf{r} , \mathbf{r}' \right) &= -\frac{e^{i \mathbf{K}_\nu^1 \cdot \mathbf{d}}}{\omega} \frac{ \hbar^2 v_F^2 G^2 }{\pi \hbar^2 \tilde{v}_F^2} K_0 \left( \frac{d}{i \ell} \right)  ,\\
\mathcal{G}_{B_1A_2}^{(0)} \left(E>0, \mathbf{K}_\nu^1, \mathbf{r} , \mathbf{r}' \right) &= \frac{e^{i \mathbf{K}_\nu^1 \cdot \mathbf{d}}}{\omega} \frac{i \hbar v_F G}{\pi \hbar^2 \tilde{v}_F^2} \left[ g_{-\nu}^{(1)}\left( \mathbf{r}' \right) w_\text{AB} e^{i \nu \phi_\mathbf{d}} K_1 \left( \frac{d}{i \ell} \right) - \nu g_{-\nu}^{(2)} \left( \mathbf{r}' \right) w_\text{AA}  K_0 \left( \frac{d}{i \ell} \right) \right],\\
\mathcal{G}_{B_1B_2}^{(0)} \left(E>0, \mathbf{K}_\nu^1, \mathbf{r} , \mathbf{r}' \right) &= \frac{e^{i \mathbf{K}_\nu^1 \cdot \mathbf{d}}}{\omega} \frac{i \hbar v_F G}{\pi \hbar^2 \tilde{v}_F^2} \left[ \nu g_{-\nu}^{(1)}\left( \mathbf{r}' \right) w_\text{AB}  K_0 \left( \frac{d}{i \ell} \right) -  g_{-\nu}^{(3)} \left( \mathbf{r}' \right) w_\text{AA} e^{i \nu \phi_\mathbf{d}} K_1 \left( \frac{d}{i \ell} \right) \right]  ,\\
\mathcal{G}_{A_2A_1}^{(0)} \left(E>0, \mathbf{K}_\nu^1, \mathbf{r} , \mathbf{r}' \right) &= \frac{e^{i \mathbf{K}_\nu^1 \cdot \mathbf{d}}}{\omega} \frac{i \hbar v_F G}{\pi \hbar^2 \tilde{v}_F^2} \left[  \nu g_{\nu}^{(1)}\left( \mathbf{r} \right) w_\text{AB}  K_0 \left( \frac{d}{i \ell} \right) -  g_{\nu}^{(2)} \left( \mathbf{r} \right) w_\text{AA} e^{i \nu \phi_\mathbf{d}} K_1 \left( \frac{d}{i \ell} \right) \right] ,\\
\mathcal{G}_{A_2B_1}^{(0)} \left(E>0, \mathbf{K}_\nu^1, \mathbf{r} , \mathbf{r}' \right) &= \frac{e^{i \mathbf{K}_\nu^1 \cdot \mathbf{d}}}{\omega} \frac{i \hbar v_F G}{\pi \hbar^2 \tilde{v}_F^2} \left[  \nu g_{\nu}^{(2)}\left( \mathbf{r} \right) w_\text{AA}  K_0 \left( \frac{d}{i \ell} \right) -   g_{\nu}^{(1)} \left( \mathbf{r} \right) w_\text{AB} e^{-i \nu \phi_\mathbf{d}} K_1 \left( \frac{d}{i \ell} \right) \right] ,\\
\mathcal{G}_{A_2A_2}^{(0)} \left(E>0, \mathbf{K}_\nu^1, \mathbf{r} , \mathbf{r}' \right) &= \frac{e^{i \mathbf{K}_\nu^1 \cdot \mathbf{d}}}{\omega} \frac{1}{\pi \hbar^2 \tilde{v}_F^2} \left[   \nu \left( g_\nu^{(1)}(\mathbf{r}) g_{-\nu}^{(2)}(\mathbf{r}') e^{-i \nu \phi_\mathbf{d}}  +  g_{\nu}^{(2)}(\mathbf{r})g_{-\nu}^{(1)}(\mathbf{r}') e^{i \nu \phi_\mathbf{d}} \right)w_\text{AA} w_\text{AB}  K_1 \left( \frac{d}{i \ell} \right) \right. \\
&\left. - \left( g_{\nu}^{(1)}(\mathbf{r}) g_{-\nu}^{(1)}(\mathbf{r}') w_\text{AB}^2 + g_{\nu}^{(2)}(\mathbf{r}) g_{-\nu}^{(2)}(\mathbf{r}')w_\text{AA}^2 \right)   K_0 \left( \frac{d}{i \ell} \right)   \right], \\
\mathcal{G}_{A_2B_2}^{(0)} \left(E>0, \mathbf{K}_\nu^1, \mathbf{r} , \mathbf{r}' \right) &= \frac{e^{i \mathbf{K}_\nu^1 \cdot \mathbf{d}}}{\omega}  \frac{1}{\pi \hbar^2 \tilde{v}_F^2} \left[   \left( g_{\nu}^{(1)}(\mathbf{r}) g_{-\nu}^{(3)}(\mathbf{r}') + g_{\nu}^{(2)}(\mathbf{r}) g_{-\nu}^{(1)}(\mathbf{r}')\right)w_\text{AA} w_\text{AB}   K_0 \left( \frac{d}{i \ell} \right) \right. \\
&\left. - \nu \left( g_\nu^{(1)}(\mathbf{r}) g_{-\nu}^{(1)}(\mathbf{r}')w_\text{AB}^2 e^{-i \nu \phi_\mathbf{d}}  +  g_{\nu}^{(2)}(\mathbf{r})g_{-\nu}^{(3)}(\mathbf{r}') w_\text{AA}^2 e^{i \nu \phi_\mathbf{d}} \right) K_1 \left( \frac{d}{i \ell} \right)    \right], \\
\mathcal{G}_{B_2A_1}^{(0)} \left(E>0, \mathbf{K}_\nu^1, \mathbf{r} , \mathbf{r}' \right) &= \frac{e^{i \mathbf{K}_\nu^1 \cdot \mathbf{d}}}{\omega} \frac{i \hbar v_F G}{\pi \hbar^2 \tilde{v}_F^2} \left[ g_{\nu}^{(1)}(\mathbf{r}) w_\text{AB} e^{i \nu \phi_\mathbf{d}} K_1 \left( \frac{d}{i \ell} \right)  - \nu g_{\nu}^{(3)}(\mathbf{r}) w_\text{AA} K_0 \left( \frac{d}{i \ell} \right) \right] ,\\
\mathcal{G}_{B_2B_1}^{(0)} \left(E>0, \mathbf{K}_\nu^1, \mathbf{r} , \mathbf{r}' \right) &= \frac{e^{i \mathbf{K}_\nu^1 \cdot \mathbf{d}}}{\omega}\frac{i \hbar v_F G}{\pi \hbar^2 \tilde{v}_F^2} \left[  g_{\nu}^{(3)}(\mathbf{r}) w_\text{AB} e^{-i \nu \phi_\mathbf{d}} K_1 \left( \frac{d}{i \ell} \right)  - \nu g_{\nu}^{(1)}(\mathbf{r}) w_\text{AA} K_0 \left( \frac{d}{i \ell} \right) \right]   ,\\
\mathcal{G}_{B_2A_2}^{(0)} \left(E>0, \mathbf{K}_\nu^1, \mathbf{r} , \mathbf{r}' \right) &= \frac{e^{i \mathbf{K}_\nu^1 \cdot \mathbf{d}}}{\omega} \frac{1}{\pi \hbar^2 \tilde{v}_F^2} \left[   \left( g_{\nu}^{(1)}(\mathbf{r}) g_{-\nu}^{(2)}(\mathbf{r}') + g_{\nu}^{(3)}(\mathbf{r}) g_{-\nu}^{(1)}(\mathbf{r}')\right)w_\text{AA} w_\text{AB}   K_0 \left( \frac{d}{i \ell} \right) \right. \\
&\left. - \nu \left( g_\nu^{(1)}(\mathbf{r}) g_{-\nu}^{(1)}(\mathbf{r}')w_\text{AB}^2 e^{i \nu \phi_\mathbf{d}}  +  g_{\nu}^{(3)}(\mathbf{r})g_{-\nu}^{(2)}(\mathbf{r}') w_\text{AA}^2 e^{-i \nu \phi_\mathbf{d}} \right) K_1 \left( \frac{d}{i \ell} \right)    \right], \\
\mathcal{G}_{B_2B_2}^{(0)} \left(E>0, \mathbf{K}_\nu^1, \mathbf{r} , \mathbf{r}' \right) &= \frac{e^{i \mathbf{K}_\nu^1 \cdot \mathbf{d}}}{\omega}  \frac{1}{\pi \hbar^2 \tilde{v}_F^2} \left[  \nu \left( g_\nu^{(3)}(\mathbf{r}) g_{-\nu}^{(1)}(\mathbf{r}')e^{-i \nu \phi_\mathbf{d}}  +  g_{\nu}^{(1)}(\mathbf{r})g_{-\nu}^{(3)}(\mathbf{r}') e^{i \nu \phi_\mathbf{d}} \right) w_\text{AA} w_\text{AB}K_1 \left( \frac{d}{i \ell} \right)   \right. \\
&\left. - \left( g_{\nu}^{(1)}(\mathbf{r}) g_{-\nu}^{(1)}(\mathbf{r}')w_\text{AB}^2 + g_{\nu}^{(3)}(\mathbf{r}) g_{-\nu}^{(3)}(\mathbf{r}')w_\text{AA}^2\right)    K_0 \left( \frac{d}{i \ell} \right)   \right], \\
\end{split}
\end{equation}
where $\omega =  \left( 6 w_\text{AA}^2+6w_\text{AB}^2+2\hbar^2 v_F^2 G^2 \right)/E$ and  $\mathbf{d} = \mathbf{r} - \mathbf{r}' \neq \mathbf{0}.$ Similarly, the bare Green's function near $\mathbf{K}_\nu^2$ is 
\begin{equation}
\mathcal{G}^{(0)}\left( z, \mathbf{K}_\nu^2, \mathbf{r}, \mathbf{r}' \right) = \frac{1}{\left( 2 \pi \right)^2} \sum_{\chi = \pm 1}\int d^2 \mathbf{q} \frac{\ket{\psi^{\chi \mathbf{q}}_{\mathbf{K}_\nu^2}(\mathbf{r})}\bra{\psi^{\chi \mathbf{q}}_{\mathbf{K}_\nu^2}(\mathbf{r}')}}{z - \chi \hbar \tilde{v}_F q}.
\end{equation}
The matrix elements are the same as those found above with the layers exchanged, $\mathbf{K}_\nu^1 \leftrightarrow \mathbf{K}_\nu^2,$ $i \leftrightarrow - i,$ $ \phi_\mathbf{k}\leftrightarrow - \phi_\mathbf{k},$  $g_\nu^{(1)}(\mathbf{r}) \leftrightarrow g_{-\nu}^{(1)}(\mathbf{r}),$  and $g_\nu^{(2)}(\mathbf{r}) \leftrightarrow g_{-\nu}^{(3)}(\mathbf{r}).$

With these matrix elements, we can compute LDOS due to a localized impurity located at $\mathbf{s}$ via
\begin{equation}
\Delta \rho(E, \mathbf{r}) = - \frac{1}{\pi} \Im \text{Tr} \left[\sum_{\nu = \pm 1}\left( \mathcal{G}^{(0)}\left( E, \mathbf{K}_\nu^1, \mathbf{r}, \mathbf{s} \right) + \mathcal{G}^{(0)}\left( E, \mathbf{K}_\nu^2, \mathbf{r}, \mathbf{s} \right) \right) \mathcal{T}_\nu\left( E, \mathbf{s} \right)\left( \mathcal{G}^{(0)}\left( E, \mathbf{K}_\nu^1, \mathbf{s}, \mathbf{r} \right) + \mathcal{G}^{(0)}\left( E, \mathbf{K}_\nu^2, \mathbf{s}, \mathbf{r} \right) \right) \right],
\end{equation}
where  $\mathcal{T}(E, \mathbf{s}) $ is the $\cal{T}$ matrix. For simplicity, we use the Born approximation with an impurity localized on an $A_1$ site at $\mathbf{s}.$ 
The impurity potential  is 
\begin{equation}
\mathcal{U}(\mathbf{r}, \mathbf{r}') =  \begin{pmatrix}
U_0 & 0 & 0 & 0 \\
0 & 0 & 0 & 0 \\
0 & 0 & 0 & 0 \\
0 & 0 & 0 & 0 
\end{pmatrix} \delta^{(2)}(\mathbf{r}-\mathbf{s}) \delta^{(2)}(\mathbf{r}'-\mathbf{s}).
\end{equation}
The LDOS can then be split into two parts, as before, with one containing intra-Dirac-cone scattering and one with inter-Dirac-cone scattering. For our purpose, we only study the latter contribution 
\begin{equation}
\label{eq: LDOS for Dirac model}
\begin{split}
\Delta \rho_\text{inter} \left( E, \mathbf{r}, \mathbf{s} \right) = - \frac{1}{\pi} \sum_{\nu = \pm 1, x = A_1,B_1,A_2,B_2}\Im &\left[ \mathcal{G}^{(0)}_{xA_1}\left(E, \mathbf{K}_\nu^1, \mathbf{r}, \mathbf{s} \right) U_0 \mathcal{G}^{(0)}_{A_1x}\left(E, \mathbf{K}_\nu^2, \mathbf{s} ,\mathbf{r}\right) \right. \\
&\left.+ \mathcal{G}^{(0)}_{xA_1}\left(E, \mathbf{K}_\nu^2, \mathbf{r}, \mathbf{s} \right)U_0 \mathcal{G}^{(0)}_{A_1x}\left(E, \mathbf{K}_\nu^1, \mathbf{s} ,\mathbf{r}\right) \right].
\end{split}
\end{equation}
equation~\eqref{eq: LDOS for Dirac model}, in general, could contain phases  $e^{\pm i \phi_\mathbf{r}}$ and $e^{\pm i 2 \phi_\mathbf{d}}.$ The interference pattern of LDOS, and in particular the presence of dislocations, depends on the relative amplitudes of these phases at a fixed radial distance $d.$ Therefore, the number of dislocations observed depends on the location of the impurity $\mathbf{s}.$ At a general $\mathbf{s},$ the change in LDOS once summed over valleys is 
\begin{equation}
\begin{split}
\Delta \rho \left(E, \mathbf{r} \right) = - \frac{4\hbar^2 v_F^2 G^2 U_0}{\pi^3 \omega^2 \hbar^4 \tilde{v}_F^4} \Im &\left[  4  w_\text{AA} w_\text{AB} K_0 \left( \frac{d}{i \ell} \right) K_1 \left( \frac{d}{i \ell} \right) \Re \left( e^{i\Delta \bar{\mathbf{K}} \cdot \mathbf{d} + i \phi_\mathbf{d}} g_{-}^{(3)}(\mathbf{s})g_{+}^{(1)}(\mathbf{r})\right)   \right.\\
&\left.  2w_\text{AB}^2 \left(  K_1^2 \left( \frac{d}{i \ell} \right)  + K_0^2 \left( \frac{d}{i \ell} \right) \right) \Re \left(e^{i\Delta \bar{\mathbf{K}} \cdot \mathbf{d}} g_{-}^{(1)}(\mathbf{s}) g_{-}^{(1)}(\mathbf{r} )\right) \right],
\end{split}
\end{equation}
where $\Delta \bar{\mathbf{K}} = \mathbf{K}_{+}^1 - \mathbf{K}_{+}^2.$ To simplify this expression slightly, we can perform a unit-cell average of the functions $g_\pm^{(i)}(\mathbf{r})$ since we are only interested in the moiré-scale interference pattern, which gives 
\begin{equation}
\label{eq: LDOS of the linear Dirac theory}
\begin{split}
\Delta \rho \left(E, \mathbf{r} \right) = - \frac{4\hbar^2 v_F^2 G^2 U_0}{\pi^3 \omega^2 \hbar^4 \tilde{v}_F^4} \Im &\left[  4   w_\text{AA} w_\text{AB} K_0 \left( \frac{d}{i \ell} \right) K_1 \left( \frac{d}{i \ell} \right) \Re \left( e^{i\Delta \bar{\mathbf{K}} \cdot \mathbf{d} + i \phi_\mathbf{d}} g_{-}^{(3)}(\mathbf{s})\right)   \right.\\
&\left.  2w_\text{AB}^2  \left(  K_1^2 \left( \frac{d}{i \ell} \right)  + K_0^2 \left( \frac{d}{i \ell} \right) \right) \Re \left(e^{i\Delta \bar{\mathbf{K}} \cdot \mathbf{d}} g_{-}^{(1)}(\mathbf{s}) \right) \right].
\end{split}
\end{equation}
From equation~\eqref{eq: LDOS of the linear Dirac theory}, we observe that the presence of a single wavefront dislocation is governed by the relative amplitudes of the terms $ e^{i\Delta \bar{\mathbf{K}} \cdot \mathbf{d} + i \phi_\mathbf{d}}$ and $ e^{i\Delta \bar{\mathbf{K}} \cdot \mathbf{d}}.$ If the amplitude of the former is greater, then we expect to observe a dislocation and not if otherwise. These two amplitudes are tuned by the interlayer hopping parameters and by the location of the impurity $\mathbf{s}.$ In practice, $w_\text{AB}$ and $w_\text{AA}$ are both non-zero and of the same order of magnitude; and they are also not easily tunable in an experiment. However, $\mathbf{s}$ is simply the location of the impurity, and can be chosen in an experiment to control the presence of a dislocation. In particular, at an $AA$ region where $g_{+}^{(1)} = 3$ and $g_{\pm}^{(2)} = g_{\pm}^{(3)} = 0,$ we expect to not observe any dislocation in the charge density. However, at an $AB$ region where $g_{+}^{(2)} = 3$ and $g_{\pm}^{(1)} = g_{\pm}^{(3)} = 0,$ we expect to observe a single dislocation. It is also interesting to note that for an impurity placed on at an $A_1$ site at a $BA$ region, the induced change in LDOS predicted by equation~\eqref{eq: LDOS of the linear Dirac theory} is precisely zero because $g_{\pm}^{(1)} = g_{\pm}^{(2)} = 0.$ This is because the density matrix at an $BA$ region does not couple to an $A_1$ site. Therefore, as long as the linear-band approximation is valid, it is predicts no change in LDOS. This does not however mean that no dislocation can exist in the full continuum theory, just that if a dislocation does exist, its amplitude will be of sub-leading order in a regime where the linear-band approximation is dominant.

We note that an atomic impurity does induce intervalley scattering because it contains wavevectors spanning the entire momentum space. These processes will induce oscillations on the atomic scale. Thus, as long as we are only probing moiré-scale interference pattern, these processes can be neglected. In the next section, we will see that by applying an appropriate Fourier transform filter, the contributions from intervalley terms can be screened. 

\section{Numerical Simulation of Friedel Oscillations}

Having studied Friedel oscillations in a truncated continuum model, we now confirm these results with simulations of the full continuum approximation. In this case, the Bloch states are written in a four-component basis of $\ket{A_1},$ $\ket{B_1},$ $\ket{A_2},$ and $\ket{B_2}$ as superposition of plane waves, as explained earlier. The wavefunctions are normalized to 
\begin{equation}
\bra{\Psi^{n'\mathbf{k}'}_{\nu'}} \ket{\Psi^{n\mathbf{k}}_{\nu}} = N^2\delta_{\nu \nu'} \delta_{n n'} \delta \left( \mathbf{k} - \mathbf{k}' \right),
\end{equation}
where the inner product is integrated over the entire crystal and $N^2$ is the number of unit cells. In our simulations, we choose $N^2 = 20 \times 20.$ The Green's function for a pristine twisted bilayer  is 
\begin{equation}
\mathcal{G}^{(0)}(z) = \frac{V_\text{cell}}{(2 \pi)^2} \int_\text{BZ} d^2 \mathbf{k} \sum_{n, \nu} \frac{\ket{\Psi^{n\mathbf{k}}_{\nu}}\bra{\Psi^{n\mathbf{k}}_{\nu}}}{z - \varepsilon_{n \mathbf{k}}^{(\nu)}}.
\end{equation}
For each $\mathbf{k},$ the eigenfunctions and energies are obtained from numerical diagonalization of the continuum Hamiltonian. Then, via the $\cal{T}$ matrix, the change in the Green's function due to the presence of a localized impurity can be calculated directly from the bare Green's function. Thus, we see that the calculation of LDOS becomes essentially a problem of evaluating the bare Green's function on a finite mesh efficiently using some numerical strategy. To this end, we employ the well-established tetrahedron method of Brillouin zone integration \cite{RF75, BJA94} to calculate the  part of the bare Green's function related to the total bare density of states. Then, we obtain the remaining part of the bare Green's function by the Hilbert transform. We first divide the irreducible Brillouin zone, which is the parallelogram whose sides are $\mathbf{G}_1^\text{M}$ and $\mathbf{G}_2^\text{M},$ into a mesh of $N \times N$  smaller parallelograms whose sides are $\mathbf{g}_1 = \mathbf{G}_1^\text{M}/N$ and $\mathbf{g}_2 = \mathbf{G}_2^\text{M}/N.$ Each parallelogram is then divided into two triangles, with each triangular having three vertices on the mesh points. There are $2N^2$ triangles on the mesh. The volume of each triangle is $V_\text{BZ}/2N^2,$ where $V_\text{BZ}$ is the volume of the Brillouin zone. We want to calculate the following quantity, 
\begin{equation}
\begin{split}
\mathcal{G}^{(0)}(E) &=  \lim_{\delta \rightarrow 0}\frac{V_\text{cell}}{(2 \pi)^2} \int_\text{BZ} d^2 \mathbf{k} \sum_{q} \frac{\mathcal{M}_{q}(\mathbf{k})}{E - \varepsilon_{q \mathbf{k}} + i \delta } \\
&=  \sum_{q} \frac{V_\text{cell}}{(2 \pi)^2} \left( \mathcal{P}\int_\text{BZ} d^2 \mathbf{k}  \frac{\mathcal{M}_{q}(\mathbf{k})}{E - \varepsilon_{q \mathbf{k}} } - \pi i \int_\text{BZ} d^2 \mathbf{k}  \mathcal{M}_{q}(\mathbf{k}) \delta \left(E - \varepsilon_{q \mathbf{k}}  \right) \right)  \\
&= \mathcal{G}_1^{(0)}(E) - \pi i \mathcal{G}_2^{(0)}(E) ,
\end{split}
\end{equation}
where $\mathcal{P}$ denotes the Cauchy principal value,  $\mathcal{M}_q(\mathbf{k})$ is some complex-valued matrix that depends on the wavevector $\mathbf{k}$ and some set of quantum numbers $q;$ In our specific case of twisted bilayer graphene, $q = \lbrace n ,\nu \rbrace$ includes both the band index and the valley index, and $\mathcal{M}_n^{(\nu)}(\mathbf{k}) = \ket{\Psi^{n\mathbf{k}}_{\nu}}\bra{\Psi^{n\mathbf{k}}_{\nu}}.$ Note that when written in a specific basis, $\mathcal{M}_n(\mathbf{k})$ may carry other variables, like spatial indices such as $\mathcal{M}_n^{(\nu)}(\mathbf{k},\mathbf{r},\mathbf{r}') = \ket{\Psi^{n\mathbf{k}}_{\nu}(\mathbf{r})}\bra{\Psi^{n\mathbf{k}}_{\nu}(\mathbf{r}')}.$ We define two matrices
\begin{equation}
\begin{split}
\mathcal{G}_1^{(0)}(E) &= \sum_{q} \frac{V_\text{cell}}{(2 \pi)^2} \mathcal{P}\int_\text{BZ} d^2 \mathbf{k}  \frac{\mathcal{M}_{q}(\mathbf{k})}{E - \varepsilon_{q \mathbf{k}} } , \\
\mathcal{G}_2^{(0)}(E) &= \sum_{q} \frac{V_\text{cell}}{(2 \pi)^2} \int_\text{BZ} d^2 \mathbf{k}  \mathcal{M}_{q}(\mathbf{k}) \delta \left(E - \varepsilon_{q \mathbf{k}}  \right).
\end{split}
\end{equation}
These two functions are related to each other by the Hilbert transform
\begin{equation}
\mathcal{G}_1^{(0)}(E) = \mathcal{P} \int_{-\infty}^\infty dE' \frac{\mathcal{G}_2^{(0)}(E')}{E-E'} .
\end{equation}
Thus we only need to calculate  $\mathcal{G}_2^{(0)}(E)$ and then obtain $\mathcal{G}_1^{(0)}(E)$ via the transform. To calculate $\mathcal{G}_2^{(0)}(E),$ we use the tetrahedron method to partition the integral over the Brillouin zone as a sum of integrals over smaller triangles
\begin{equation}
\mathcal{G}_2^{(0)}(E) = \sum_{q} \frac{V_\text{cell}}{(2 \pi)^2} \int_\text{BZ} d^2 \mathbf{k}  \mathcal{M}_{q}(\mathbf{k}) \delta \left(E - \varepsilon_{q \mathbf{k}}  \right) =  \sum_{q} \sum_{\tau = 1}^{2N^2} \frac{V_\text{cell}}{(2 \pi)^2} \int_{\Omega_\tau} d^2 \mathbf{k}  \mathcal{M}_{q}(\mathbf{k}) \delta \left(E - \varepsilon_{q \mathbf{k}}  \right),
\end{equation}
then  because $\mathcal{M}_q(\mathbf{k})$ and $\varepsilon_{q\mathbf{k}}$ are not exactly known everywhere within each triangle, we approximate them as linear functions of $\mathbf{k}$ under the assumption that they converge to the true quantities of the band structure as the mesh is made finer and finer since they are expected to be smooth functions. Using this approximation, we can write 
\begin{equation}
\label{eq: the imaginary part}
\mathcal{G}_2^{(0)}(E) = \sum_{q} \sum_{\tau = 1}^{2N^2} \sum_{i = 1}^3 \mathcal{M }_q(\mathbf{k}_i^\tau) \mathcal{W}_{q, \tau, i}(E),
\end{equation}
where $\mathbf{k}^{\tau}_i$ are the wavevectors at the vertices of the $\tau$ triangular arranged in such a way that $\varepsilon_q(\mathbf{k}_1^\tau) \leq \varepsilon_q(\mathbf{k}_2^\tau) \leq \varepsilon_q(\mathbf{k}_3^\tau),$ and the weight factors are defined piecewise as follows  \cite{SY16}
\begin{equation}
\begin{split}
\mathcal{W}_{q, \tau, 1}(E) &= \begin{cases}
0 & \text{if } E < \varepsilon_{q,1}\\
\frac{1}{2N^2} \frac{(E - \varepsilon_{q,1})}{\left( \varepsilon_{q,3}-\varepsilon_{q,1} \right)\left( \varepsilon_{q,2}-\varepsilon_{q,1} \right)} \left( \frac{\varepsilon_{q,2}-E}{\varepsilon_{q,2}-\varepsilon_{q,1}} +  \frac{\varepsilon_{q,3}-E}{\varepsilon_{q,3}-\varepsilon_{q,1}} \right)& \text{if } \varepsilon_{q,1} \leq  E < \varepsilon_{q,2}\\
\frac{1}{2N^2} \frac{(\varepsilon_{q,3} - E )}{\left( \varepsilon_{q,3}-\varepsilon_{q,1} \right)\left( \varepsilon_{q,3}-\varepsilon_{q,2} \right)}\frac{\varepsilon_{q,3}-E}{\varepsilon_{q,3}-\varepsilon_{q,1}}  &\text{if } \varepsilon_{q,2} \leq  E < \varepsilon_{q,3}\\ 
0 &\text{if } \varepsilon_{q,3} \leq  E
  \end{cases} , \\
  \mathcal{W}_{q, \tau, 2}(E) &= \begin{cases}
0 & \text{if } E < \varepsilon_{p,1}\\
\frac{1}{2N^2} \frac{(E - \varepsilon_{q,1})}{\left( \varepsilon_{q,3}-\varepsilon_{q,1} \right)\left( \varepsilon_{q,2}-\varepsilon_{q,1} \right)} \frac{E - \varepsilon_{q,1}}{\varepsilon_{q, 2}-\varepsilon_{q,1}} \text{ }\quad \quad \quad \quad \quad \quad & \text{if } \varepsilon_{q,1} \leq  E < \varepsilon_{q,2}\\
\frac{1}{2N^2} \frac{(\varepsilon_{q,3} - E )}{\left( \varepsilon_{q,3}-\varepsilon_{q,1} \right)\left( \varepsilon_{q,3}-\varepsilon_{q,2} \right)}\frac{\varepsilon_{q,3}-E}{\varepsilon_{q,3}-\varepsilon_{q,2}}  &\text{if } \varepsilon_{q,2} \leq  E < \varepsilon_{q,3}\\ 
0 &\text{if } \varepsilon_{q,3} \leq  E
  \end{cases},\\
  \mathcal{W}_{q, \tau, 3}(E) &= \begin{cases}
0 & \text{if } E < \varepsilon_{q,1}\\
\frac{1}{2N^2} \frac{(E - \varepsilon_{q,1})}{\left( \varepsilon_{q,3}-\varepsilon_{q,1} \right)\left( \varepsilon_{q,2}-\varepsilon_{q,1} \right)} \frac{E-\varepsilon_{q,1}}{\varepsilon_{q,3}-\varepsilon_{q,1}}& \text{if } \varepsilon_{q,1} \leq  E < \varepsilon_{q,2}\\
\frac{1}{2N^2} \frac{(\varepsilon_{q,3} - E )}{\left( \varepsilon_{q,3}-\varepsilon_{q,1} \right)\left( \varepsilon_{q,3}-\varepsilon_{q,2} \right)}\left( \frac{E- \varepsilon_{q,1}}{\varepsilon_{q,3}-\varepsilon_{q,1}}+\frac{E-\varepsilon_{q,2}}{\varepsilon_{q,3}-\varepsilon_{q,2}} \right)  &\text{if } \varepsilon_{q,2} \leq  E < \epsilon_{q,3}\\ 
0 &\text{if } \varepsilon_{q,3} \leq  E
  \end{cases}.
  \end{split}
\end{equation}
Here, we adopt the notation that $\varepsilon_q(\mathbf{k}_i^\tau) = \varepsilon_{q,i}.$ From these explicit functions, we can also obtain $\mathcal{G}_1^{(0)}(E)$ via the Hilbert transform 
\begin{equation}
\label{eq: the real part}
\mathcal{G}_1^{(0)}(E) = \sum_{q} \sum_{\tau = 1}^{2N^2} \sum_{i = 1}^3 \mathcal{M }_q(\mathbf{k}_i^\tau) \tilde{\mathcal{W}}_{q, \tau, i}(E),
\end{equation}
where 
\begin{equation}
\tilde{\mathcal{W}}_{q, \tau, i}(E) = \mathcal{P} \int_{-\infty}^\infty dE' \frac{\mathcal{W}_{q, \tau, i}(E')}{E-E'}.
\end{equation}
The principal value is defined as long as $E \neq \varepsilon_{q,1},$ $E \neq \varepsilon_{q,2},$ and $E \neq \varepsilon_{q,3},$ which are marginal cases that can be avoided in numerical calculation by small changes to the energies. These are given explicitly as 
\begin{equation}
\tilde{\mathcal{W}}_{q, \tau, 1}(E) = \begin{cases} 
\text{if } E < \varepsilon_{q,1} &\begin{cases}
 \frac{\frac{\left(\varepsilon _{q,1}-\varepsilon _{q,3}\right) \left(\varepsilon _{q,1}-3 \varepsilon _{q,2}+2 E\right)}{\varepsilon _{q,1}-\varepsilon _{q,2}}-3 \varepsilon _{q,1}+\varepsilon _{q,2}+2 E}{4 N^2 \left(\varepsilon _{q,1}-\varepsilon _{q,3}\right){}^2} \\
 + \frac{\left(\varepsilon _{q,1}-E\right) \left(E \varepsilon _{q,3}+\varepsilon _{q,1} \left(\varepsilon _{q,2}+\varepsilon _{q,3}-2 E\right)+\varepsilon _{q,2} \left(E-2 \varepsilon _{q,3}\right)\right) \log \left(\frac{E-\varepsilon _{q,1}}{E-\varepsilon _{q,2}}\right)}{2 N^2 \left(\varepsilon _{q,1}-\varepsilon _{q,2}\right){}^2 \left(\varepsilon _{q,1}-\varepsilon _{q,3}\right){}^2} & \text{if } \varepsilon_{q,1} <  \varepsilon_{q,2} = \varepsilon_{q,3}\\
\frac{\left(\varepsilon _{q,2}-\varepsilon _{q,3}\right) \left(\varepsilon _{q,2}-3 \varepsilon _{q,3}+2 E\right)+2 \left(E-\varepsilon _{q,3}\right){}^2 \log \left(\frac{E-\varepsilon _{q,2}}{E-\varepsilon _{q,3}}\right)}{4 N^2 \left(\varepsilon _{q,1}-\varepsilon _{q,3}\right){}^2 \left(\varepsilon _{q,3}-\varepsilon _{q,2}\right)} & \text{if } \varepsilon_{q,2} = \varepsilon_{q,2} < \varepsilon_{q,3}  \\
\frac{\left(\varepsilon _{q,1}-\varepsilon _{q,3}\right) \left(\varepsilon _{q,3}-\varepsilon _{q,2}\right) \left(\varepsilon _{q,1}-E\right)-\left(\varepsilon _{q,1}-\varepsilon _{q,2}\right) \left(E-\varepsilon _{q,3}\right){}^2 \log \left(\frac{E-\varepsilon _{q,2}}{E-\varepsilon _{q,3}}\right)}{2 N^2 \left(\varepsilon _{q,1}-\varepsilon _{q,2}\right) \left(\varepsilon _{q,1}-\varepsilon _{q,3}\right){}^2 \left(\varepsilon _{q,2}-\varepsilon _{q,3}\right)} \\
+\frac{\left(\varepsilon _{q,1}-E\right) \left(E \varepsilon _{q,3}+\varepsilon _{q,1} \left(\varepsilon _{q,2}+\varepsilon _{q,3}-2 E\right)+\varepsilon _{q,2} \left(E-2 \varepsilon _{q,3}\right)\right) \log \left(\frac{E-\varepsilon _{q,1}}{E-\varepsilon _{q,2}}\right)}{2 N^2 \left(\varepsilon _{q,1}-\varepsilon _{q,2}\right){}^2 \left(\varepsilon _{q,1}-\varepsilon _{q,3}\right){}^2} & \text{if } \varepsilon _{q,1}< \varepsilon _{q,2} < \varepsilon _{q,3} \\
0 & \text{otherwise}
\end{cases} \\
\text{if } \varepsilon_{q,1} < E < \varepsilon_{q,2} &\begin{cases}
\frac{\frac{\left(\varepsilon _{q,1}-\varepsilon _{q,3}\right) \left(\varepsilon _{q,1}-3 \varepsilon _{q,2}+2 E\right)}{\varepsilon _{q,1}-\varepsilon _{q,2}}-3 \varepsilon _{q,1}+\varepsilon _{q,2}+2 E}{4 N^2 \left(\varepsilon _{q,1}-\varepsilon _{q,3}\right){}^2}\\
-\frac{\left(\varepsilon _{q,1}-E\right) \left(E \varepsilon _{q,3}+\varepsilon _{q,1} \left(\varepsilon _{q,2}+\varepsilon _{q,3}-2 E\right)+\varepsilon _{q,2} \left(E-2 \varepsilon _{q,3}\right)\right) \log \left(\frac{\varepsilon _{q,2}-E}{E-\varepsilon _{q,1}}\right)}{2 N^2 \left(\varepsilon _{q,1}-\varepsilon _{q,2}\right){}^2 \left(\varepsilon _{q,1}-\varepsilon _{q,3}\right){}^2} & \text{if } \varepsilon_{q,2} = \varepsilon_{q,3}\\
-\frac{\left(\varepsilon _{q,1}-E\right) \left(E \varepsilon _{q,3}+\varepsilon _{q,1} \left(\varepsilon _{q,2}+\varepsilon _{q,3}-2 E\right)+\varepsilon _{q,2} \left(E-2 \varepsilon _{q,3}\right)\right) \log \left(\frac{\varepsilon _{q,2}-E}{E-\varepsilon _{q,1}}\right)}{2 N^2 \left(\varepsilon _{q,1}-\varepsilon _{q,2}\right){}^2 \left(\varepsilon _{q,1}-\varepsilon _{q,3}\right){}^2}\\
+\frac{\left(E-\varepsilon _{q,3}\right){}^2 \log \left(\frac{E-\varepsilon _{q,2}}{E-\varepsilon _{q,3}}\right)}{2 N^2 \left(\varepsilon _{q,1}-\varepsilon _{q,3}\right){}^2 \left(\varepsilon _{q,3}-\varepsilon _{q,2}\right)} - \frac{\left( \varepsilon_{q,1} - E \right) }{2 N^2 \left(\varepsilon _{q,1}-\varepsilon _{q,2}\right) \left(\varepsilon _{q,1}-\varepsilon _{q,3}\right)}  & \text{otherwise} \\
\end{cases} \\
\text{if } \varepsilon_{q,2} < E < \varepsilon_{q,3} &\begin{cases}
\frac{\left(\varepsilon _{q,2}-\varepsilon _{q,3}\right) \left(\varepsilon _{q,2}-3 \varepsilon _{q,3}+2 E\right)-2 \left(E-\varepsilon _{q,3}\right){}^2 \log \left(\frac{\varepsilon _{q,3}-E}{E-\varepsilon _{q,2}}\right)}{4 N^2 \left(\varepsilon _{q,1}-\varepsilon _{q,3}\right){}^2 \left(\varepsilon _{q,3}-\varepsilon _{q,2}\right)} & \text{if } \varepsilon_{q, 1} = \varepsilon_{q,2} \\
\frac{\left(\varepsilon _{q,1}-E\right) \left(E \varepsilon _{q,3}+\varepsilon _{q,1} \left(\varepsilon _{q,2}+\varepsilon _{q,3}-2 E\right)+\varepsilon _{q,2} \left(E-2 \varepsilon _{q,3}\right)\right) \log \left(\frac{E-\varepsilon _{q,1}}{E-\varepsilon _{q,2}}\right)}{2 N^2 \left(\varepsilon _{q,1}-\varepsilon _{q,2}\right){}^2 \left(\varepsilon _{q,1}-\varepsilon _{q,3}\right){}^2} \\
+ \frac{\left(\varepsilon _{q,1}-\varepsilon _{q,3}\right) \left(\varepsilon _{q,3}-\varepsilon _{q,2}\right) \left(\varepsilon _{q,1}-E\right)+\left(\varepsilon _{q,1}-\varepsilon _{q,2}\right) \left(E-\varepsilon _{q,3}\right){}^2 \log \left(\frac{\varepsilon _{q,3}-E}{E-\varepsilon _{q,2}}\right)}{2 N^2 \left(\varepsilon _{q,1}-\varepsilon _{q,2}\right) \left(\varepsilon _{q,1}-\varepsilon _{q,3}\right){}^2 \left(\varepsilon _{q,2}-\varepsilon _{q,3}\right)} & \text{otherwise } \\
\end{cases} \\
\text{if } \varepsilon_{q,3} < E  &\begin{cases}
\frac{\frac{\left(\varepsilon _{q,1}-\varepsilon _{q,3}\right) \left(\varepsilon _{q,1}-3 \varepsilon _{q,2}+2 E\right)}{\varepsilon _{q,1}-\varepsilon _{q,2}}-3 \varepsilon _{q,1}+\varepsilon _{q,2}+2 E}{4 N^2 \left(\varepsilon _{q,1}-\varepsilon _{q,3}\right){}^2}\\
+\frac{\left(\varepsilon _{q,1}-E\right) \left(E \varepsilon _{q,3}+\varepsilon _{q,1} \left(\varepsilon _{q,2}+\varepsilon _{q,3}-2 E\right)+\varepsilon _{q,2} \left(E-2 \varepsilon _{q,3}\right)\right) \log \left(\frac{E-\varepsilon _{q,1}}{E-\varepsilon _{q,2}}\right)}{2 N^2 \left(\varepsilon _{q,1}-\varepsilon _{q,2}\right){}^2 \left(\varepsilon _{q,1}-\varepsilon _{q,3}\right){}^2} & \text{if } \varepsilon_{q,1} < \varepsilon_{q,2}  = \varepsilon_{q, 3}\\
\frac{\left(\varepsilon _{q,2}-\varepsilon _{q,3}\right) \left(\varepsilon _{q,2}-3 \varepsilon _{q,3}+2 E\right)-2 \left(E-\varepsilon _{q,3}\right){}^2 \log \left(\frac{E-\varepsilon _{q,3}}{E-\varepsilon _{q,2}}\right)}{4 N^2 \left(\varepsilon _{q,1}-\varepsilon _{q,3}\right){}^2 \left(\varepsilon _{q,3}-\varepsilon _{q,2}\right)} & \text{if } \varepsilon_{q, 1} = \varepsilon_{q, 2} < \varepsilon_{q, 3}\\
 \frac{\left(\varepsilon _{q,1}-\varepsilon _{q,3}\right) \left(\varepsilon _{q,3}-\varepsilon _{q,2}\right) \left(\varepsilon _{q,1}-E\right)+\left(\varepsilon _{q,1}-\varepsilon _{q,2}\right) \left(E-\varepsilon _{q,3}\right){}^2 \log \left(\frac{E-\varepsilon _{q,3}}{E-\varepsilon _{q,2}}\right)}{2 N^2 \left(\varepsilon _{q,1}-\varepsilon _{q,2}\right) \left(\varepsilon _{q,1}-\varepsilon _{q,3}\right){}^2 \left(\varepsilon _{q,2}-\varepsilon _{q,3}\right)} \\
 + \frac{\left(\varepsilon _{q,1}-E\right) \left(E \varepsilon _{q,3}+\varepsilon _{q,1} \left(\varepsilon _{q,2}+\varepsilon _{q,3}-2 E\right)+\varepsilon _{q,2} \left(E-2 \varepsilon _{q,3}\right)\right) \log \left(\frac{E-\varepsilon _{q,1}}{E-\varepsilon _{q,2}}\right)}{2 N^2 \left(\varepsilon _{q,1}-\varepsilon _{q,2}\right){}^2 \left(\varepsilon _{q,1}-\varepsilon _{q,3}\right){}^2} & \text{if } \varepsilon_{q, 1} < \varepsilon_{q, 2} < \varepsilon_{q, 3}\\
0 & \text{otherwise } \\
\end{cases} 
\end{cases},
\end{equation}

\begin{equation}
\tilde{\mathcal{W}}_{q, \tau, 2}(E) = \begin{cases}
\text{if } E < \varepsilon_{q, 1} &\begin{cases}
\frac{2 \left(E-\varepsilon _{q,1}\right){}^2 \log \left(\frac{E-\varepsilon _{q,2}}{E-\varepsilon _{q,1}}\right)-\left(\varepsilon _{q,1}-\varepsilon _{q,2}\right) \left(-3 \varepsilon _{q,1}+\varepsilon _{q,2}+2 E\right)}{4 N^2 \left(\varepsilon _{q,1}-\varepsilon _{q,2}\right){}^2 \left(\varepsilon _{q,1}-\varepsilon _{q,3}\right)} & \text{if } \varepsilon_{q, 1} < \varepsilon_{q, 2} = \varepsilon_{q, 3}\\
\frac{\left(\varepsilon _{q,3}-\varepsilon _{q,2}\right) \left(\varepsilon _{q,2}-3 \varepsilon _{q,3}+2 E\right)+2 \left(E-\varepsilon _{q,3}\right){}^2 \log \left(\frac{E-\varepsilon _{q,3}}{E-\varepsilon _{q,2}}\right)}{4 N^2 \left(\varepsilon _{q,1}-\varepsilon _{q,3}\right) \left(\varepsilon _{q,2}-\varepsilon _{q,3}\right){}^2} & \text{if } \varepsilon_{q, 1} = \varepsilon_{q, 2} < \varepsilon_{q, 3} \\
 \frac{\left(E-\varepsilon _{q,1}\right){}^2 \log \left(\frac{E-\varepsilon _{q,2}}{E-\varepsilon _{q,1}}\right)}{2 N^2 \left(\varepsilon _{q,1}-\varepsilon _{q,2}\right){}^2 \left(\varepsilon _{q,1}-\varepsilon _{q,3}\right)} \\
 + \frac{\frac{\left(\varepsilon _{q,2}-\varepsilon _{q,3}\right) \left(\varepsilon _{q,2}-E\right)}{\varepsilon _{q,1}-\varepsilon _{q,2}}+\frac{\left(E-\varepsilon _{q,3}\right){}^2 \log \left(\frac{E-\varepsilon _{q,3}}{E-\varepsilon _{q,2}}\right)}{\varepsilon _{q,1}-\varepsilon _{q,3}}}{2 N^2 \left(\varepsilon _{q,2}-\varepsilon _{q,3}\right){}^2} & \text{if } \varepsilon_{q, 1} < \varepsilon_{q, 2} < \varepsilon_{q, 3} \\
 0 & \text{otherwise}
\end{cases}\\
\text{if } \varepsilon_{q, 1} < E < \varepsilon_{q, 2} &\begin{cases}
\frac{2 \left(E-\varepsilon _{q,1}\right){}^2 \log \left(\frac{\varepsilon _{q,2}-E}{E-\varepsilon _{q,1}}\right)-\left(\varepsilon _{q,1}-\varepsilon _{q,2}\right) \left(-3 \varepsilon _{q,1}+\varepsilon _{q,2}+2 E\right)}{4 N^2 \left(\varepsilon _{q,1}-\varepsilon _{q,2}\right){}^2 \left(\varepsilon _{q,1}-\varepsilon _{q,3}\right)} & \text{if } \varepsilon_{q, 2} = \varepsilon_{q, 3}\\
\frac{\left(E-\varepsilon _{q,3}\right){}^2 \log \left(\frac{E-\varepsilon _{q,3}}{E-\varepsilon _{q,2}}\right)}{2 N^2 \left(\varepsilon _{q,1}-\varepsilon _{q,3}\right) \left(\varepsilon _{q,2}-\varepsilon _{q,3}\right){}^2} \\
+ \frac{\frac{\left(\varepsilon _{q,1}-\varepsilon _{q,2}\right) \left(\varepsilon _{q,2}-E\right)}{\varepsilon _{q,2}-\varepsilon _{q,3}}+\frac{\left(E-\varepsilon _{q,1}\right){}^2 \log \left(\frac{\varepsilon _{q,2}-E}{E-\varepsilon _{q,1}}\right)}{\varepsilon _{q,1}-\varepsilon _{q,3}}}{2 N^2 \left(\varepsilon _{q,1}-\varepsilon _{q,2}\right){}^2} & \text{otherwise }
\end{cases}\\
\text{if } \varepsilon_{q, 2} < E < \varepsilon_{q, 3} &\begin{cases}
\frac{\left(\varepsilon _{q,3}-\varepsilon _{q,2}\right) \left(\varepsilon _{q,2}-3 \varepsilon _{q,3}+2 E\right)+2 \left(E-\varepsilon _{q,3}\right){}^2 \log \left(\frac{\varepsilon _{q,3}-E}{E-\varepsilon _{q,2}}\right)}{4 N^2 \left(\varepsilon _{q,1}-\varepsilon _{q,3}\right) \left(\varepsilon _{q,2}-\varepsilon _{q,3}\right){}^2} & \text{if } \varepsilon_{q,1} = \varepsilon_{q, 2}\\
\frac{\left(E-\varepsilon _{q,1}\right){}^2 \log \left(\frac{E-\varepsilon _{q,2}}{E-\varepsilon _{q,1}}\right)}{2 N^2 \left(\varepsilon _{q,1}-\varepsilon _{q,2}\right){}^2 \left(\varepsilon _{q,1}-\varepsilon _{q,3}\right)} \\
+ \frac{\frac{\left(\varepsilon _{q,2}-\varepsilon _{q,3}\right) \left(\varepsilon _{q,2}-E\right)}{\varepsilon _{q,1}-\varepsilon _{q,2}}+\frac{\left(E-\varepsilon _{q,3}\right){}^2 \log \left(\frac{\varepsilon _{q,3}-E}{E-\varepsilon _{q,2}}\right)}{\varepsilon _{q,1}-\varepsilon _{q,3}}}{2 N^2 \left(\varepsilon _{q,2}-\varepsilon _{q,3}\right){}^2} & \text{otherwise}
\end{cases} \\
\text{if } \varepsilon_{q, 3} < E  &\begin{cases}
\frac{2 \left(E-\varepsilon _{q,1}\right){}^2 \log \left(\frac{E-\varepsilon _{q,2}}{E-\varepsilon _{q,1}}\right)-\left(\varepsilon _{q,1}-\varepsilon _{q,2}\right) \left(-3 \varepsilon _{q,1}+\varepsilon _{q,2}+2 E\right)}{4 N^2 \left(\varepsilon _{q,1}-\varepsilon _{q,2}\right){}^2 \left(\varepsilon _{q,1}-\varepsilon _{q,3}\right)} & \text{if } \varepsilon_{q, 1} < \varepsilon_{q, 2} = \varepsilon_{q, 3} \\
\frac{\left(\varepsilon _{q,3}-\varepsilon _{q,2}\right) \left(\varepsilon _{q,2}-3 \varepsilon _{q,3}+2 E\right)+2 \left(E-\varepsilon _{q,3}\right){}^2 \log \left(\frac{E-\varepsilon _{q,3}}{E-\varepsilon _{q,2}}\right)}{4 N^2 \left(\varepsilon _{q,1}-\varepsilon _{q,3}\right) \left(\varepsilon _{q,2}-\varepsilon _{q,3}\right){}^2} & \text{if } \varepsilon_{q, 1} = \varepsilon_{q, 2} < \varepsilon_{q, 3} \\
\frac{\left(E-\varepsilon _{q,1}\right){}^2 \log \left(\frac{E-\varepsilon _{q,2}}{E-\varepsilon _{q,1}}\right)}{2 N^2 \left(\varepsilon _{q,1}-\varepsilon _{q,2}\right){}^2 \left(\varepsilon _{q,1}-\varepsilon _{q,3}\right)} \\
+ \frac{\frac{\left(\varepsilon _{q,2}-\varepsilon _{q,3}\right) \left(\varepsilon _{q,2}-E\right)}{\varepsilon _{q,1}-\varepsilon _{q,2}}+\frac{\left(E-\varepsilon _{q,3}\right){}^2 \log \left(\frac{E-\varepsilon _{q,3}}{E-\varepsilon _{q,2}}\right)}{\varepsilon _{q,1}-\varepsilon _{q,3}}}{2 N^2 \left(\varepsilon _{q,2}-\varepsilon _{q,3}\right){}^2} & \text{if }  \varepsilon_{q, 1} < \varepsilon_{q, 2} < \varepsilon_{q, 3} \\
0 & \text{otherwise}
\end{cases}
\end{cases},
\end{equation}

\begin{equation}
\tilde{\mathcal{W}}_{q, \tau, 3}(E) = \begin{cases}
\text{if } E < \varepsilon_{q, 1} &\begin{cases}
\frac{2 \left(E-\varepsilon _{q,1}\right){}^2 \log \left(\frac{E-\varepsilon _{q,2}}{E-\varepsilon _{q,1}}\right)-\left(\varepsilon _{q,1}-\varepsilon _{q,2}\right) \left(-3 \varepsilon _{q,1}+\varepsilon _{q,2}+2 E\right)}{4 N^2 \left(\varepsilon _{q,1}-\varepsilon _{q,2}\right) \left(\varepsilon _{q,1}-\varepsilon _{q,3}\right){}^2} & \text{if } \varepsilon_{q, 1} < \varepsilon_{q, 2} = \varepsilon_{q, 3} \\
-\frac{-\varepsilon _{q,3} \left(\varepsilon _{q,2}+4 E\right)+\varepsilon _{q,2} \left(\varepsilon _{q,2}+2 E\right)+\varepsilon _{q,1} \left(-3 \varepsilon _{q,2}+\varepsilon _{q,3}+2 E\right)+2 \varepsilon _{q,3}^2}{4 N^2 \left(\varepsilon _{q,1}-\varepsilon _{q,3}\right){}^2 \left(\varepsilon _{q,3}-\varepsilon _{q,2}\right)} \\
+ \frac{\left(E-\varepsilon _{q,3}\right) \left(-2 E \varepsilon _{q,3}+\varepsilon _{q,2} \left(\varepsilon _{q,3}+E\right)+\varepsilon _{q,1} \left(-2 \varepsilon _{q,2}+\varepsilon _{q,3}+E\right)\right) \log \left(\frac{E-\varepsilon _{q,2}}{E-\varepsilon _{q,3}}\right)}{2 N^2 \left(\varepsilon _{q,1}-\varepsilon _{q,3}\right){}^2 \left(\varepsilon _{q,2}-\varepsilon _{q,3}\right){}^2} & \text{if } \varepsilon_{q, 1} = \varepsilon_{q, 2} < \varepsilon_{q, 3}\\
\frac{\left(E-\varepsilon _{q,1}\right){}^2 \log \left(\frac{E-\varepsilon _{q,2}}{E-\varepsilon _{q,1}}\right)}{2 N^2 \left(\varepsilon _{q,1}-\varepsilon _{q,2}\right) \left(\varepsilon _{q,1}-\varepsilon _{q,3}\right){}^2} 
+ \frac{E-\varepsilon _{q,3}}{2 N^2 \left(\varepsilon _{q,3}-\varepsilon _{q,1}\right) \left(\varepsilon _{q,3}-\varepsilon _{q,2}\right)} \\
+ \frac{\left(E-\varepsilon _{q,3}\right) \left(-2 E \varepsilon _{q,3}+\varepsilon _{q,2} \left(\varepsilon _{q,3}+E\right)+\varepsilon _{q,1} \left(-2 \varepsilon _{q,2}+\varepsilon _{q,3}+E\right)\right) \log \left(\frac{E-\varepsilon _{q,2}}{E-\varepsilon _{q,3}}\right)}{2 N^2 \left(\varepsilon _{q,1}-\varepsilon _{q,3}\right){}^2 \left(\varepsilon _{q,2}-\varepsilon _{q,3}\right){}^2} & \text{if } \varepsilon_{q, 1} < \varepsilon_{q, 2} < \varepsilon_{q, 3}\\
0 & \text{otherwise}
\end{cases}\\
\text{if } \varepsilon_{q, 1} < E < \varepsilon_{q, 2} &\begin{cases}
\frac{2 \left(E-\varepsilon _{q,1}\right){}^2 \log \left(\frac{\varepsilon _{q,2}-E}{E-\varepsilon _{q,1}}\right)-\left(\varepsilon _{q,1}-\varepsilon _{q,2}\right) \left(-3 \varepsilon _{q,1}+\varepsilon _{q,2}+2 E\right)}{4 N^2 \left(\varepsilon _{q,1}-\varepsilon _{q,2}\right) \left(\varepsilon _{q,1}-\varepsilon _{q,3}\right){}^2} & \text{if }\varepsilon_{q, 2} = \varepsilon_{q, 3} \\
\frac{\left(E-\varepsilon _{q,3}\right) \left(-2 E \varepsilon _{q,3}+\varepsilon _{q,2} \left(\varepsilon _{q,3}+E\right)+\varepsilon _{q,1} \left(-2 \varepsilon _{q,2}+\varepsilon _{q,3}+E\right)\right) \log \left(\frac{E-\varepsilon _{q,2}}{E-\varepsilon _{q,3}}\right)}{2 N^2 \left(\varepsilon _{q,1}-\varepsilon _{q,3}\right){}^2 \left(\varepsilon _{q,2}-\varepsilon _{q,3}\right){}^2} \\
+ \frac{\left(\varepsilon _{q,3}-\varepsilon _{q,2}\right) \left(\left(\varepsilon _{q,1}-\varepsilon _{q,2}\right) \left(\varepsilon _{q,1}-\varepsilon _{q,3}\right) \left(\varepsilon _{q,3}-E\right)+\left(\varepsilon _{q,3}-\varepsilon _{q,2}\right) \left(E-\varepsilon _{q,1}\right){}^2 \log \left(\frac{\varepsilon _{q,2}-E}{E-\varepsilon _{q,1}}\right)\right)}{2 N^2 \left(\varepsilon _{q,1}-\varepsilon _{q,2}\right) \left(\varepsilon _{q,1}-\varepsilon _{q,3}\right){}^2 \left(\varepsilon _{q,2}-\varepsilon _{q,3}\right){}^2} & \text{otherwise }
\end{cases}\\
\text{if } \varepsilon_{q, 2} < E < \varepsilon_{q, 3} &\begin{cases}
-\frac{-\varepsilon _{q,3} \left(\varepsilon _{q,2}+4 E\right)+\varepsilon _{q,2} \left(\varepsilon _{q,2}+2 E\right)+\varepsilon _{q,1} \left(-3 \varepsilon _{q,2}+\varepsilon _{q,3}+2 E\right)+2 \varepsilon _{q,3}^2}{4 N^2 \left(\varepsilon _{q,1}-\varepsilon _{q,3}\right){}^2 \left(\varepsilon _{q,3}-\varepsilon _{q,2}\right)} \\
+ \frac{\left(\varepsilon _{q,3}-E\right) \left(-2 E \varepsilon _{q,3}+\varepsilon _{q,2} \left(\varepsilon _{q,3}+E\right)+\varepsilon _{q,1} \left(-2 \varepsilon _{q,2}+\varepsilon _{q,3}+E\right)\right) \log \left(\frac{\varepsilon _{q,3}-E}{E-\varepsilon _{q,2}}\right)}{2 N^2 \left(\varepsilon _{q,1}-\varepsilon _{q,3}\right){}^2 \left(\varepsilon _{q,2}-\varepsilon _{q,3}\right){}^2} & \text{if } \varepsilon_{q, 1} = \varepsilon_{q, 2}\\
\frac{\frac{\left(\varepsilon _{q,2}-\varepsilon _{q,1}\right) \left(\varepsilon _{q,1}-\varepsilon _{q,3}\right) \left(\varepsilon _{q,3}-E\right)}{\varepsilon _{q,2}-\varepsilon _{q,3}}+\left(E-\varepsilon _{q,1}\right){}^2 \log \left(\frac{E-\varepsilon _{q,2}}{E-\varepsilon _{q,1}}\right)}{2 N^2 \left(\varepsilon _{q,1}-\varepsilon _{q,2}\right) \left(\varepsilon _{q,1}-\varepsilon _{q,3}\right){}^2} \\
+ \frac{\left(\varepsilon _{q,3}-E\right) \left(-2 E \varepsilon _{q,3}+\varepsilon _{q,2} \left(\varepsilon _{q,3}+E\right)+\varepsilon _{q,1} \left(-2 \varepsilon _{q,2}+\varepsilon _{q,3}+E\right)\right) \log \left(\frac{\varepsilon _{q,3}-E}{E-\varepsilon _{q,2}}\right)}{2 N^2 \left(\varepsilon _{q,1}-\varepsilon _{q,3}\right){}^2 \left(\varepsilon _{q,2}-\varepsilon _{q,3}\right){}^2} & \text{otherwise }
\end{cases} \\
\text{if } \varepsilon_{q, 3} < E  &\begin{cases}
\frac{2 \left(E-\varepsilon _{q,1}\right){}^2 \log \left(\frac{E-\varepsilon _{q,2}}{E-\varepsilon _{q,1}}\right)-\left(\varepsilon _{q,1}-\varepsilon _{q,2}\right) \left(-3 \varepsilon _{q,1}+\varepsilon _{q,2}+2 E\right)}{4 N^2 \left(\varepsilon _{q,1}-\varepsilon _{q,2}\right) \left(\varepsilon _{q,1}-\varepsilon _{q,3}\right){}^2} & \text{if } \varepsilon_{q, 1} < \varepsilon_{q, 2} = \varepsilon_{q, 3}\\
-\frac{-\varepsilon _{q,3} \left(\varepsilon _{q,2}+4 E\right)+\varepsilon _{q,2} \left(\varepsilon _{q,2}+2 E\right)+\varepsilon _{q,1} \left(-3 \varepsilon _{q,2}+\varepsilon _{q,3}+2 E\right)+2 \varepsilon _{q,3}^2}{4 N^2 \left(\varepsilon _{q,1}-\varepsilon _{q,3}\right){}^2 \left(\varepsilon _{q,3}-\varepsilon _{q,2}\right)}\\ 
+ \frac{\left(E-\varepsilon _{q,3}\right) \left(-2 E \varepsilon _{q,3}+\varepsilon _{q,2} \left(\varepsilon _{q,3}+E\right)+\varepsilon _{q,1} \left(-2 \varepsilon _{q,2}+\varepsilon _{q,3}+E\right)\right) \log \left(\frac{E-\varepsilon _{q,2}}{E-\varepsilon _{q,3}}\right)}{2 N^2 \left(\varepsilon _{q,1}-\varepsilon _{q,3}\right){}^2 \left(\varepsilon _{q,2}-\varepsilon _{q,3}\right){}^2} & \text{if } \varepsilon_{q, 1} = \varepsilon_{q, 2} < \varepsilon_{q, 3}\\
\frac{\frac{\left(\varepsilon _{q,2}-\varepsilon _{q,1}\right) \left(\varepsilon _{q,1}-\varepsilon _{q,3}\right) \left(\varepsilon _{q,3}-E\right)}{\varepsilon _{q,2}-\varepsilon _{q,3}}+\left(E-\varepsilon _{q,1}\right){}^2 \log \left(\frac{E-\varepsilon _{q,2}}{E-\varepsilon _{q,1}}\right)}{2 N^2 \left(\varepsilon _{q,1}-\varepsilon _{q,2}\right) \left(\varepsilon _{q,1}-\varepsilon _{q,3}\right){}^2} \\
+ \frac{\left(E-\varepsilon _{q,3}\right) \left(-2 E \varepsilon _{q,3}+\varepsilon _{q,2} \left(\varepsilon _{q,3}+E\right)+\varepsilon _{q,1} \left(-2 \varepsilon _{q,2}+\varepsilon _{q,3}+E\right)\right) \log \left(\frac{E-\varepsilon _{q,2}}{E-\varepsilon _{q,3}}\right)}{2 N^2 \left(\varepsilon _{q,1}-\varepsilon _{q,3}\right){}^2 \left(\varepsilon _{q,2}-\varepsilon _{q,3}\right){}^2} &\text{if } \varepsilon_{q, 1} < \varepsilon_{q, 2} < \varepsilon_{q, 3}\\
0 & \text{otherwise} 
\end{cases}
\end{cases}.
\end{equation}

Equations~\eqref{eq: the imaginary part} and \eqref{eq: the real part} together give a numerical approximate to the full Green's function. The continuum theory is formally an infinite-band theory, and so the Green's function calculated from it should, in principle, contain a sum over an infinite number of bands. However, since only states near the Fermi surface contribute significantly to the Green's function, in all of our calculations, unless otherwise noted, we only include the two flat bands. This is because we are working with energies close to the Dirac cones. Once we have the bare Green's function, we calculate the change to LDOS for a particular impurity potential. To observe  the wavefront dislocation in the change in LDOS, we need to filter out only wavevector components near to the relevant Dirac cones. We do so numerically using the fast Fourier transform (FFT) and applying a filter that picks out conjugate momentum pairs around the relevant Dirac cones \cite{DG19}.

\begin{figure}
\includegraphics[scale=0.7]{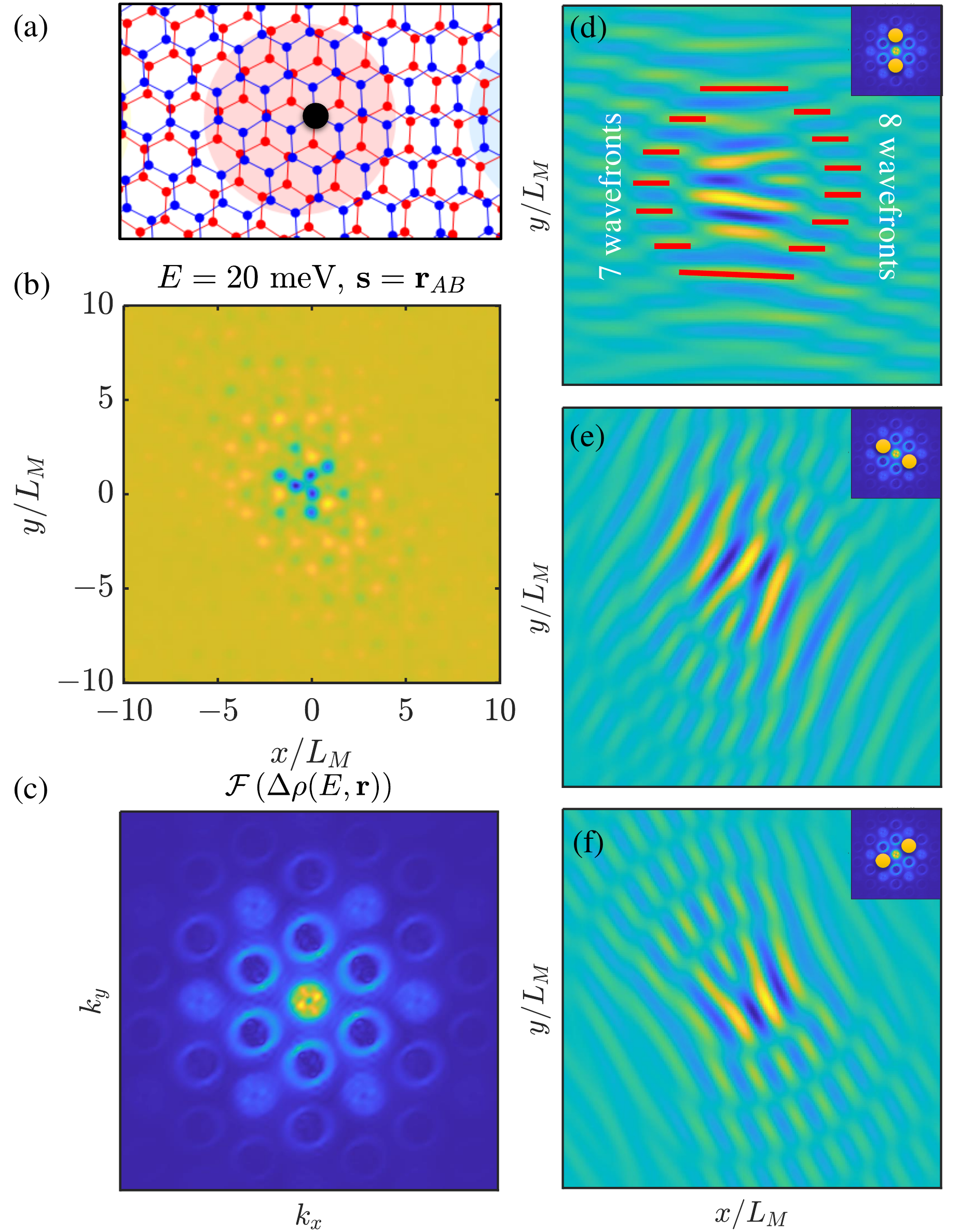}
\caption{(a) Location of an atomic impurity placed on an $A_1$ site in an $AB$ region. (b) Simulation of the change in LDOS due to the presence of an impurity at a bias energy of $E = 20$ meV. (c) The absolute value of the FFT of the interference pattern in (b). The approximate threefold rotational symmetry in (b) is reflected in the amplitude pattern of the Fourier transform. (d)-(f) The LDOS profiles once an FFT has been applied to pick out only the conjugate pairs of momentum indicated by the bright dots in the insets located at the top right corners. In all three directions of momentum scattering, we observe a single wavefront dislocation. The magnitude of the density oscillations is recorded in arbitrary units, chosen to give the appropriate contrast. The parameters for the simulation are: $\theta = 2^\circ,$ $w_\text{AA} = 79.7$ meV, $w_\text{AB} = 97.5$ meV, and $U_0 = 1$ eV, and $V = 0$ meV. }
\label{fig: Friedel 1}
\end{figure}

\begin{figure}
\includegraphics[scale=0.7]{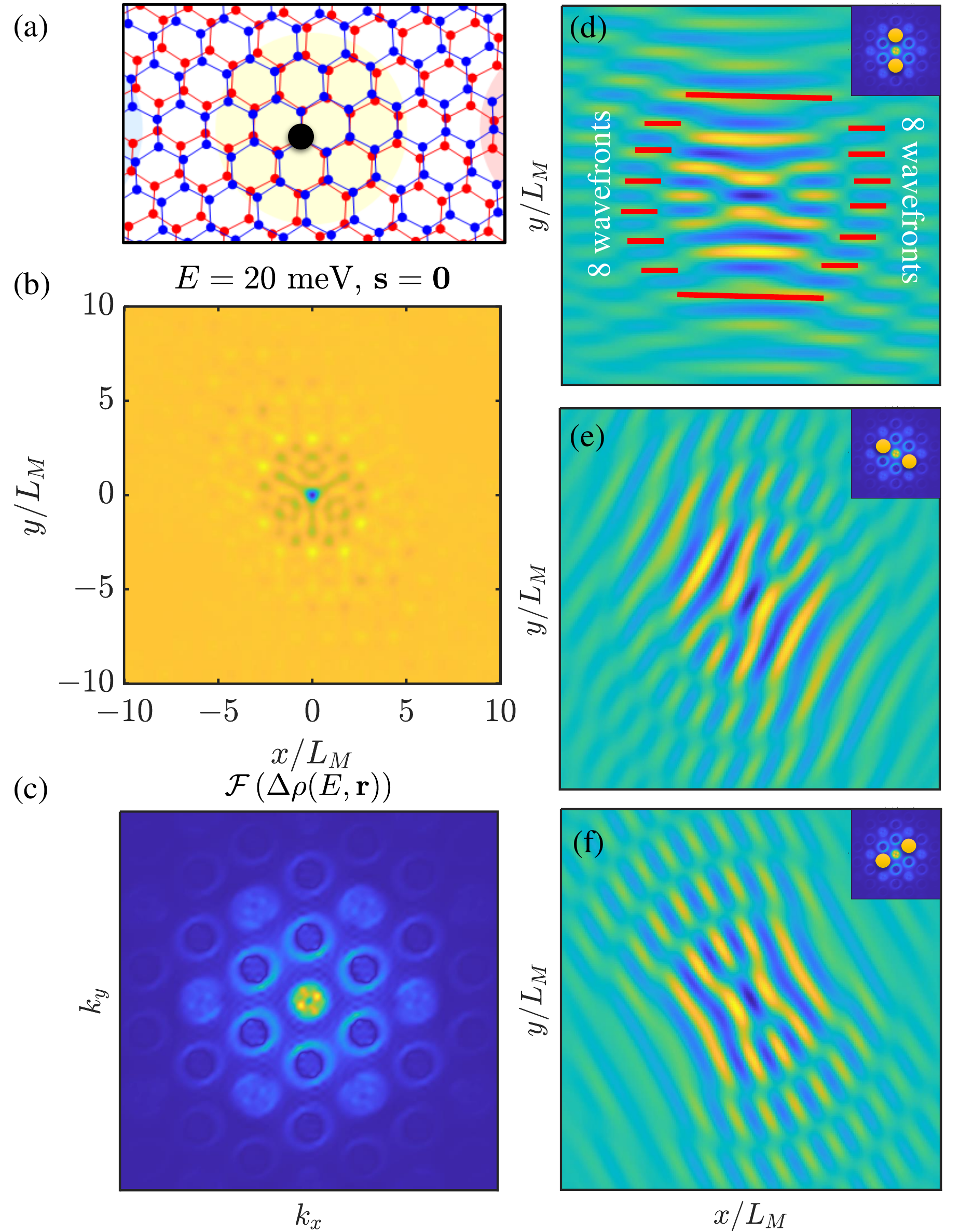}
\caption{(a) Location of an atomic impurity placed on an $A_1$ site in an $AA$ region. (b) Simulation of the change in LDOS due to the presence of an impurity at a bias energy of $E = 20$ meV. (c) The absolute value of the FFT of the interference pattern in (b). The approximate threefold rotational symmetry in (b) is reflected in the amplitude pattern of the Fourier transform. (d)-(f) The LDOS profiles once an FFT has been applied to pick out only the conjugate pairs of momentum indicated by the bright dots in the insets located at the top right corners. In all three directions of momentum scattering, we observe no wavefront dislocation. The magnitude of the density oscillations is recorded in arbitrary units, chosen to give the appropriate contrast. The parameters for the simulation are: $\theta = 2^\circ,$ $w_\text{AA} = 79.7$ meV, $w_\text{AB} = 97.5$ meV, and $U_0 = 1$ eV, and $V = 0$ meV.}
\label{fig: Friedel 2}
\end{figure}

\begin{figure}
\includegraphics[scale=0.7]{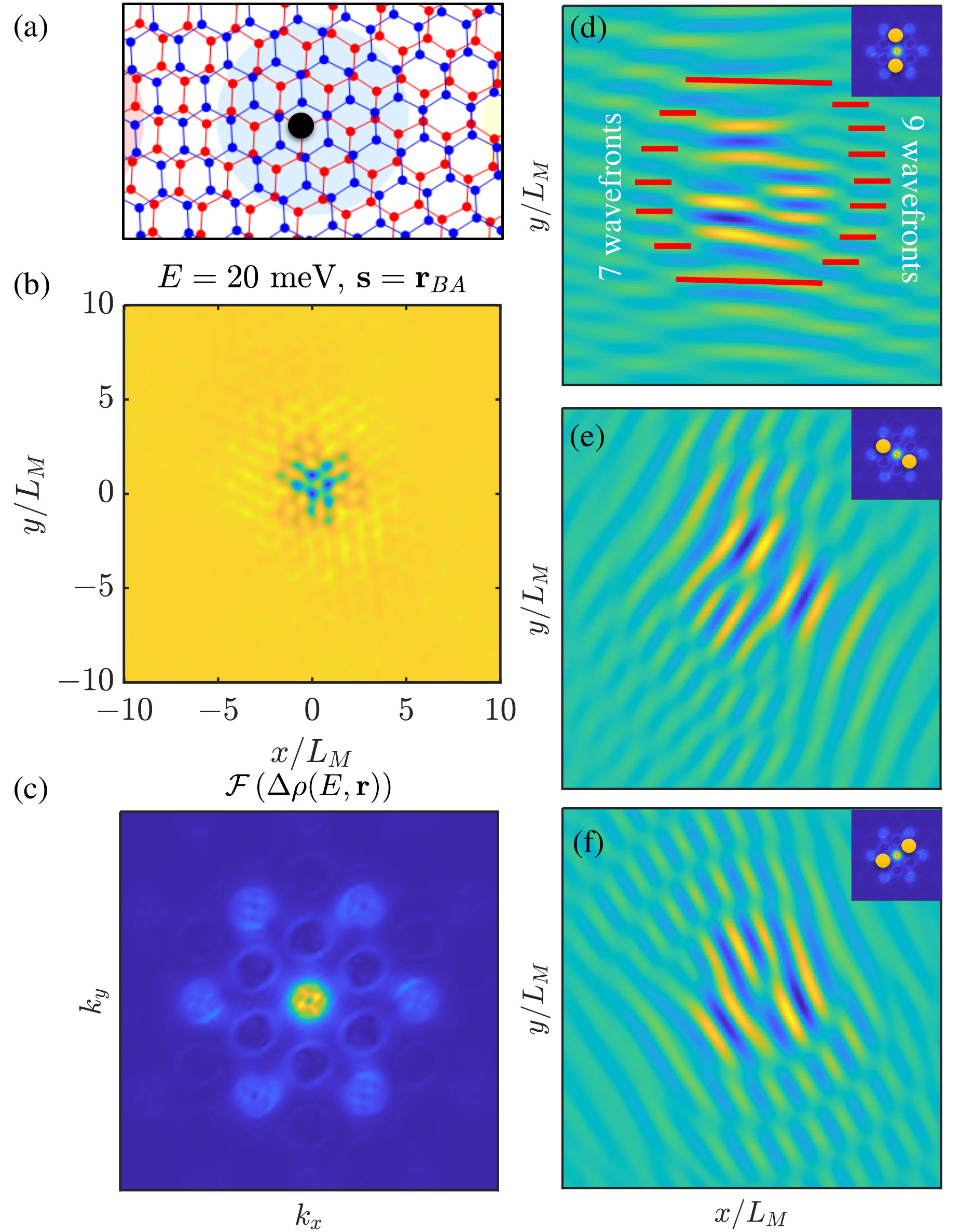}
\caption{(a) Location of an atomic impurity placed on an $A_1$ site in an $BA$ region. (b) Simulation of the change in LDOS due to the presence of an impurity at a bias energy of $E = 20$ meV. (c) The absolute value of the FFT of the interference pattern in (b). The approximate threefold rotational symmetry in (b) is reflected in the amplitude pattern of the Fourier transform. (d)-(f) The LDOS profiles once an FFT has been applied to pick out only the conjugate pairs of momentum indicated by the bright dots in the insets located at the top right corners. In all three directions of momentum scattering, we observe two  wavefront dislocations. The magnitude of the density oscillations is recorded in arbitrary units, chosen to give the appropriate contrast. However, the magnitude of the oscillations is at least an order of magnitude smaller than those observed for an impurity placed near the $AB$ or $AA$ region. The parameters for the simulation are: $\theta = 2^\circ,$ $w_\text{AA} = 79.7$ meV, $w_\text{AB} = 97.5$ meV, and $U_0 = 1$ eV, and $V = 0$ meV.}
\label{fig: Friedel 3}
\end{figure}

\begin{figure}
\includegraphics[scale=0.7]{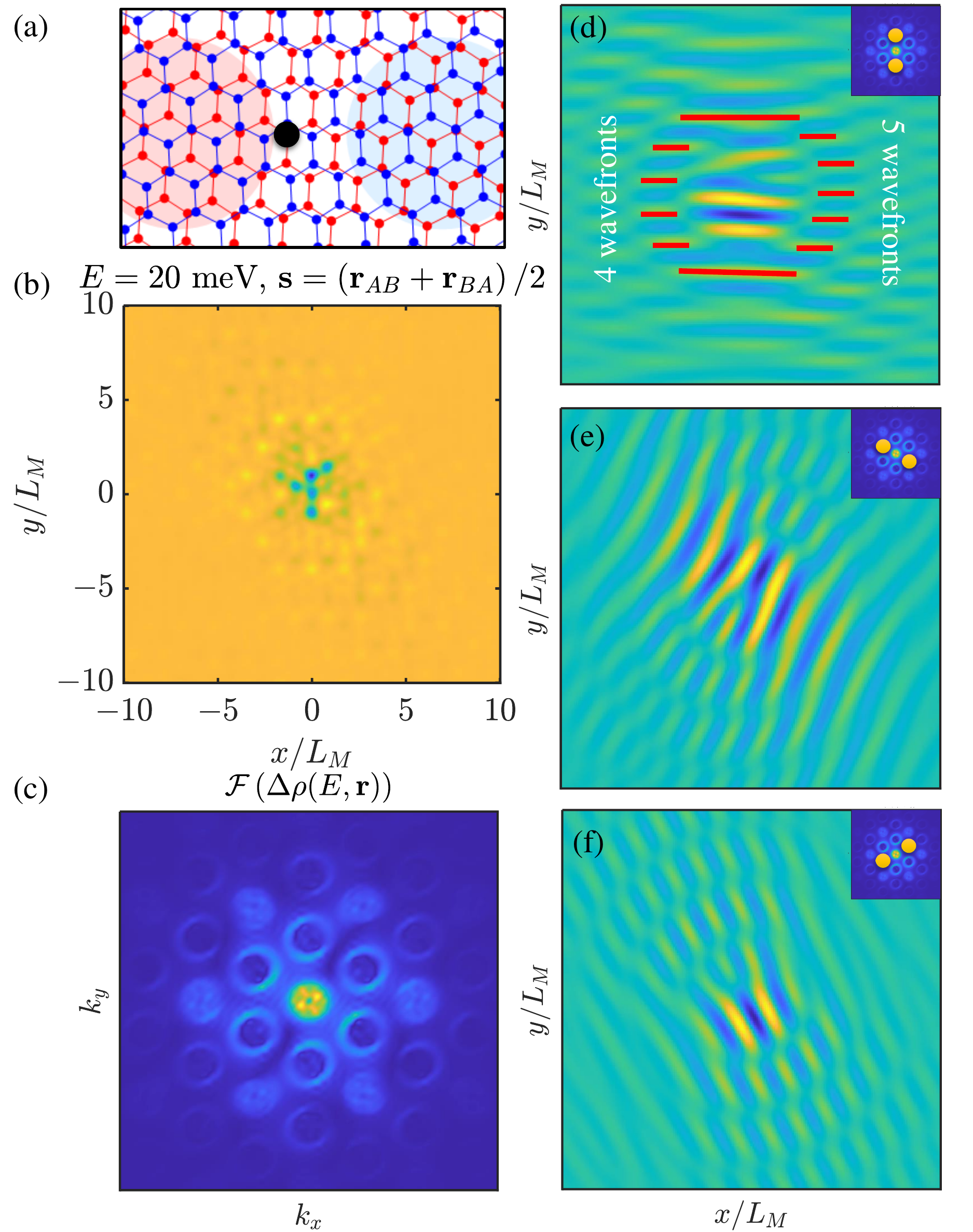}
\caption{(a) Location of an atomic impurity placed on an $A_1$ site in a saddle-point region, halfway between its adjacent $AB$ and $BA$ regions. (b) Simulation of the change in LDOS due to the presence of an impurity at a bias energy of $E = 20$ meV. (c) The absolute value of the FFT of the interference pattern in (b). The approximate threefold rotational symmetry in (b) is reflected in the amplitude pattern of the Fourier transform. (d)-(f) The LDOS profiles once an FFT has been applied to pick out only the conjugate pairs of momentum indicated by the bright dots in the insets located at the top right corners. In all three directions of momentum scattering, we observe a single wavefront dislocation. The magnitude of the density oscillations is recorded in arbitrary units, chosen to give the appropriate contrast. The parameters for the simulation are: $\theta = 2^\circ,$ $w_\text{AA} = 79.7$ meV, $w_\text{AB} = 97.5$ meV, $U_0 = 1$ eV, and $V = 0$ meV.}
\label{fig: Friedel 4}
\end{figure}

\begin{figure}
\includegraphics[scale=0.5]{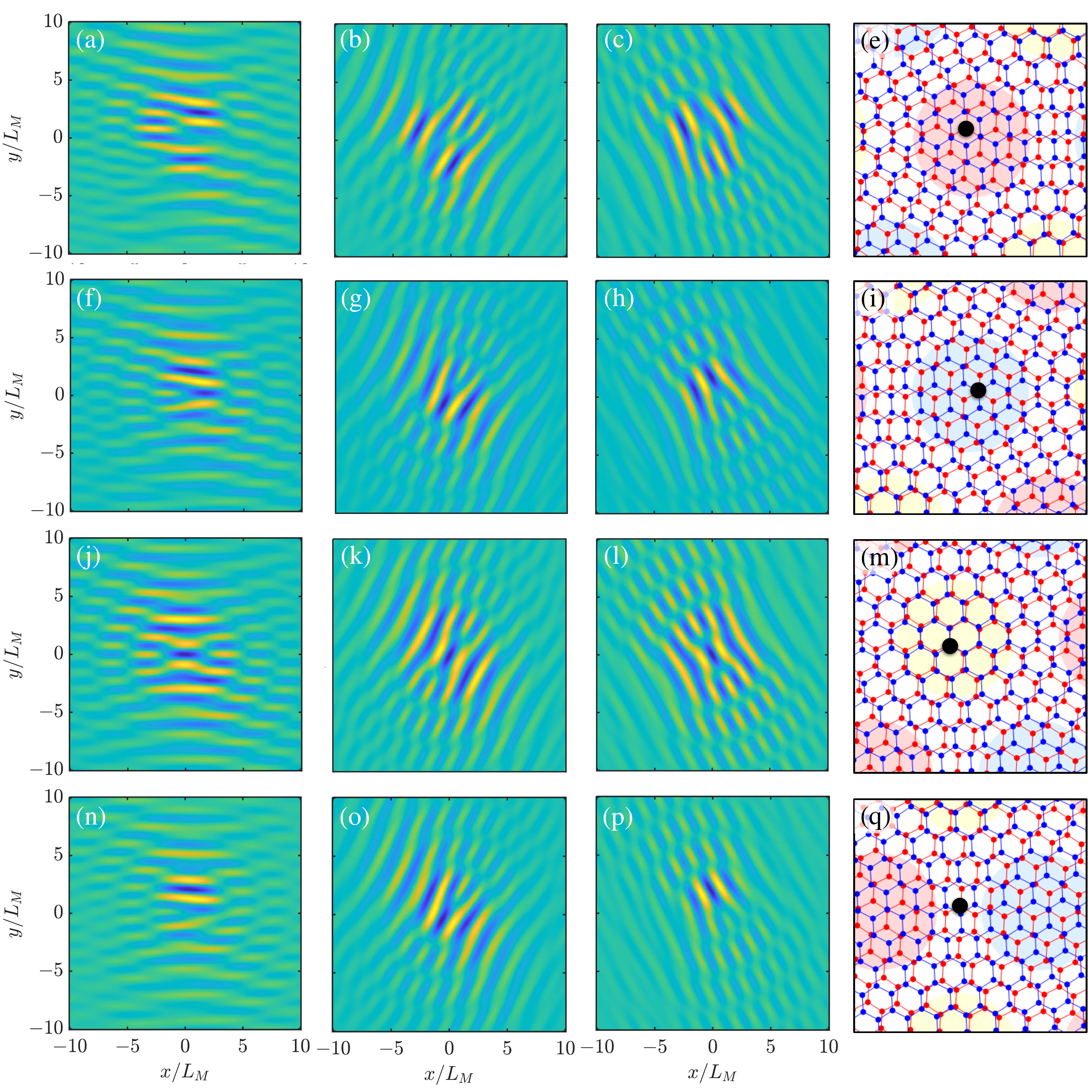}
\caption{The FFT-filtered LDOS for an impurity placed on a $B_1$ site at a bias energy of $E = 20$ meV. Each row corresponds to results obtained for a different location within a unit cell on which the impurity is located. (e), (i), (m), and (q) show the location of the impurity for each row. (a)-(c) Simulation of the change in LDOS due to the presence of an impurity at an $AB$ site. We observe two dislocations in LDOS. (f)-(h) Simulation of the change in LDOS due to the presence of an impurity at a $BA$ site. We observe one dislocation in LDOS for this case. (j)-(l) Simulation of the change in LDOS due to the presence of an impurity at an $AA$ site. We observe no dislocation in LDOS. Finally, (n)-(p) Simulation of the change in LDOS due to the presence of an impurity at a saddle-point location. Here, we observe one dislocation in LDOS. The magnitude of the density oscillations is recorded in arbitrary units, chosen to give the appropriate contrast. The parameters for the simulation are: $\theta = 2^\circ,$ $w_\text{AA} = 79.7$ meV, $w_\text{AB} = 97.5$ meV, $U_0 = 1$ eV, and $V = 0$ meV.}
\label{fig: Friedel 5}
\end{figure}

The results for the induced LDOS, summed over both valleys, when the impurity is placed on an $A_1$ site at various locations within a moiré unit cell are shown in Figs. \ref{fig: Friedel 1}, \ref{fig: Friedel 2}, \ref{fig: Friedel 3}, and \ref{fig: Friedel 4}. In Fig. \ref{fig: Friedel 1}b, we show the interference pattern of the induced change in LDOS when the atomic impurity resides within an $AB$ region with magnitude $U_0 = 1$ eV. We then apply the FFT to LDOS and plot the magnitude of the resulting Fourier components in Fig.~\ref{fig: Friedel 1}c. The approximate threefold rotation symmetry is reflected in the hexagonal distribution of the Fourier components in reciprocal space. To observe the presence of dislocations in LDOS, we filter out only conjugate pairs of momentum states near the relevant Dirac cones. This procedure restricts the domain of scattering wavevectors, and describes scattering processes that occur only between momentum states near pairs of Dirac cones that are related by time reversal. There are three such pairs that give rise to wavefronts traveling in three inequivalent directions, as shown in Fig. \ref{fig: Friedel 1}d, e, and f. So far, we have only been concerned with scattering processes that occur between Dirac cones which reside within the same valley. A sufficiently localized impurity in principle can induce intervalley scattering events. However, these processes will induce oscillations in LDOS that vary on the atomic scale. Thus, if we are only concerned about oscillations that occur on the moiré scale, then the intervalley terms can be neglected. The FFT filter once applied to only Dirac cones near to $\Gamma$ will eliminate large wavevectors that correspond to atomic-scale oscillations in real space. This justifies our neglect of intervalley scattering. For the case of an $A_1$ impurity placed in an $AB$ region, the FFT-filtered LDOS, as Fig. \ref{fig: Friedel 1}d, e, and f, features a single wavefront dislocation in all three scattering directions, consistent with the prediction of the effective low-energy Dirac theory previously studied. The same analysis is applied to an $A_1$ impurity placed at an $AA$ region in Fig.~\ref{fig: Friedel 2}. Here, we observe no dislocation, as predicted by the Dirac theory. In Fig.~\ref{fig: Friedel 3}, we show the results for an impurity placed in an $BA$ region. In this case, the Dirac theory predicts that LDOS oscillations are exactly zero. Meanwhile, in numerical simulation, we find two dislocations in the interference pattern. However, the amplitude of oscillations are numerically found to be at least one order of magnitude smaller than those obtained from the impurity placed at the $AB$ and $AA$ regions. This means that the wavefront dislocationsn	 in Fig.~\ref{fig: Friedel 3} is due to contributions to the Green's function beyond the linear-dispersion regime. Finally, we also analyze the situation for an $A_1$ impurity placed at a saddle-point region between adjacent $AB$ and $BA$ regions. The LDOS hosts a single dislocation in this case. To confirm our numerical simulation, we also simulate the LDOS induced by an impurity placed on the $B_1$ sublattice. The results are shown in Fig.~\ref{fig: Friedel 5}.

\end{document}